\documentclass[%
reprint,%
superscriptaddress,%
showkeys,%
aps,%
pra,%
longbibliography,%
floatfix,%
nofootinbib%
]{revtex4-2}

\usepackage[utf8]{inputenc}

\usepackage[hyperref]{xcolor}
	\definecolor{goethe-blau}{cmyk}{1.0,0.2,0.0,0.4}
	\definecolor{hellgrau}{cmyk}{0.04,0.04,0.05,0.02}
	\definecolor{sandgrau}{cmyk}{0.12,0.09,0.13,0.0}
	\definecolor{dunkelgrau}{cmyk}{0.25,0.25,0.30,0.75}
	\definecolor{emo-rot}{cmyk}{0.04,1.0,0.8,0.07}
	\definecolor{purple}{cmyk}{0.08,1.0,0.3,0.36}
	\definecolor{senfgelb}{cmyk}{0.01,0.25,1.0,0.05}
	\definecolor{gruen}{cmyk}{0.62,0.4,0.87,0.09}
	\definecolor{magenta}{cmyk}{0.08,0.86,0.12,0.12}
	\definecolor{orange}{cmyk}{0.0,0.7,1.0,0.04}
	\definecolor{sonnengelb}{cmyk}{0.0,0.12,0.95,0.0}
	\definecolor{helles-gruen}{cmyk}{0.4,0.17,0.81,0.07}
	\definecolor{lichtblau}{cmyk}{0.8,0.0,0.06,0.04}

\usepackage[%
colorlinks,%
pdfpagelabels,%
breaklinks,%
pdfstartview=FitH,%
bookmarksopen=true,%
bookmarksnumbered=true,%
bookmarksopenlevel=2,%
plainpages=false,%
hypertexnames=false,%
citecolor=emo-rot,%
linkcolor=goethe-blau,%
urlcolor=purple,%
pdftitle={RG flows in a multi-dimensional field space},%
pdfauthor={Niklas Zorbach, Adrian Koenigstein, Jens Braun},%
pdfkeywords={FRG, numerical fluid dynamics, finite volume method, advection-diffusion equation, field space}%
]{hyperref}

\usepackage[acronym]{glossaries-extra}
\glssetcategoryattribute{acronym}{nohyperfirst}{true}
\setabbreviationstyle[acronym]{long-short}

\newacronym{cfd}{CFD}{computational fluid dynamics}
\newacronym{rg}{RG}{Renormalization Group}
\newacronym{frg}{FRG}{Functional Renormalization Group}
\newacronym{fv}{FV}{finite volume}
\newacronym{qft}{QFT}{Quantum Field Theory}
\newacronym{uv}{UV}{ultraviolet}
\newacronym{ir}{IR}{infrared}
\newacronym{lpa}{LPA}{local potential approximation}
\newacronym{qcd}{QCD}{Quantum Chromodynamics}
\newacronym{kt}{KT}{Kurganov-Tadmor}
\newacronym{pde}{PDE}{partial differential equation}
\newacronym{ode}{ODE}{ordinary differential equation}
\newacronym{tvd}{TVD}{total variation diminishing}
\newacronym{tvni}{TVNI}{total variation nonincreasing}
\newacronym{wrt}{w.r.t.}{with respect to}
\newacronym{wlog}{w.l.o.g.}{without loss of the generality}
\newacronym{rhs}{r.h.s.}{right hand side}
\newacronym{lhs}{l.h.s.}{left hand side}

\usepackage{amsmath} 
\usepackage{amssymb} 
\usepackage{bm} 
\usepackage{dsfont}
\usepackage{slashed} 
\usepackage{tensor} 
\usepackage{upgreek} 

\usepackage{bookmark} 
\usepackage{comment} 
\usepackage{graphicx} 

\usepackage{placeins} 

\usepackage{booktabs} 
\usepackage{dcolumn} 

\usepackage[caption=false]{subfig} 

\usepackage{orcidlink}

\usepackage{tikz}
\usetikzlibrary{decorations.pathreplacing}

\allowdisplaybreaks 

\usepackage[capitalise,sort,compress]{cleveref}

	\newcommand{\vdistance}{\vphantom{\bigg(\bigg)}}
	\newcommand{\Vdistance}{\vphantom{\Bigg(\Bigg)}}
	
	\newcommand{\slice}{\!:\!}
	
	\newcommand{\etc}{\textit{etc.}}
	\newcommand{\ie}{\textit{i.e.}}
	\newcommand{\eg}{\textit{e.g.}}
	\newcommand{\cf}{\textit{cf.}}

	\newcommand{\reff}{Ref.~}
	\newcommand{\reffs}{Refs.~}

	\DeclareMathOperator{\adj}{adj}

	\newcommand{\dd}{\mathrm{d}}
	\newcommand{\ee}{\mathrm{e}}

\begin{document}

\title{Functional Renormalization Group meets Computational Fluid Dynamics:\texorpdfstring{\\}{ }RG flows in a multi-dimensional field space}

	\author{Niklas Zorbach \orcidlink{0000-0002-8434-5641}}
	\email{niklas.zorbach@tu-darmstadt.de}
	\affiliation{
		Technische Universität Darmstadt, Department of Physics, Institut für Kernphysik, Theoriezentrum,\\
		Schlossgartenstraße 2, D-64289 Darmstadt, Germany
	}

	\author{Adrian Koenigstein \orcidlink{0000-0001-7482-2195}}
	\email{adrian.koenigstein@uni-jena.de}
	\affiliation{
		Theoretisch-Physikalisches Institut, Friedrich-Schiller-Universität Jena
		\\
		Max-Wien-Platz 1, D-07743 Jena, Germany
	}

	\author{Jens Braun \orcidlink{0000-0003-4655-9072}}
	\email{jens.braun@tu-darmstadt.de}
	\affiliation{
		Technische Universität Darmstadt, Department of Physics, Institut für Kernphysik, Theoriezentrum,\\
		Schlossgartenstraße 2, D-64289 Darmstadt, Germany
	}
	\affiliation{
		Helmholtz Research Academy Hesse for FAIR, Campus Darmstadt,\\
		D-64289 Darmstadt, Germany
	}
	\affiliation{
		ExtreMe Matter Institute EMMI, GSI, Planckstra{\ss}e 1, D-64291 Darmstadt, Germany
	}

\begin{abstract}
	Within the Functional Renormalisation Group (FRG) approach, we present a fluid-dynamical approach to solving flow equations for models living in a multi-dimensional field space.
	To this end, the underlying exact flow equation of the effective potential is reformulated as a set of nonlinear advection-diffusion-type equations which can be solved using the Kurganov-Tadmor central scheme, a modern finite-volume discretization from computational fluid dynamics (CFD).
	We demonstrate the effectiveness of our approach by performing explicit benchmark tests using zero-dimensional models with two discretized field space directions or two symmetry invariants.
	Our techniques can be directly applied to flow equations of effective potentials of general (fermion-)boson systems with multiple invariants or condensates, as we also demonstrate for two concrete examples in three spacetime dimensions.
\end{abstract}


\maketitle

\tableofcontents

\section{Introduction}
\label{sec:introduction}

	Within the last decades lattice Monte-Carlo simulations turned out to be one of the most successful methods to study nonperturbative problems in \gls{qft}, such as \gls{qcd}.
	However, also Monte-Carlo simulations come with limitations.
	Especially in the context of critical phenomena, the study of phase transitions, and calculations at moderate and high densities, other nonperturbative approaches are required to complement lattice Monte-Carlo simulations.
	One of these methods is the \gls{frg} approach which is based on a flow equation for the quantum effective action, the Wetterich equation~\cite{Wetterich:1992yh}. 
	The formal derivation of this equation immediately triggered further field-theoretic developments~\cite{Ellwanger:1993mw,Morris:1993qb,Morris:1994ie}. 
	By now, this approach has been successfully applied to a wide range of problems in high-energy, condensed matter, statistical, and gravitational physics~\cite{Wetterich:2001kra,Berges:2000ew,Pawlowski:2005xe,Gies:2006wv,Kopietz:2010zz,Braun:2011pp,Dupuis:2020fhh}.
	One of the main advantages of the \gls{frg} is that it provides direct access to the effective action, which is directly linked to the vertex functions and thermodynamic observables of a system.

	A particularly important quantity is the effective potential, which represents the lowest-order contribution of an expansion of the effective action in derivatives of the fields and comprises all orders of local (point-like) interactions of the (effective) degrees of freedom of the system. 
	Usually, within \gls{frg} calculations, the effective potential is calculated by taking advantage of the symmetries of the system in field space.
	Hence, it oftentimes suffices to solve the \gls{rg} flow equations for the effective potential as a function of a single field variable or field invariant.
	However, in many cases the effective potential exhibits a lower degree of symmetry and the \gls{rg} flow equations have to be solved in a multi-dimensional field space.
	Typical situations arise in systems with multiple order parameters, condensates, and/or field invariants.
	Selected relevant examples from high-energy physics that have already been addressed with the \gls{frg} are mesonic models with chiral and diquark condensates \cite{Strodthoff:2011tz,Strodthoff:2013cua,Khan:2015puu,Lakaschus2021}, models that comprise strange quarks and their condensation in addition to light quarks \cite{Schaefer:2008hk,Mitter:2013fxa,Otto:2019zjy,Otto:2020hoz}, models that allow for pion and sigma condensation \cite{Kamikado:2012bt} and many more.
	Also in the context of condensed-matter physics, the effective potentials of many systems cannot be reduced to a single field-space direction.
	Some examples from \gls{frg} studies are systems that exhibit inverted phase diagrams and exhibit the Pomeranchuk effect \cite{Hawashin:2024dpp} or (frustrated) magnetic systems \cite{Tissier:1999hv,Tissier:2000tz,Tissier:2001de,Tissier:2001uk,Delamotte:2003dw,Delamotte:2003dw,Delamotte:2015zgf,Debelhoir:2016yfp,Yabunaka:2024tmc,Sanchez-Villalobos:2024vmd}.
	Other examples are models for $(2+1)$-dimensional Dirac materials such as graphene \cite{Classen:2015mar,Torres:2018jij}. 
	All of them can be approximately described via $O(N) \times O(M)$-symmetric models or in general models with two invariants.
	However, it should also be mentioned that probably the first \gls{frg} calculations with several invariants originated from the field of statistical physics, in particular studies of matrix models should be mentioned here \cite{Berges:1996ja}, see also \reffs\cite{Eichhorn:2013zza,Eichhorn:2016hdi,Eichhorn:2018phj} for a more recent analysis.

\subsection{Motivation and contextualization}
\label{sec:motivation_and_contextualization}

	Given these many examples, the reader might object that the problem of solving \gls{frg} flow equations in multi-dimensional field space is already solved and it seems unnecessary to address this problem in a separate work again.
	Indeed, several methods have been developed to study the \gls{rg} flow equations in multi-dimensional field space:
	The most common approach is to use a multi-dimensional Taylor expansion of the effective potential around a fixed or moving point in field space which is usually chosen to be the global \gls{ir} or flowing minimum of the effective potential, respectively.
	Here, one effectively reduces the problem to a set of coupled \glspl{ode} for the Taylor coefficients.
	Another approach is to use a grid-based method where the field space is discretized and the \gls{rg} flow equations are solved on this grid.
	Often, ``naive'' multi-dimensional finite difference schemes have been used to solve the \gls{pde} for the effective potential in the past. 
	However, it should also be mentioned that other approaches have been developed, such as a discretization of field space via spectral methods in terms of Chebyshev polynomials or by approximating the effective potential in terms of splines or other basis functions or a matching of Taylor expansions, see the references above for concrete examples.

	Mostly, these approaches are suited to address the respective problem at hand.
	Especially in the context of fixed-point searches and analyses or studies of second-order phase transitions, some of these techniques are very powerful.
	Though, in the context of first-order phase transitions and multiple competing minima in the effective potential, most of the methods are no longer unconditionally stable.
	For example, Taylor expansions certainly fail, if the correct expansion point jumps from one minimum to another during the flow or if the radius of convergence is limited by nonanalyticities in the effective potential.
	The latter can easily be caused by chemical potentials or other external fields.
	Also, grid-based methods can fail in such situations, if the discretization scheme is for example fundamentally based on the assumption of analyticity and smoothness of the effective potential.
	However, since all of these challenging situations are of high physical relevance, we decided to address the construction of a powerful and stable multi-purpose discretization scheme for \gls{frg} flow equations in multi-dimensional field space again.

	In this spirit, this paper addresses researchers who are interested in the challenges and the numerical techniques which are required to solve \gls{frg} flow equations of bosonic potentials with more than one background field.
	Furthermore, the manuscript addresses readers who are interested in the general connection between \gls{frg} flow equations and fluid dynamics.
	For a more detailed introduction to \gls{frg} via zero-dimensional \gls{qft} and recent developments on the connections between \gls{frg} and numerical fluid dynamics we refer the interested reader to \reffs\cite{Koenigstein:2021syz,Koenigstein:2021rxj,Steil:2021cbu} and some directly related works \cite{Grossi:2019urj,Grossi:2021ksl,Stoll:2021ori,Ihssen:2022xjv,Ihssen:2022xkr,Batini:2023nan,Ihssen:2023dmx,Ihssen:2023qaq,Ihssen:2023nqd,Ihssen:2023xlp,Ihssen:2024miv,Ihssen:2024ihp}.

\subsection{Research objective}

	Within this work, we propose to use modern numerical fluid-dynamic techniques to solve \gls{frg} flow equations in multi-dimensional field space.
	The idea of looking at flow equations from a fluid dynamic point was proposed in \reff\cite{Grossi:2019urj} (based on earlier observations by \reffs\cite{Tetradis:1995br,Litim:1995ex,Aoki:2014,Aoki:2017rjl}) and further worked out in \reffs\cite{Koenigstein:2021syz,Koenigstein:2021rxj,Steil:2021cbu,Grossi:2019urj,Grossi:2021ksl,Stoll:2021ori,Ihssen:2022xjv,Ihssen:2022xkr,Batini:2023nan,Ihssen:2023dmx,Ihssen:2023qaq,Ihssen:2023nqd,Ihssen:2023xlp,Ihssen:2024miv,Ihssen:2024ihp}.
	However, so far, the literature mostly focuses on solving \gls{frg} problems that can be reduced to a single field-space direction.
	Here, we extend this approach to the case of multi-dimensional field space.
	To this end, the present work might be considered an extension of the ideas and methods presented in the aforementioned references.
	In particular, we shall demonstrate how to apply a modern \gls{fv} discretization scheme -- the \gls{kt} (central) scheme -- to \gls{frg} flow equations of effective potentials living in a multi-dimensional field space.
	We shall show that this method is stable and accurate by performing multiple benchmark tests.
	As a testing ground, we shall use zero-dimensional models with two field-space directions, with and without $O(2)$ symmetry as well as $O(N) \times O(M)$-symmetric models.
	Note that studies in zero spacetime dimensions come with the advantage that \gls{rg}-time stepping towards the \gls{ir} is not a relevant problem, see also \reff\cite{Ihssen:2023qaq}, such that we can focus on the spatial discretization.
	Additionally, we can directly compare our results to exact reference values which can be computed exactly from the corresponding partition function.
	Apart from these formal tests, we shall also discuss general aspects of \gls{rg} flow equations, such as possible restrictions on initial conditions or the general structure of flow equations in multi-dimensional field space and their reformulation in terms of a fluid-dynamic diffusion problem.

	Ultimately, we shall also demonstrate that our method is indeed applicable to the study of realistic models in nonzero spacetime dimensions.
	Because the main use case of our method is to consider the actual \gls{rg} flow and time stepping from the \gls{uv} to the \gls{ir}, we consider the following test cases:
	As a first test model, we use an $O(2)$-symmetric model in three spatial dimensions in the \gls{lpa} in the symmetry broken phase where the initial condition, i.e., the \gls{uv} potential, is chosen such that the system comprises competing minima and nonanalyticities.
	As a second test model, we consider an $O(N) \times O(M)$-symmetric model in three spatial dimensions in \gls{lpa}. 
	In this case, the \gls{uv} initial conditions are also chosen such that the system describes a phase with a broken $O(N) \times O(M)$ symmetry in the ground state in the \gls{ir} limit.

	To summarize, with our present work, we aim to provide a powerful numerical toolbox to solve \gls{frg} flow equations for effective potentials which are multi-dimensional in field space. 
	To be more concrete, our goal is to provide a ``black-box solver" for the \gls{rg} time evolution that is based on robust and stable numerical techniques from \gls{cfd} and tested against exact reference values and can therefore be included in the toolbox of \gls{frg} practitioners.
	Furthermore, we aim to convince the reader through example calculations that our method is directly applicable to realistic models.
	Last but not least, the here discussed zero-dimensional models may also be of interest to researchers who are developing new methods to solve \gls{frg} flow equations in multi-dimensional field space and want to benchmark their schemes.

	Our present work is organized as follows:
	In \cref{sec:a_zero-dimensional_qft_of_scalar_fields}, we introduce the zero-dimensional \gls{qft} of scalar fields which serves as a testing ground for our numerical method.
	We provide a brief overview of the key aspects of \gls{qft} in zero spacetime dimensions and introduce the effective action and vertex functions.
	In \cref{sec:the_zero-dimensional_frg_and_fluid_dynamics}, we present the zero-dimensional version of the Wetterich equation for scalar fields and demonstrate how this equation can be turned into a fluid-dynamic problem in a two-dimensional field space (or the space spanned by suitably chosen field invariants).
	We then show that our result can be consistently reduced to a flow equation with a single field-space direction, if $O ( 2 )$ symmetry is assumed.
	After that, we shall also discuss some additional restrictions on initial conditions for \gls{rg} flows in two-dimensional field space.
	Finally, we present the generalization to the case of $O ( N ) \times O ( M )$-symmetric models which can be reduced to a flow equation on a two-dimensional domain spanned by the corresponding field invariants.
	In \cref{sec:numeric_approach}, we introduce the \gls{kt} scheme, a modern \gls{fv} discretization scheme from \gls{cfd} which we use to solve the \gls{frg} flow equations of the various models considered in our present work.
	We provide a brief overview of the \gls{kt} scheme and discuss its adaption to \gls{frg} problems.
	
	These more theoretical sections are followed by considerations of appropriate test setups, see \cref{sec:testing_setup}.
	After that, we discuss various toy models and the challenges they pose to our numerical approach and numerical methods in general.
	To be specific, in \cref{sec:o_2_symmetric_model}, we introduce our $O(2)$ symmetric test models, whereas \cref{sec:non-symmetric_model} is dedicated to non-symmetric models.
	A model with a $O(N) \times O(M)$ symmetry is then considered in \cref{sec:a_test_model_with_on_om_symmetry}.
	Actual numerical results from our test models are presented in \cref{sec:results_of_tests_in_zero_dimensions} where we also discuss the performance of the \gls{kt} scheme in the context of \gls{frg} flow equations.
	Before we finally conclude in \cref{sec:conclusion_and_outlook}, we shall discuss selected models in three spacetime dimensions in \cref{sec:higher-dimensional_models}, namely an $O(2)$ and an $O(N) \times O(M)$-symmetric model in their symmetry broken phases.

\section{A zero-dimensional QFT of scalar fields}
\label{sec:a_zero-dimensional_qft_of_scalar_fields}

	Since \gls{qft} in zero spacetime dimensions provides the testing ground for our numerical method, we give a brief overview of its key aspects in this section.
	To be more specific, we primarily introduce the fundamental definitions of a zero-dimensional model of $N$ real scalar (interacting) fields.
	Additionally, we provide some formulae for correlation and vertex functions which can be utilized to compute high-precision reference values through straightforward numerical integration.
	For a more comprehensive discussion of \gls{qft} in zero spacetime dimensions, we refer the reader to \reffs\cite{Rivasseau:2009pi,Keitel:2011pn,SkinnerScript,Koenigstein:2021syz,Fraboulet:2021amf} and references therein.
	
\subsection{Correlation functions} 

	A zero-dimensional \gls{qft} model is defined by an action $\mathcal{S} ( \vec{\phi} \, ) = \mathcal{S} ( \phi_1, \ldots, \phi_N )$ of the fields $\vec \phi = (\phi_1, \dots, \phi_N )^{T}$.
	The fields are plain real numbers without spacetime-dependence of any kind as there is no spacetime.
	The entire field theory can be considered in terms of interacting quantum fields in a single point.	
	Hence, there is also no notion of energies, momenta \etc\ and the generating functional of correlation functions reduces to an ordinary function that is defined in terms of an $N$-dimensional ordinary integral,
		\begin{align}
			\mathcal{Z} ( \vec{J} \, ) = \mathcal{N} \int_{- \infty}^{\infty} \mathrm{d}^N \phi \,  \ee^{- \mathcal{S} ( \vec{\phi} \, ) + \vec{J}^{\, T} \cdot \vec{\phi}} \, ,	\label{eq:generating_funcitonal_of_correlation_funcitons}
		\end{align}
	where $\vec{J}$ is the vector of source fields and $\mathcal{N}$ is the normalization of the probability distribution.
	
	In complete analogy to higher-dimensional \glspl{qft} and statistical physics, correlation functions can be derived from \cref{eq:generating_funcitonal_of_correlation_funcitons} by taking derivatives \gls{wrt}\ the components of $\vec{J}$.
	Also these correlation functions reduce to ordinary integrals, \ie,
		\begin{align}
			\langle \phi_{i_n} \cdots \phi_{i_1} \rangle = \, & \frac{1}{\mathcal{Z} ( \vec{J} \, )} \frac{\partial^n \mathcal{Z} ( \vec{J} \, )}{\partial J_{i_n} \cdots \partial J_{i_1}} \bigg|_{\vec{J} = 0} =	\Vdistance	\label{eq:correlation_functions}
			\\
			= \, & \frac{\int_{- \infty}^{\infty} \mathrm{d}^N \phi \, \phi_{i_n} \cdots \phi_{i_1} \, \ee^{- \mathcal{S} ( \vec{\phi} \, )}}{\int_{- \infty}^{\infty} \mathrm{d}^N \phi \,  \ee^{- \mathcal{S} ( \vec{\phi} \, )}} \, .	\Vdistance	\nonumber
		\end{align}
	It only remains to choose some $\mathcal{S} ( \vec{\phi} \, )$ and we have a full-fledged theory.
	Here, we are basically free to choose any function of the fields which is
		\begin{enumerate}
			\item	bounded from below and grows asymptotically at least quadratically in every field-space direction, and
			
			\item	continuous.
		\end{enumerate}
	The first restriction ensures convergence of the integrals in \cref{eq:generating_funcitonal_of_correlation_funcitons,eq:correlation_functions}. 
	The role of the second requirement is solely to maintain contact with higher-dimensional models where potentials are continuous functions of the fields. 
	In zero spacetime dimensions the classical \gls{uv} action is identical to a \gls{uv} potential.
	Apart from this, there is no need to restrict $\mathcal{S}$ to smooth or even analytic functions.

\subsection{The effective action and vertex functions}

	Of course, we are not interested in calculating the $N$-dimensional integrals \labelcref{eq:correlation_functions} numerically, but in discussing and testing novel fluid-dynamic methods for \gls{frg} flow equations.
	
	The \gls{frg} approach is formulated on the level of the effective action  \cite{Wetterich:1991be,Wetterich:1992yh,Ellwanger:1993mw,Morris:1993qb,Morris:1994ie,Wetterich:2001kra,Berges:2000ew,Pawlowski:2005xe}. 
	 Therefore, we introduce this quantity in terms of its definition as the Legendre transform of the logarithm of \cref{eq:generating_funcitonal_of_correlation_funcitons},
		\begin{align}
			\Gamma ( \vec{\varphi} \, ) \equiv \underset{\vec{J}}{\mathrm{sup}} \big\{  \vec{J}^{\, T} \cdot \vec{\varphi} - \ln \mathcal{Z} ( \vec{J} \, ) \big\} \, ,	\label{eq:legendre_transformation}
		\end{align}
	where $\vec{\varphi}$ are the mean fields.
	The corresponding observables are the vertex functions which can be directly calculated from the effective action by taking  derivatives \gls{wrt}\ the components of $\vec{\varphi}$ at the minimum of $\Gamma ( \vec{\varphi} \, )$,
			\begin{align}
				\Gamma^{( n )}_{\varphi_{i_n} \ldots \varphi_{i_1}} = \frac{\partial^n \Gamma ( \vec{\varphi} \, )}{\partial \varphi_{i_n} \cdots \partial \varphi_{i_1}} \bigg|_{\vec{\varphi} = \vec{\varphi}_{\min}} \, ,	\label{eq:vertex_functions}
			\end{align}
	once $\Gamma ( \vec{\varphi} \, )$ is known.
	Here, $\vec{\varphi}_{\min}$ denotes its minimum.
	
	In zero spacetime dimensions it is of course possible to calculate $\mathcal{Z} ( \vec{J} \, )$ for a given $\mathcal{S} ( \vec{\phi} \, )$ and perform the Legendre transformation (numerically) to extract the vertex functions.
	Alternatively, it is straightforward to derive direct relations between (specific) correlation functions~\labelcref{eq:correlation_functions} and vertex functions \labelcref{eq:vertex_functions}.
	For example, for the two-point vertex function, which is the only relevant quantity in this work, one finds \cite{Keitel:2011pn,Koenigstein:2021syz,Fraboulet:2021amf}
		\begin{align}
			\big( \Gamma^{( 2 )}_{\vec{\varphi} \vec{\varphi}} \big)^{- 1}_{j i} = \langle \phi_j \, \phi_i \rangle - \langle \phi_j \rangle \, \langle \phi_i \rangle \, .	\label{eq:two-point_vertex}
		\end{align}
	This can be evaluated (numerically) using \cref{eq:correlation_functions}.

	In higher-dimensional \glspl{qft}, this is no longer trivial and $\Gamma ( \vec{\varphi} \, )$ and ultimately the vertex functions are often calculated with some other method, \eg, with the \gls{frg} approach.
	In this work, we are doing exactly this:
	We use the \gls{frg} to obtain $\Gamma ( \vec{\varphi} \, )$ including the vertex functions, especially \cref{eq:two-point_vertex}.
	This is still a challenging task in zero spacetime dimensions because we have to solve a highly nonlinear \gls{pde}.
	Though, in zero spacetime dimensions, we have easily accessible reference values from exact (numeric) computations of the correlation functions \labelcref{eq:correlation_functions}.
	Therefore, we can benchmark the (numeric) solution strategy for the \gls{pde} in the \gls{frg} formalism via \cref{eq:two-point_vertex}.
	This is rarely possible in higher-dimensional \glspl{qft}.\footnote{For example, exceptions are \glspl{qft} in certain limits, such as the infinite-$N$ limit, where theories sometimes become integrable, \cf\ \reffs\cite{Wolff:1985av,Schnetz:2005ih,DAttanasio:1997yph,Grossi:2019urj,tHooft:1974pnl}.}

\section{The zero-dimensional FRG and fluid dynamics}
\label{sec:the_zero-dimensional_frg_and_fluid_dynamics}

	In this section, we briefly present the zero-dimensional version of the \gls{frg} and the corresponding zero-dimensional formulation of the Wetterich equation for our \gls{qft} of $N$ real scalar fields.
	Afterwards, we demonstrate how this equation is turned into a fluid-dynamic problem on a two-dimensional spatial domain.
	We then show that the resulting equation can be consistently reduced to a flow equation in a single dimension, if $O(2)$ symmetry is assumed.
	Afterwards, we discuss some additional restrictions on initial conditions for \gls{rg} flows in two-dimensional field space.
	Before we close the section, we also present the generalization of our considerations to the case of $O(\bar{N}) \times O(\bar{M})$-symmetric models (with $\bar{N}+\bar{M}=N$), which can be reduced to a flow equation on a two-dimensional spatial domain spanned by the field invariants.
	
	For comprehensive general introductions to the \gls{frg} approach, we refer to \reffs\cite{Berges:2000ew,Wetterich:2001kra,Pawlowski:2005xe,Gies:2006wv,Delamotte:2007pf,Kopietz:2010zz,Braun:2011pp,Dupuis:2020fhh,Blaizot:2021ikl}. 
	Detailed introductions to and applications of zero-dimensional systems within the \gls{frg} framework can be found in \reffs\cite{Keitel:2011pn,Pawlowski:talk,Fl_rchinger_2010,Kemler:2013yka,Catalano:2019,Fraboulet:2021amf,Koenigstein:2021syz,Koenigstein:2021rxj,Steil:2021cbu,Ihssen:2022xjv,Ihssen:2022xkr}.

\subsection{The zero-dimensional Wetterich equation for scalar fields}
\label{subsec:the_zero-dimensional_wetterich_equation_for_scalar_fields}
	
	The zero-dimensional version of the Wetterich equation \cite{Wetterich:1992yh} for $N$ real scalar fields simply reads,
		\begin{align}
			\partial_t \bar{\Gamma} ( t, \vec{\varphi} \, ) = \mathrm{tr} \big[ \big( \tfrac{1}{2} \, \partial_t R ( t ) \big) \big( \bar{\Gamma}^{( 2 )} ( t, \vec{\varphi} \, ) + R ( t ) \big)^{-1} \big] \, .	\label{eq:wetterich_equation_zero_dimensions}
		\end{align}
	It is an exact evolution equation for the effective average action $\bar{\Gamma} ( t, \vec{\varphi} \, )$ in field space, which is spanned by $\vec{\varphi} \in \mathbb{R}^N$, and the \gls{rg} time $t \in [ 0, \infty )$.
	We define the \gls{rg} time $t$ to be manifestly positive. 
	It runs from $t = 0$, which we refer to as the \gls{uv}, to $t \to \infty$, which is the zero-dimensional analogue of the \gls{ir} limit.
	In the \gls{ir} limit, the effective average action $\bar{\Gamma} ( t, \vec{\varphi} \, )$ approaches the quantum effective action $\Gamma ( \vec{\varphi} \, )$, that we are interested in, whereas the \gls{uv} initial condition is given by the classical action $\mathcal{S} ( \vec{\varphi} \, )$ evaluated on the mean fields $\vec{\varphi}$, \ie,
		\begin{align}
			& \bar{\Gamma} ( t = 0, \vec{\varphi} \, )  = \mathcal{S} ( \vec{\varphi} \, ) \, ,	&&	\lim\limits_{t \to \infty} \bar{\Gamma} ( t, \vec{\varphi} \, ) = \Gamma ( \vec{\varphi} \, ) \, .	\label{eq:limits_effective_average_action}
		\end{align}
	The matrix valued function $R ( t )$ is the regulator, which we choose\footnote{If there are no symmetries in the system, we could in principle choose a regulator that is nondiagonal.} to be diagonal in field space with an exponentially monotonically decreasing regulator-shape function $r ( t )$,
		\begin{align}
			&	R ( t ) = \openone_{N \times N} \, r ( t ) \, ,	&&	r ( t ) = \Lambda \, \ee^{-t} \, .	\label{eq:regulator_definition}
		\end{align}
	Here, $\Lambda$ is the zero-dimensional version of the \gls{uv} cutoff, which needs to be much greater than the typical ``scales'' in the classical action $\mathcal{S} ( \vec{\varphi} \, )$.\footnote{%
		For a detailed discussion of suitable choices of the \gls{uv} cutoff in the context of zero-dimensional \gls{qft}, we refer to \reff\cite{Koenigstein:2021syz} and to \reff\cite{Braun:2018svj} and \reffs therein for higher-dimensional systems.
	}
	
	Typically, in higher-dimensional \gls{frg} applications it is not possible to solve \cref{eq:wetterich_equation_zero_dimensions} exactly and approximations are necessary.
	In zero dimensions, there is no need for truncations and one finds that the only possible ``ansatz'' for the effective average action is an \gls{rg}-time dependent potential,\footnote{%
		Of course, it is still possible to artificially impose truncations in zero dimensions, \cf\ \reffs\cite{Keitel:2011pn,Pawlowski:talk,Moroz:2011thesis,Fraboulet:2021amf}.
		For example, in \reff\cite{Koenigstein:2021syz}, the convergence of the \gls{frg} vertex/Taylor expansion was tested in zero spacetime dimensions.
	}
		\begin{align}
			\bar{\Gamma} ( t, \vec{\varphi} \, ) = U ( t, \vec{\varphi} \, ) \, .	\label{eq:ansatz_effective_average_action}
		\end{align}
	Still, \cref{eq:wetterich_equation_zero_dimensions} constitutes a highly nonlinear \gls{pde} for $U ( t, \vec{\varphi} \, )$, whose spatial domain is the $N$-dimensional field space and whose temporal domain is the \gls{rg} time.
	It is therefore not feasible to solve this equation for some large $N$ without imposing additional constraints.
	
	Luckily, we can often employ symmetries of the classical action $\mathcal{S} ( \vec{\varphi} \, )$ in field space, which transfer to $U ( t, \vec{\varphi} \, )$ by construction, to reduce the computational domain to a subspace of $\mathbb{R}^N$.
	For example, for an $O(N)$-symmetric model, the minimally required subspace is $\mathbb{R}_{(\geq 0)}$ and one-dimensional.
	This is achieved by choosing the $O(N)$-invariant $\varrho = \frac{1}{2} \, \vec{\varphi}^{\, 2}$ or functions thereof as the spatial domain.
	Similarly, for $O(\bar{N}) \times O(\bar{M})$-symmetric models, with $\bar{N} + \bar{M} = N$, the $N$-dimensional computational domain $\mathbb{R}^N$ can be reduced to $\mathbb{R}_{(\geq 0)} \times \mathbb{R}_{(\geq 0)}$, that is spanned by the two invariants $\varrho_1 = \frac{1}{2} \, \vec{\varphi}_1^{\, 2}$ and $\varrho_2 = \frac{1}{2} \, \vec{\varphi}_2^{\, 2}$, where $\vec{\varphi}_1 = ( \varphi_1, \ldots, \varphi_{\bar{N}} )^T$ and $\vec{\varphi}_2 = ( \varphi_{\bar{N} + 1}, \ldots, \varphi_N )^T$.
	
	Oftentimes, however, for more complicated problems or symmetry groups, one remains with a multi-dimensional domain, \cf\ \cref{sec:introduction}, such that a \gls{pde} for $U ( t, \vec{\varphi} \, )$ is still a flow equation in more than one/two field-space dimensions.
	Thus, it is also useful to artificially simulate this situation by simply considering a model of two interacting real scalar fields with -- a priori -- no additional symmetry assumptions.
	The above mentioned cases are certainly the minimal setup to test \gls{frg} flow equations in more than one field-space dimension.
	This is the main topic of this work.

\subsection{Flow equation of a zero-dimensional model with two fields}
\label{sec:flow_equation_zero-dim_model_two_fields}

	We now derive the flow equation for the effective potential of a zero-dimensional interacting \gls{qft} of two ($N = 2$) real scalar fields from the Wetterich equation \labelcref{eq:wetterich_equation_zero_dimensions}. 
	Recall that the effective potential is identical to the effective action in zero spacetime dimensions. 
	In a subsequent step we then show how this flow equation can be recast as a fluid dynamic problem in terms of two nonlinear diffusion-type equations.

\subsubsection{The flow equation of the effective potential}
\label{subsubsec:flow_equation_of_the_effective_potential}

	First, we set $N = 2$ and therefore consider a two-dimensional Euclidean field space with coordinates $\varphi_1, \varphi_2 \in \mathbb{R}$ and $\vec{\varphi} = ( \varphi_1, \varphi_2 )^T$.
	For the sake of the readability, we use the following abbreviations:
		\begin{align}
			&	U = U ( t, \vec{\varphi} \, ) \, ,	&&	r = r ( t ) \, .
		\end{align}
	In general, we only have to insert our ansatz \labelcref{eq:ansatz_effective_average_action} in the Wetterich \cref{eq:wetterich_equation_zero_dimensions} and evaluate the trace on the \gls{rhs}.
	For the sake of clarity, we proceed step-by-step and start with the calculation of the full two-point function,
		\begin{align}
			\bar{\Gamma}^{( 2 )} ( t, \vec{\varphi} \, ) + R ( t ) = \, &
			\begin{pmatrix}
				r +\partial_{\varphi_1}^2 U					&	\partial_{\varphi_1} \partial_{\varphi_2} U
				\\
				\partial_{\varphi_2} \partial_{\varphi_1} U	&	r + \partial_{\varphi_2}^2 U
			\end{pmatrix} \, ,	\label{eq:full_two-point_function}
		\end{align}
	which is simply a two-dimensional matrix in field space.
	This matrix needs to be inverted to obtain the full propagator.
	The inverse is
		\begin{align}
			\big( \bar{\Gamma}^{( 2 )} ( t, \vec{\varphi} \, ) + R ( t ) \big)^{-1} = \frac{\adj \big( \bar{\Gamma}^{( 2 )} ( t, \vec{\varphi} \, ) + R ( t ) \big)}{\det \big( \bar{\Gamma}^{( 2 )} ( t, \vec{\varphi} \, ) + R ( t ) \big)} \, ,
		\end{align}
	where $\mathrm{adj}$ denotes the adjugate matrix,
		\begin{align}
			& \adj \big( \bar{\Gamma}^{( 2 )} ( t, \vec{\varphi} \, ) + R ( t ) \big) =	\Vdistance
			\\
			= \, &
			\begin{pmatrix}
				r + \partial_{\varphi_2}^2 U					&	- \partial_{\varphi_1} \partial_{\varphi_2} U
				\\
				- \partial_{\varphi_2} \partial_{\varphi_1} U	&	r + \partial_{\varphi_1}^2 U
			\end{pmatrix} \, 	\nonumber
		\end{align}
	and the determinant is 
			\begin{align}
				& \det \big( \bar{\Gamma}^{(2)} ( t, \vec{\varphi} \, ) + r ( t ) \big) =	\vdistance	\label{eq:determinant_two-point_function}
				\\
				= \, & \big( r + \partial_{\varphi_1}^2 U \big) \big( r + \partial_{\varphi_2}^2 U \big) - \big( \partial_{\varphi_1} \partial_{\varphi_2} U \big) \big( \partial_{\varphi_2} \partial_{\varphi_1} U \big) \, .	\vdistance	\nonumber
			\end{align} 
	It should be noted that the inversion is only possible and the Wetterich equation is only well defined if this determinant is non-zero for all \gls{rg} times $t$ and all points in field space.
	We return to this delicate issue below in \cref{eq:initial_conditions}.
	In addition, we emphasize that we do not necessarily have $\partial_{\varphi_i} \partial_{\varphi_j} U = \partial_{\varphi_j} \partial_{\varphi_i} U$ since $U$ does not need to be analytic.
	
	For the moment, we combine our intermediate results and arrive at the flow equation for the effective potential:
		\begin{align}
			& \partial_t U =	\Vdistance	\label{eq:flow_equation_effective_potential}
			\\
			= \, & \frac{\big( \frac{1}{2} \, \partial_t r \big) \big( 2 r + \partial_{\varphi_1}^2 U + \partial_{\varphi_2}^2 U \big)}{\big( r + \partial_{\varphi_1}^2 U \big) \big( r + \partial_{\varphi_2}^2 U \big) - \big( \partial_{\varphi_1} \partial_{\varphi_2} U \big) \big( \partial_{\varphi_2} \partial_{\varphi_1} U \big)} \, .	\Vdistance	\nonumber
		\end{align}
	This is a nonlinear \gls{pde}, up to second order in spatial/field derivatives and first order in time, on the two-dimensional noncompact domain $\mathbb{R}^2$.
	Structurally, flow equations for effective bosonic potentials in higher-dimensional fermion-boson systems with for example two possible condensate directions are usually rather similar, see our discussion in \cref{sec:introduction}.
	
	Note that only second derivatives of $U$ with respect to the fields appear on the  \gls{rhs}\ of \cref{eq:flow_equation_effective_potential}.
	In addition, the actual value of a potential has no relevance and physical quantities correspond to relative differences or derivatives of potentials.
	These observations already suggest to study the field-space derivatives of $U$ instead of $U$ itself.

\subsubsection{A fluid-dynamic reformulation} 

	It was shown in \reffs\cite{Aoki:2014,Aoki:2017rjl,Grossi:2019urj} that taking a field-space derivative of the flow equation of an effective potential recasts the corresponding \gls{pde} into its conservative formulation.
	This reformulation and its consequences were further worked out in \reffs\cite{Grossi:2021ksl,Koenigstein:2021syz,Koenigstein:2021rxj,Steil:2021cbu,Stoll:2021ori,Ihssen:2022xjv,Ihssen:2022xkr,Ihssen:2023qaq}.
	Here, we shall only demonstrate how this reformulation is done for the \gls{rg} flow equation at hand.
	
	Taking spatial derivatives \gls{wrt}\ each spatial coordinate of \cref{eq:flow_equation_effective_potential}, without executing these derivatives on the \gls{rhs}, we obtain, 
		\begin{align}
			&	\partial_t \big( \partial_{\varphi_i} U \big) =	\Vdistance	\label{eq:derivative_rg_flow_equation_potential}
			\\
			= \, & \frac{\mathrm{d}}{\mathrm{d} \varphi_i} \Bigg[ \frac{\big( \frac{1}{2} \, \partial_t r \big) \big( 2 r + \partial_{\varphi_1}^2 U + \partial_{\varphi_2}^2 U \big)}{\big( r + \partial_{\varphi_1}^2 U \big) \big( r + \partial_{\varphi_2}^2 U \big) - \big( \partial_{\varphi_1} \partial_{\varphi_2} U \big) \big( \partial_{\varphi_2} \partial_{\varphi_1} U \big)} \Bigg] \, .	\Vdistance	\nonumber
		\end{align}
	Next, we rename the two orthogonal field-space derivatives of $U$ as follows:
		\begin{align}
			&	u = \partial_{\varphi_1} U \, ,	&&	v = \partial_{\varphi_2} U \, ,
		\end{align}
	where we again use shorthand notations $u = u ( t, \vec{\varphi} \, )$ and $v = v ( t, \vec{\varphi} \, )$.
	Expressing \cref{eq:derivative_rg_flow_equation_potential} in terms of these new variables, we find a set of two coupled \glspl{pde}, \ie,
		\begin{align}
			&	\partial_t u =	\Vdistance	\label{eq:diffusion_equation_u}
			\\
			= \, & \frac{\mathrm{d}}{\mathrm{d} \varphi_1} \Bigg[ \frac{\big( \frac{1}{2} \, \partial_t r \big) \big( 2 r + \partial_{\varphi_1} u + \partial_{\varphi_2} v \big)}{\big( r + \partial_{\varphi_1} u \big) \big( r + \partial_{\varphi_2} v \big) - \big( \partial_{\varphi_1} v \big) \big( \partial_{\varphi_2} u \big)} \Bigg] \, ,	\Vdistance	\nonumber
			\\
			&	\partial_t v =	\Vdistance	\label{eq:diffusion_equation_v}
			\\
			= \, & \frac{\mathrm{d}}{\mathrm{d} \varphi_2} \Bigg[ \frac{\big( \frac{1}{2} \, \partial_t r \big) \big( 2 r + \partial_{\varphi_1} u + \partial_{\varphi_2} v \big)}{\big( r + \partial_{\varphi_1} u \big) \big( r + \partial_{\varphi_2} v \big) - \big( \partial_{\varphi_1} v \big) \big( \partial_{\varphi_2} u \big)} \Bigg] \, .	\Vdistance	\nonumber
		\end{align}
	This system can be rearranged in vector notation and we arrive at a ``conservation law''
		\begin{align}
			\partial_t
			\begin{pmatrix}
				u
				\\
				v
			\end{pmatrix}
			= \partial_{\varphi_1}
			\begin{pmatrix}
				Q
				\\
				0
			\end{pmatrix}
			+ \partial_{\varphi_2}
			\begin{pmatrix}
				0
				\\
				Q
			\end{pmatrix} \, .	\label{eq:conservation_law_u_v}
		\end{align}
	In compact notation, this equation reads
		\begin{align}
			&	\partial_t \vec{u}^{\, T} = \vec{\nabla}^{\, T} \cdot \mathbf{Q} \, ,
			&&	\mathbf{Q} \equiv
				\begin{pmatrix}
					Q	&	0
					\\
					0	&	Q
				\end{pmatrix} \, ,	\label{eq:conservation_law_u_vec}
		\end{align}
	where $\vec{u} \equiv ( u, v )^T$, $\vec{\nabla} \equiv ( \partial_{\varphi_1}, \partial_{\varphi_2} )^T$ and the nonlinear diffusion flux is defined as
		\begin{align}
			Q = \frac{\big( \frac{1}{2} \, \partial_t r \big) \big( 2 r + \partial_{\varphi_1} u + \partial_{\varphi_2} v \big)}{\big( r + \partial_{\varphi_1} u \big) \big( r + \partial_{\varphi_2} v \big) - \big( \partial_{\varphi_1} v \big) \big( \partial_{\varphi_2} u \big)} \, .	\label{eq:diffusion_flux}
		\end{align}
	Some comments are in order:
		\begin{enumerate}
			\item	In this conservative form, we can identify $u$ and $v$ with two fluid fields, which are evolving in the two-dimensional field space with \gls{rg} time.
			
			\item	The denomination of the equations as nonlinear diffusion-type equations with a diffusion flux becomes clear when the derivatives on the \gls{rhs}\ of \cref{eq:diffusion_equation_u,eq:diffusion_equation_v} are executed.
			One then observes terms which are proportional to second derivatives of $u$ and $v$.
			Interpreting their coefficients as highly nonlinear diffusion coefficients, we are confronted with a quasi parabolic problem -- a nonlinear diffusion equation.
			
			\item	Structurally, \cref{eq:conservation_law_u_vec} is a conservation law.
			Therefore, methods from the field of \gls{cfd} are the appropriate choice to approach it numerically \cite{Hesthaven2007,RezzollaZanotti:2013,LeVeque:1992,LeVeque:2002}.
			Because of the highly nonlinear character of the equation, nonanalytic behavior may emerge during the \gls{rg} flow, such that expansion schemes or numerical methods that cannot cope with nonanalyticities are inappropriate.
			In \cref{sec:numeric_approach}, we briefly sketch a possible choice for a numeric \gls{fv} scheme which can be used to solve equations of the type of \cref{eq:conservation_law_u_v}.
		\end{enumerate}

\subsection{\texorpdfstring{$O(2)$}{O(2)} symmetry in the Wetterich equation -- the one-dimensional reduction as a consistency check}
\label{subsec:one-dimensional_reduction}

	As already stated above, a multi-dimensional formulation of the field dependence is not required for models that exhibit an additional symmetry in field space.
	For example, if the action in the path integral of our field theory with $N = 2$ is symmetric under $O(2)$ transformations, the effective average action inherits this symmetry by construction.
	This implies that the \gls{rg}-time dependent potential is only a function of the $O(2)$-invariant $\varrho = \frac{1}{2} \, \vec{\varphi}^{\, 2}$ constructed from the background fields rather than a function of two independent fields $\varphi_1$ and $\varphi_2$, \ie,
		\begin{align}
			U ( t, \vec{\varphi} \, ) = \tilde{U} ( t, \varrho ) \, .	\label{eq:o_2_invariant_potential}
		\end{align} 
	It follows that the field-space derivatives on the \gls{rhs}\ of \cref{eq:flow_equation_effective_potential} should be rewritten in terms of $\varrho$.
	We simply use the chain rule and find,
		\begin{align}
			\partial_{\varphi_i} U ( t, \vec{\varphi} \, ) = \varphi_i \, \partial_\varrho \tilde{U} ( t, \varrho ) \, , \label{eq:chain_rule_U}
		\end{align}
	and
		\begin{align}
			\partial_{\varphi_i} \partial_{\varphi_j} U ( t, \vec{\varphi} \, ) = \delta_{i j} \, \partial_\varrho \tilde{U} ( t, \varrho ) + \varphi_i \, \varphi_j \, \partial_\varrho^2 \tilde{U} ( t, \varrho ) \, ,
		\end{align}
	where $i,j \in \{ 1, 2 \}$.
	Inserting this explicitly in \cref{eq:flow_equation_effective_potential} and using \cref{eq:o_2_invariant_potential} on the \gls{lhs}, we obtain the \gls{rg} flow equation of the potential expressed in terms of $\varrho$,
		\begin{align}
			& \partial_t \tilde{U}  
			=\frac{\frac{1}{2} \, \partial_t r}{r + \partial_\varrho \tilde{U}} + \frac{\frac{1}{2} \, \partial_t r}{r + \partial_\varrho \tilde{U} + 2 \varrho \, \partial_\varrho^2 \tilde{U}} \, .	\Vdistance 
		\end{align}
	Readers, who are familiar with common \gls{frg} literature on $O(N)$-models in higher dimensions, will immediately recognize the generic structure of this flow equation as a \gls{lpa}, which is however not an approximation in zero dimensions but exact.
	
	Thus, we have successfully demonstrated that the version of the flow equation in two field-space dimensions, where we do not use the symmetry to restrict the spatial domain of the \gls{pde}, correctly reduces to the one-dimensional \gls{pde}, where the symmetry is encoded in field space.
	Here, the spatial domain is parameterized by the coordinate $\varrho \in \mathbb{R}_{\geq 0}$.
	
	In \reff\cite{Koenigstein:2021syz}, it is discussed in great detail why another formulation of this \gls{pde} may be better suited for practical numerical implementations.
	We do not repeat this discussion here but only show how we can arrive at this formulation.
	To this end, we introduce the background field $\sigma$ which can be viewed without loss of generality as a field configuration $\vec{\varphi} = ( \sigma, 0 )$, such that
		\begin{align}
		&	\varrho = \tfrac{1}{2} \, \sigma^2 \, ,	&&	\Leftrightarrow &&	\sigma = \pm \sqrt{2 \varrho} \, .
		\end{align}
	This coordinate transformation results in the following flow equation:
		\begin{align}
			\partial_t {U} = \, & \frac{\frac{1}{2} \, \partial_t r}{r + \frac{1}{\sigma} \, \partial_\sigma {U}} + \frac{\frac{1}{2} \, \partial_t r}{r + \partial_\sigma^2 {U}} \, ,	\label{eq:flow_equation_potential_sigma}
		\end{align}
	which has a spatial domain spanned by $\sigma \in \mathbb{R}$ and where ${U} ( t, \sigma ) = {U} ( t, - \sigma )$.

\subsection{A one-dimensional advection-diffusion equation}

	It was also shown in \reff\cite{Koenigstein:2021syz} how \cref{eq:flow_equation_potential_sigma} can be recast into conservative form and solved numerically.
	Again, defining ${u} = \partial_\sigma {U}$ with ${u} = {u} ( t, \sigma ) = - {u} ( t, - \sigma )$, we obtain a conservation law,
		\begin{align}
			\partial_t {u} = \, & \frac{\mathrm{d}}{\mathrm{d} \sigma} \bigg( \frac{\frac{1}{2} \, \partial_t r}{r + \frac{1}{\sigma} \, {u}} + \frac{\frac{1}{2} \, \partial_t r}{r + \partial_\sigma {u}} \bigg) \, ,	\label{eq:one_dim_advection_diffusion}
		\end{align}
	which presents as a highly nonlinear advection-diffusion equation.
	Again, some comments are in order:
		\begin{enumerate}
			\item	It was explicitly demonstrated in \reffs\cite{Koenigstein:2021syz,Ihssen:2022xkr} that this equation can be solved with modern schemes from \gls{cfd}.
			To be specific,  \reff\cite{Ihssen:2022xkr} uses discontinuous Galerkin methods whereas \reff\cite{Koenigstein:2021syz} presents benchmark tests for a \gls{fv} method -- the \gls{kt} central scheme -- as also presented below.
			
			\item	It is remarkable that the formal description of a system of two highly nonlinear diffusion-type equations \labelcref{eq:diffusion_equation_u,eq:diffusion_equation_v} in a two-dimensional spatial domain with initial conditions that include the $O(2)$-symmetry in some way is equivalent to a highly nonlinear advection-diffusion equation \labelcref{eq:one_dim_advection_diffusion}, where the $O(2)$-symmetry is imprinted in the spatial coordinate.
			In any case, since the mathematical concepts are clear, it only remains to compare performance of the two approaches on the numerical level, see below.
		\end{enumerate}

\subsection{Initial conditions and well-posedness}
\label{eq:initial_conditions}
	
	Before we present the numeric implementation and study explicit examples, let us revisit the derivation of the flow equation in \cref{subsubsec:flow_equation_of_the_effective_potential}.
	Specifically, we analyze the invertibility of the full two-point function \labelcref{eq:full_two-point_function}.
	If this matrix is not invertible for all \gls{rg} times $t \in [ 0, \infty )$ and positions in field space $\vec{\varphi} \in \mathbb{R}^2$, the Wetterich equation becomes ill-defined.
	In particular, it has to be invertible in the \gls{uv} for the initial potential.\footnote{In general, this should be sufficient to also guarantee invertibility and well-posedness for the entire \gls{rg} flow, at least for untruncated flows.}
	Invertibility of this matrix is guaranteed when its determinant~\labelcref{eq:determinant_two-point_function} is nonzero.
	Moreover, both eigenvalues and the determinant of \cref{eq:full_two-point_function} should always be positive.
	This is because the scale-dependent version of the Legendre transformation \labelcref{eq:legendre_transformation} is only well-defined for a convex scale-dependent effective action
		\begin{align}
			\Gamma ( t, \vec{\varphi} \, ) = \bar{\Gamma} ( t, \vec{\varphi} \, ) + \tfrac{1}{2} \, \vec{\varphi}^{\, T} \, R ( t ) \, \vec{\varphi} \, .
		\end{align}
	Positivity of the determinant also ensures a well-behaved diffusion flux \labelcref{eq:diffusion_flux}.
	Loosely speaking, in \gls{qft} language, we should never ``overshoot" the pole of the propagator in \cref{eq:flow_equation_effective_potential} at any \gls{rg} time and any position in field space.
	
	An obvious question is whether these constraints are always fulfilled for any initial potentials that are continuous, bounded from below, and come with at least a quadratic asymptotic behavior in both field directions. 
	The answer to this question is not straightforward. 
	In any case, this class of potentials leads to well-defined converging integrals and field expectation values when we directly operate on the level of the partition function. 
	
	Before we present an explicit counterexample, which violates the above constraints, let us also provide the eigenvalues of the full two-point function \labelcref{eq:full_two-point_function}, assuming $\partial_{\varphi_1} \partial_{\varphi_2} U = \partial_{\varphi_2} \partial_{\varphi_1} U$ for the moment.
	These are
		\begin{align}
			\lambda_{1/2} = \, & \tfrac{1}{2} \, \Big( 2 r + \partial_{\varphi_1}^2 U + \partial_{\varphi_2}^2 U +	\vdistance	\label{eq:eigenvalues_general}
			\\
			& \quad \pm\sqrt{4 \, ( \partial_{\varphi_1} \partial_{\varphi_2} U )^2 + ( \partial_{\varphi_1}^2 U - \partial_{\varphi_2}^2 U )^2} \Big) \, .	\vdistance	\nonumber
		\end{align}
	Considering now the initial potential
		\begin{align}
			U ( \vec{\varphi} \, ) = \tfrac{1}{2} \, \varphi_1 \, \varphi_2^2 + \tfrac{1}{4!} \, ( \varphi_1^4 + \varphi_2^4 ) \, ,	\label{eq:potential_counter_example}
		\end{align}
	it can be checked by numeric integration that this potential leads to well-defined correlation functions \labelcref{eq:correlation_functions} and vertex functions, \eg, via \cref{eq:two-point_vertex}.
	However, the determinant of the full \gls{uv} two-point function is
		\begin{align}
			& \det \big( \bar{\Gamma}^{(2)} ( 0, \vec{\varphi} \, ) + r ( 0 ) \big) =	\vdistance
			\\
			= \, & \big( \Lambda + \tfrac{1}{2} \, \varphi_1^2 \big) \big( \Lambda + \varphi_1 + \tfrac{1}{2} \, \varphi_2^2 \big) - \varphi_2^2 \, ,	\vdistance	\nonumber
		\end{align}
	and the corresponding \gls{uv} eigenvalues of the two-point function are
		\begin{align}
			\lambda_{1/2} = \, & \tfrac{1}{2} \, \Big( 2 \Lambda + \varphi_1 + \tfrac{1}{2} \, ( \varphi_1^2 + \varphi_2^2 ) +	\vdistance
			\\
			& \quad \pm \sqrt{4 \, \varphi_2^2 + \tfrac{1}{4} \, (\varphi_2^2 - \varphi_1 (\varphi_1 - 2)  )^2} \Big) \, .	\vdistance	\nonumber
		\end{align}
	From this, we deduce that, for $\varphi_2 = 0$, the determinant and the eigenvalue $\lambda_2$ become negative at $\varphi_1 = - \Lambda$.
	This implies that this eigenvalue as well as the determinant cannot be regularized by increasing $\Lambda$. 
	The problematic region in field space is simply moved to smaller $\varphi_1$.
	Additionally, employing the one-loop effective action
		\begin{align}
			\Gamma_\Lambda ( \vec{\varphi} \, ) = \mathcal{S} ( \vec{\varphi} \, ) + \tfrac{1}{2} \ln \det \mathcal{S}^{(2)} ( \vec{\phi} \, ) \big|_{\vec{\phi} = \vec{\varphi}}
		\end{align}
	as the initial condition at the \gls{uv} scale $\Lambda$, as is oftentimes done while working with a finite \gls{uv} cutoff, will not solve the issue since the logarithm of the determinant suffers from these problems.
	The same applies to choosing another regulator function that is still quadratic in the fields, but, \eg, involves off-diagonal terms in \cref{eq:regulator_definition}.
	Field-dependent regulators, \cf\ \reffs \cite{Reuter:1993kw,Litim:2002hj,Gies:2002af}, might solve the problem of the unregulated eigenvalue but generically lead to modifications of the Wetterich equation. 
	
	Hence, we come to the conclusion that there are \gls{uv} actions/initial potentials which in general lead to an ill-posed initial value problem for the \gls{pde} \labelcref{eq:wetterich_equation_zero_dimensions} with standard mass-type regulators, while being well-defined on the level of the path integral (also in the presence of the regulator).\footnote{It is straightforward to construct additional counter examples.}
	The same problem is also present in (truncated) \gls{frg} flow equations in higher dimensions. 
	
	We refrain from discussing this issue any further here and defer it to future work. In our present work, we  focus on well-posed problems and only remark that one should always explicitly monitor the determinant/eigenvalues of the full two-point function at the beginning and during \gls{rg} flows over the entire field space.
	In addition, we note that potentials with an $O(2)$- or at least an $\mathbb{Z}_2 \times \mathbb{Z}_2$-symmetric asymptotic behavior do not feature the above problems and it is always possible to start with a sufficiently large $\Lambda$ that ensures positivity of both eigenvalues for all \gls{rg} times.\footnote{%
		It is even possible to construct potentials which are defined piecewise with an outer symmetric region and an inner region without symmetry that do not violate the above conditions, see below.} 
	As can be seen from \cref{eq:determinant_two-point_function,eq:eigenvalues_general}, explicit symmetry breaking terms linear in $\varphi_1$ or $\varphi_2$ do not cause any problems.
	We conjecture that the problematic initial conditions are directly linked to higher-order interaction terms that break the $\mathbb{Z}_2 \times \mathbb{Z}_2$ symmetry and are finite at scales~$k\neq \Lambda$. 
  	We add that this problem might therefore be solved by suitable modifications of the initial conditions of the Wetterich equation in cases where the  \gls{rg} flows are initialized at finite \gls{uv} scales.

\subsection{Generalization to \texorpdfstring{$O(\bar{N}) \times O(\bar{M})$}{O(N) times O(M)}-symmetric RG flow equations}
\label{sec:generalization_to_on_om} 

	Let us now also consider situations where the $N$-dimensional field space can be divided into two subspaces of dimension $\bar{N}$ and $\bar{M}$, such that $\bar{N} + \bar{M} = N$ and where the model is separately invariant under $O(\bar{N})$ and $O(\bar{M})$ transformations of the fields in the respective subspaces. 
	Hence, we study a zero-dimensional model whose classical action, (scale-dependent) effective (average) action and the (scale-dependent) potential are functions of the two corresponding invariants:
		\begin{align}
			& \rho_1 = \tfrac{1}{2} \, \vec{\phi}_1^{\, 2} \, ,	&&	\rho_2 = \tfrac{1}{2} \, \vec{\phi}_2^{\, 2} \, ,
		\end{align}
	where the entire field-space vector is split into two field-space vectors $\vec{\phi} = ( \phi_1, \ldots, \phi_N )^T = ( \vec{\phi}_1, \vec{\phi}_2 )^T$,
		\begin{align}
			& \vec{\phi}_1 = ( \phi_1, \ldots, \phi_{\bar{N}} )^T \, ,	&&	\vec{\phi}_2 = ( \underbrace{\phi_{\bar{N} + 1}, \ldots, \phi_N}_{\bar{M}} )^T \, .
		\end{align}
	From the perspective of the generating functional, the calculation of (connected) correlation functions and also vertex functions is still possible via simple numerical integration, see \cref{eq:correlation_functions} and \cref{sec:a_zero-dimensional_qft_of_scalar_fields}.
	This allows us to compute exact reference values for our flow equation studies. 

\subsubsection{Flow equation of the effective potential} 

	From the perspective of our \gls{rg} approach, the most general ansatz for the Wetterich \cref{eq:wetterich_equation_zero_dimensions} is the scale-dependent effective potential
		\begin{align}
			\tilde{\bar{\Gamma}} ( t, \varrho_1, \varrho_2 ) = \tilde{U} ( t, \varrho_1, \varrho_2 ) \,.
		\end{align}
	The initial condition is given by $\tilde{\bar{\Gamma}} ( t = 0, \varrho_1, \varrho_2 ) = \tilde{\mathcal{S}} ( t, \varrho_1, \varrho_2 )$. 
	Theoretically, we have the option to choose two different regulator-shape functions for the two subspaces, such that the regulator insertion is also only $O(\bar{N}) \times O(\bar{M})$-symmetric.
	Instead, for the sake of the simplicity, we work with an $O(N)$-symmetric regulator insertion which provides the same regulator-shape function for all fields, see \cref{eq:regulator_definition}.
	Basically, these are all ingredients that are required to derive the \gls{rg} flow equation for the potential $\tilde{U} ( t, \varrho_1, \varrho_2 )$ and we can follow the same steps as in \cref{subsubsec:flow_equation_of_the_effective_potential}. 

	It is convenient to derive all quantities in terms of the field-space invariants $\varrho_1$ and $\varrho_2$ instead of the individual fields $\vec{\varphi}_1$ and $\vec{\varphi}_2$.
	Furthermore, \gls{wlog}, we can choose a particular background field configuration which simplifies the calculation of the flow equation.
	For example, we can choose $\vec{\varphi}_1 = ( 0, \ldots, \sigma_1 )$ and $\vec{\varphi}_2 = ( \sigma_2, \ldots, 0 )$, such that the invariants are $\varrho_1 = \tfrac{1}{2} \, \sigma_1^2$ and $\varrho_2 = \tfrac{1}{2} \, \sigma_2^2$.
	It follows that the field-dependent two-point function (evaluated on this background-field configuration) reads
		\begin{align}
			\tilde{\bar{\Gamma}}^{( 2 )} ( t ) + R ( t ) =
			\begin{pmatrix}
				A_1	&	0	&	0
				\\
				0	&	B	&	0
				\\
				0 	&	0	&	A_2
			\end{pmatrix} \, ,
		\end{align}
	with
		\begin{align}
			A_{1/2} = \mathrm{diag} ( \underbrace{r + \partial_{\varrho_{1/2}} \tilde{U}, \ldots, r + \partial_{\varrho_{1/2}} \tilde{U}}_{\bar{N}/\bar{M}} ) \, ,
		\end{align}
	and
		\begin{align}
			B =
			\begin{pmatrix}
				r + \partial_{\varrho_1} \tilde{U} + 2 \varrho_1 \partial_{\varrho_1}^2 \tilde{U}	&	\sigma_1 \sigma_2 \, \partial_{\varrho_1} \partial_{\varrho_2} \tilde{U}
				\\
				\sigma_2 \sigma_1 \, \partial_{\varrho_2} \partial_{\varrho_1} \tilde{U}	&	r + \partial_{\varrho_2} \tilde{U} + 2 \varrho_2 \partial_{\varrho_2}^2 \tilde{U}
			\end{pmatrix}\,.
		\end{align}
	Because of the block-diagonal structure, an inversion of the regularized and field-dependent two-point function is possible and the Wetterich equation is well-defined, provided that the determinant of the block matrices are separately nonzero for all $t$ and $\varrho_{1/2}$.
	In the following, we shall assume that this is the case for the considered initial conditions.
	Inverting this two-point function and inserting the result together with the regulator \labelcref{eq:regulator_definition} in the Wetterich equation~\labelcref{eq:wetterich_equation_zero_dimensions}, we obtain the flow equation for the effective potential:
\begin{widetext}
		\begin{align}
			& \partial_t \tilde{U} =	\Vdistance	\label{eq:flow_equation_U_on_om}
			\\
			= \, & \frac{( \bar{N} - 1 ) \, \big( \frac{1}{2} \, \partial_t r \big)}{r + \partial_{\varrho_1} \tilde{U}} + \frac{( \bar{M} - 1 ) \, \big( \frac{1}{2} \, \partial_t r \big)}{r + \partial_{\varrho_2} \tilde{U}} + \frac{\big( \frac{1}{2} \, \partial_t r \big) \big( 2 r + \partial_{\varrho_1} \tilde{U} + 2 \varrho_1 \partial_{\varrho_1}^2 \tilde{U} + \partial_{\varrho_2} \tilde{U} + 2 \varrho_2 \partial_{\varrho_2}^2 \tilde{U} \big)}{\big( r + \partial_{\varrho_1} \tilde{U} + 2 \varrho_1 \partial_{\varrho_1}^2 \tilde{U} \big) \big( r + \partial_{\varrho_2} \tilde{U} + 2 \varrho_2 \partial_{\varrho_2}^2 \tilde{U} \big) - 4 \varrho_1 \varrho_2 \big( \partial_{\varrho_1} \partial_{\varrho_2} \tilde{U} \big) \big( \partial_{\varrho_2} \partial_{\varrho_1} \tilde{U} \big)} \, .	\Vdistance	\nonumber
		\end{align}
\end{widetext}
	Reformulated in terms of the background fields $\sigma_1$ and $\sigma_2$, which are associated with the directions of possibly existing condensates in higher-dimensional problems, we have
		\begin{align}
			& \partial_t U =	\Vdistance	\label{eq:flow_equation_U_on_om_sigma}
			\\
			= \, & \frac{( \bar{N} - 1 ) \, \big( \frac{1}{2} \, \partial_t r \big)}{r + \frac{1}{\sigma_1} \, \partial_{\sigma_1} U} + \frac{( \bar{M} - 1 ) \, \big( \frac{1}{2} \, \partial_t r \big)}{r + \frac{1}{\sigma_2} \, \partial_{\sigma_2} U} +	\Vdistance	\nonumber
			\\
			& + \frac{\big( \frac{1}{2} \, \partial_t r \big) \big( 2 r + \partial_{\sigma_1}^2 U + \partial_{\sigma_2}^2 U \big)}{\big( r + \partial_{\sigma_1}^2 U \big) \big( r + \partial_{\sigma_2}^2 U \big) - \big( \partial_{\sigma_1} \partial_{\sigma_2} U \big) \big( \partial_{\sigma_2} \partial_{\sigma_1} U \big)} \, .	\Vdistance	\nonumber
		\end{align}
	The reason for choosing the background fields~$\sigma_1$ and~$\sigma_2$ rather than the invariants~$\varrho_1$ and~$\varrho_2$ to span our spatial computational domain is to ensure a more straightforward handling of the boundary conditions in field space~\cite{Koenigstein:2021syz}, see also our discussion above.
	Note that, for $\bar{N} = \bar{M} = 1$, we recover the flow equation~\labelcref{eq:flow_equation_effective_potential} for the $\mathbb{Z}_2 \times \mathbb{Z}_2$-symmetric case with two fields.

	In general, the situation is now the same as in \cref{sec:flow_equation_zero-dim_model_two_fields}:
	We find a \gls{pde} for $U ( t, \sigma_1, \sigma_2 )$ whose spatial domain is $\sigma_{1/2} \in [ 0, \infty )$, $\mathbb{R}_{(\geq0)} \times \mathbb{R}_{(\geq0)}$ with a temporal evolution from $t = 0$ to $t \to \infty$.
	In total, this constitutes a $(2 + 1)$-dimensional \gls{pde} problem.
	Again, we shall not solve this problem for given \gls{uv} initial conditions for $U$ but consider the derivatives of $U$ \gls{wrt} the two field directions as evolving fluid fields.

\subsubsection{A fluid-dynamic reformulation}

	As above, we start our fluid-dynamic reformulation by defining 
		\begin{align}
			&	u = \partial_{\sigma_1} U \, ,	&&	v = \partial_{\sigma_2} U \, ,
		\end{align}
	and taking derivatives of \cref{eq:flow_equation_U_on_om_sigma} \gls{wrt} $\sigma_1$ and $\sigma_2$ to obtain two ``conservation'' laws for $u$ and $v$,
		\begin{align}
			& \partial_t
			\begin{pmatrix}
				u
				\\
				v
			\end{pmatrix} +
			\partial_{\sigma_1}
			\begin{pmatrix}
				f^x
				\\
				0
			\end{pmatrix}
			+ \partial_{\sigma_2}
			\begin{pmatrix}
				0
				\\
				f^y
			\end{pmatrix}
			=	\Vdistance	\label{eq:conservation_law_u_v_on_om}
			\\
			= \, & \partial_{\sigma_1}
			\begin{pmatrix}
				Q^x
				\\
				0
			\end{pmatrix}
			+ \partial_{\sigma_2}
			\begin{pmatrix}
				0
				\\
				Q^y
			\end{pmatrix} \, .	\Vdistance	\nonumber
		\end{align}
	Here,
		\begin{align}
			Q^{x/y} = \frac{\big( \frac{1}{2} \, \partial_t r \big) \big( 2 r + \partial_{\sigma_1} u + \partial_{\sigma_2} v \big)}{\big( r + \partial_{\sigma_1} u \big) \big( r + \partial_{\sigma_2} v \big) - \big( \partial_{\sigma_1} v \big) \big( \partial_{\sigma_2} u \big)}	\label{eq:diffusion_flux_u_v_on_om}
		\end{align}
	is again identified as a diffusion flux, see above, while
		\begin{align}
			f^{x/y} = \, & - \frac{( \bar{N} - 1 ) \, \big( \frac{1}{2} \, \partial_t r \big)}{r + \frac{1}{\sigma_1} \, u} - \frac{( \bar{M} - 1 ) \, \big( \frac{1}{2} \, \partial_t r \big)}{r + \frac{1}{\sigma_2} \, v} \, ,	\label{eq:advection_flux_u_v_on_om}
		\end{align}
	is identified as an advection flux.
	The explicit field dependence, which is an explicit position-dependence on the level of the \glspl{pde}, causes the advection flux to contain some contributions from the source. 
	Both is seen best, by executing the $\sigma_{1/2}$-derivatives in \cref{eq:conservation_law_u_v_on_om} and comparing the result to standard advection-diffusion equations with sources/sinks.
	For a detailed discussion we refer to \reff\cite{Koenigstein:2021syz}.
	
\section{Numeric approach}
\label{sec:numeric_approach} 

	Once the \gls{pde} problem is formulated in a conservative form, see \cref{eq:conservation_law_u_vec,eq:conservation_law_u_v_on_om}, we are free to choose the numerical method as long as we stick to the highly developed toolbox of numerical fluid dynamics and ensure the applicability of the respective scheme.
	
	For example, modern discontinuous Galerkin methods (and first-order upwind schemes for particular applications) have been successfully applied to flow equations formulated in a conservative form~\cite{Grossi:2019urj,Grossi:2021ksl,Ihssen:2022xjv,Ihssen:2022xkr,Ihssen:2023qaq}. 
	Loosely speaking, these methods are a variant of so-called finite element methods. 
	Here, we opt for the also well-established and related finite volume methods in terms of the \gls{kt} central scheme \cite{KTO2-0}, which was already used and tested in detail in our previous works~\cite{Koenigstein:2021syz,Koenigstein:2021rxj,Steil:2021cbu,Stoll:2021ori}.
	In particular, we apply the two-dimensional version of this scheme to \cref{eq:conservation_law_u_v,eq:conservation_law_u_v_on_om}.

	Readers, who are not interested in the explicit implementation of this scheme, may skip the remainder of this section. 

\subsection{The Kurganov-Tadmor central scheme}\label{sec:KT-central-scheme} 

	Here, we briefly recapitulate the two-dimensional \gls{kt} central scheme as well as some minor but relevant modifications of the scheme.
	For details, we refer to the original work by Kurganov and Tadmor \cite{KTO2-0} and to \reff\cite{Koenigstein:2021syz} for its one-dimensional version in the context of the \gls{frg} approach. 
	All details of its explicit two-dimensional implementation, which is used in this work, are presented in \cref{app:implementation_of_the_multi-dimensional_kt_central_scheme}, while all necessary underlying formulae are provided in the following, amended by short explanations but without derivation.

	In general, the two-dimensional version of the \gls{kt} central scheme is taylor-made for the solution of fluid-dynamic \glspl{pde} of the advection-diffusion type,
		\begin{align}
			& \partial_t \vec{u} + \tfrac{\mathrm{d}}{\mathrm{d} x} \, \vec{f}^x \, [ \vec{u} \, ] + \tfrac{\mathrm{d}}{\mathrm{d} y} \, \vec{f}^y \, [ \vec{u} \, ] =	\vdistance	\label{eq:general_pd_kt_scheme}
			\\
			= \, & \tfrac{\mathrm{d}}{\mathrm{d} x} \, \vec{Q}^x [ \vec{u}, \partial_x \vec{u}, \partial_y \vec{u} \, ] + \tfrac{\mathrm{d}}{\mathrm{d} y} \, \vec{Q}^y [ \vec{u}, \partial_x \vec{u}, \partial_y \vec{u} \, ] \, ,	\vdistance	\nonumber
		\end{align}
	where $\vec{u}$ is the vector of fluid fields, $t$ is the temporal coordinate and $x$ and $y$ denote the Cartesian spatial coordinates.\footnote{ %
		In principle, this can be straightforwardly generalized to higher-dimensional systems.
	}
	Furthermore, $\vec{f}^x$ and $\vec{f}^y$ are the advection fluxes and $\vec{Q}^x$ and $\vec{Q}^y$ are diffusion-type fluxes.
	All fluxes can be (highly) nonlinear functions of their arguments.
	Equations of this type therefore tend to form nonanalytic structures in the (weak) solution for $\vec{u}$ \cite{Ames:1992,Hesthaven2007,LeVeque:1992,LeVeque:2002,RezzollaZanotti:2013}, such as shock waves.
	These have to be handled by a numerical scheme which can for example be done via modern \gls{fv} methods.
	
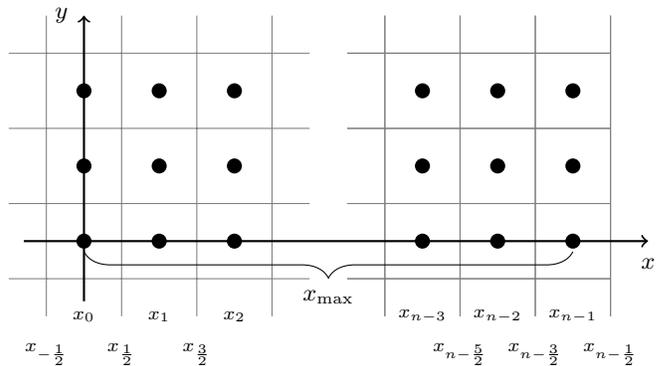
\begin{figure}
			\centering
			\begin{tikzpicture}[scale=2]
				\draw[step=0.5cm,gray, xshift=-0.25cm, yshift=-0.25cm] (-0.25,-0.25) grid (1.75,1.75);
				\draw[step=0.5cm,gray, xshift=2cm, yshift=-0.25cm] (-0.25,-0.25) grid (1.5,1.75);
		
				\draw[step=0.5cm,gray, xshift=2cm, yshift=-0.25cm, mark=*, mark size=1.5mm, only marks] (-0.25,-0.25) grid (1.5,1.75);
		
				\foreach \y in {0, 0.5, 1} {
					\foreach \x in {0,0.5,1,2.25,2.75,3.25} {
						\fill[color=black] (\x,\y) circle (0.05);
					}
				}
		
				\foreach \x/\y in {0/$x_0$,0.5/$x_1$,1/$x_2$,2.25/$x_{n - 3}$,2.75/$x_{n - 2}$,3.25/$x_{n - 1}$} {
					\draw (\x,-0.5) node {\scriptsize \y};
				}
		
				\foreach \x/\y in {-0.25/$x_{-\frac{1}{2}}$,0.25/$x_{\frac{1}{2}}$,0.75/$x_{\frac{3}{2}}$,2.5/$x_{n - \frac{5}{2}}$,3.0/$x_{n - \frac{3}{2}}$,3.5/$x_{n - \frac{1}{2}}$} {
					\draw (\x,-0.75) node {\scriptsize \y};
				}
		
				\draw [thick, ->] (-0.4,0) -- (3.75,0) node [black,yshift=-0.3cm] {$x$};
				\draw [thick, ->] (0,-0.4) -- (0,1.5) node [black,xshift=-0.3cm] {$y$};
				\draw [decorate, decoration={brace,amplitude=10pt, mirror},yshift=-0.07cm]
				(0,0) -- (3.25,0) node [black,midway,yshift=-0.6cm] {$x_{\mathrm{max}}$};
		
			\end{tikzpicture}
			\caption{\label{fig:finite_volume_discretization}%
				Sketch of the two-dimensional \gls{fv} discretization of the computational domain.
				Cell centers are marked with bullets.
			}
		\end{figure}

	\gls{fv} methods are based on a partitioning of the spatial computational domain of the \gls{pde} into small cells of finite volumes.
	In our calculations, we shall use a rectangular regular mesh of equally sized volume cells of size $\Delta x \cdot \Delta y$.
	The cell centers are located at positions $( x_{j_x}, y_{j_y} )$ with cell boundaries at 
	\begin{align}
	x_{j_x \pm \frac{1}{2}} = x_{j_x} \pm \frac{\Delta x}{2}\quad\ \text{and}\quad x_{j_y \pm \frac{1}{2}} = x_{j_y} \pm \frac{\Delta y}{2}\,, 
	\end{align}
	see \cref{fig:finite_volume_discretization}.
	This partitioning of the computational domain can be used to turn the \gls{pde} \labelcref{eq:general_pd_kt_scheme} into its weak integral form for each volume cell and time step.
	To this end, one first computes the spatial volume integral over both sides of \cref{eq:general_pd_kt_scheme} and defines the cell-average of the fluid fields $\vec{u}$ in a cell with cell center $( x_{j_x}, y_{j_y} )$ as follows:
		\begin{align}
			\vec{\bar{u}}_{j_x, j_y} ( t ) = \tfrac{1}{\Delta x \, \Delta y} \int_{x_{j_x - \frac{1}{2}}}^{x_{j_x + \frac{1}{2}}} \mathrm{d} x \int_{y_{j_y - \frac{1}{2}}}^{y_{j_y + \frac{1}{2}}} \mathrm{d} y \, \vec{u} ( t, x, y ) \, .	\label{eq:cell_average}
		\end{align}
	This already turns the \gls{pde} in Eq.~\labelcref{eq:general_pd_kt_scheme} into a coupled set of \glspl{ode} for the cell averages.
	However, in order to perform a single time step of the \gls{ode} system from $t^i$ to some $t^{i + 1} = t^i + \Delta t$ with some small $\Delta t$ via some \gls{ode} integrator, one has to evaluate the spatial integrals and the temporal integration from $t^i$ to $t^{i + 1}$ over the advection and diffusion fluxes.
	This evaluation of the flux integrals or their approximate evaluation is at the heart of any \gls{fv} scheme and constitutes a so-called Riemann problem at each cell interface for each time step which has to be solved \cite{Ames:1992,LeVeque:1992,LeVeque:2002,Hesthaven2007,RezzollaZanotti:2013}.
	
	There are \gls{fv} schemes which are based on the exact or approximate solution of these Riemann problems.
	Interestingly, the \gls{kt} scheme does not rely on such a Riemann solver or the characteristic decomposition of the (advection) fluxes.
	The only information, which is required is the spectral radius of the Jacobian of the advection fluxes to estimate the characteristic velocities at each cell interface. 
	
	Luckily, we do not have to construct the \gls{kt} scheme here or recapitulate its entire construction.
	We can simply use the results from \reff\cite{KTO2-0}.
	First of all, the result is a fully discrete scheme for the computation of time steps of size $\Delta t$, which is second order accurate in $\Delta x$ and $\Delta y$.
	The particular appealing aspect of the \gls{kt} scheme is that there is a semi-discrete reduction of the scheme (a controlled limit $\Delta t \to 0$) -- being continuous in time, though discrete and still second-order accurate in space.
	This makes the scheme a ready-made black-box solver which can be combined with basically any \gls{ode} time steppers.\footnote{For recent advances on the problem of time stepping of \gls{rg} flow equations at late \gls{rg} times (the deep \gls{ir}), we again refer to \reff\cite{Ihssen:2023qaq}.}
	Furthermore, there exist several improved or adapted versions of this semi-discrete scheme, which yield higher-order accuracy under certain circumstances \cite{KTO2-1,KTO3-0,KTO5-0,KT-MP5,KTmovingMesh}.
	However, these are not discussed in this work.
	
	For this work, we simply need the following formulae which are all continuous in time and discrete in space.
	The corresponding semi-discrete version of the \gls{pde} \labelcref{eq:general_pd_kt_scheme}, \ie, the \glspl{ode} for the cell-averages read	
		\begin{align}
			& \partial_t \vec{\bar{u}}_{j_x, j_y} =	\Vdistance	\label{eq:semi_discrete_pd_kt_scheme}
			\\
			= \, & - \frac{\vec{H}^x_{j_x + \frac{1}{2}, j_y} - \vec{H}^x_{j_x - \frac{1}{2}, j_y}}{\Delta x} - \frac{\vec{H}^y_{j_x, j_y + \frac{1}{2}} - \vec{H}^y_{j_x, j_y - \frac{1}{2}}}{\Delta y} +	\Vdistance	\nonumber
			\\
			& + \frac{\vec{P}^x_{j_x + \frac{1}{2}, j_y} - \vec{P}^x_{j_x - \frac{1}{2}, j_y}}{\Delta x} - \frac{\vec{P}^y_{j_x, j_y + \frac{1}{2}} - \vec{P}^y_{j_x, j_y - \frac{1}{2}}}{\Delta y} \, .	\Vdistance	\nonumber
		\end{align}
	The corresponding numerical advection fluxes to the adjacent cells in $x$- and $y$-direction are
\begin{widetext}
		\begin{align}
			\vec{H}^x_{j_x + \frac{1}{2}, j_y} = \, & \tfrac{1}{2} \Big( \vec{f}^x \, \Big[ \vec{u}^{\, +}_{j_x + \frac{1}{2}, j_y} \Big] + \vec{f}^x \, \Big[ \vec{u}^{\, -}_{j_x + \frac{1}{2}, j_y} \Big] \Big) - \tfrac{1}{2} \, a^x_{j_x + \frac{1}{2}, j_y} \cdot \Big( \vec{u}^{\, +}_{j_x + \frac{1}{2}, j_y} - \vec{u}^{\, -}_{j_x + \frac{1}{2}, j_y} \Big) \, ,	\Vdistance	\label{eq:advection_fluxes_discrete_kt_x}
			\\
			\vec{H}^y_{j_x, j_y + \frac{1}{2}} = \, & \tfrac{1}{2} \, \Big( \vec{f}^y \, \Big[ \vec{u}^{\, +}_{j_x, j_y + \frac{1}{2}} \Big] + \vec{f}^y \, \Big[ \vec{u}^{\, -}_{j_x, j_y + \frac{1}{2}} \Big] \Big) - \tfrac{1}{2} \, a^y_{j_x, j_y + \frac{1}{2}} \cdot \Big( \vec{u}^{\, +}_{j_x, j_y + \frac{1}{2}} - \vec{u}^{\, -}_{j_x, j_y + \frac{1}{2}} \Big) \, ,	\Vdistance	\label{eq:advection_fluxes_discrete_kt_y}
		\end{align}
	whereas the $x$- and $y$-numerical diffusion fluxes are given by
		\begin{align}
			\vec{P}^x_{j_x + \frac{1}{2}, j_y} = \, & \tfrac{1}{2} \, \Big( \vec{Q}^x \, \Big[ \vec{\bar{u}}_{j_x, j_y}, \tfrac{\vec{\bar{u}}_{j_x + 1, j_y} - \vec{\bar{u}}_{j_x, j_y}}{\Delta x}, ( \partial_y \vec{u} \, )_{j_x, j_y} \Big] + \vec{Q}^x \, \Big[ \vec{\bar{u}}_{j_x + 1, j_y}, \tfrac{\vec{\bar{u}}_{j_x + 1, j_y} - \vec{\bar{u}}_{j_x, j_y}}{\Delta x}, ( \partial_y \vec{u} \, )_{j_x + 1, j_y} \Big] \Big) \, ,	\Vdistance	\label{eq:diffusion_fluxes_discrete_kt_x}
			\\
			\vec{P}^y_{j_x, j_y + \frac{1}{2}} = \, & \tfrac{1}{2} \, \Big( \vec{Q}^y \, \Big[ \vec{\bar{u}}_{j_x, j_y}, ( \partial_x \vec{u} \, )_{j_x, j_y}, \tfrac{\vec{\bar{u}}_{j_x, j_y + 1} - \vec{\bar{u}}_{j_x, j_y}}{\Delta y} \Big] + \vec{Q}^y \, \Big[ \vec{\bar{u}}_{j_x + 1, j_y}, ( \partial_x \vec{u} \, )_{j_x, j_y + 1}, \tfrac{\vec{\bar{u}}_{j_x, j_y + 1} - \vec{\bar{u}}_{j_x, j_y}}{\Delta y} \Big] \Big) \, .	\Vdistance	\label{eq:diffusion_fluxes_discrete_kt_y}
		\end{align}
\end{widetext}
	The calculation of these numerical fluxes first requires the value of the fluid vector $\vec{u}$ on the cell interfaces.
	This is reconstructed from the cell averages via a piecewise linear reconstruction from both sides of the corresponding cell interface:
		\begin{align}
			\vec{u}^{\, \pm}_{j_x + \frac{1}{2}, j_y} = \, & \vec{\bar{u}}_{j_x + 1, j_y} \mp \tfrac{\Delta x}{2} \, ( \partial_x \vec{u} \, )_{j_x + \frac{1}{2} \pm \frac{1}{2}, j_y} \, ,	\vdistance	\label{eq:reconstruction_x}
			\\
			\vec{u}^{\, \pm}_{j_x, j_y + \frac{1}{2}} = \, & \vec{\bar{u}}_{j_x, j_y + 1} \mp \tfrac{\Delta y}{2} \, ( \partial_y \vec{u} \, )_{j_x, j_y + \frac{1}{2} \pm \frac{1}{2}} \, .	\vdistance	\label{eq:reconstruction_y}
		\end{align}
	Here and for the calculation of the diffusion fluxes, one needs to estimate the gradient of each fluid field in each cell, $( \partial_{x/y} \vec{u} \, )_{j_x, j_y}$.
	This estimate is also based on the cell averages. 
	Componentwise we find
		\begin{align}
			& ( \partial_x u^\alpha  )_{j_x, j_y} =	\Vdistance	\label{eq:gradient_estimation_x}
			\\
			= \, & f_{\mathrm{limiter}}\bigg(\frac{\bar{u}^{\, \alpha}_{j_x + 1, j_y} - \bar{u}^{\, \alpha}_{j_x, j_y}}{\Delta x}, \frac{\bar{u}^{\, \alpha}_{j_x, j_y} - \bar{u}^{\, \alpha}_{j_x - 1, j_y}}{\Delta x}\bigg) \, ,	\Vdistance	\nonumber
			\\
			& ( \partial_y u^\alpha  )_{j_x, j_y} =	\Vdistance	\label{eq:gradient_estimation_y}
			\\
			= \, & f_{\mathrm{limiter}}\bigg(\frac{\bar{u}^{\, \alpha}_{j_x, j_y+1} - \bar{u}^{\, \alpha}_{j_x, j_y}}{\Delta y}, \frac{\bar{u}^{\, \alpha}_{j_x, j_y} - \bar{u}^{\, \alpha}_{j_x, j_y-1}}{\Delta y}\bigg) \, .	\Vdistance	\nonumber
		\end{align}
		Here, $f_{\mathrm{limiter}}$ is a flux limiter function which avoids an over- or underestimate of the slopes which would lead to spurious oscillations in the solution.\footnote{Sometimes one also uses limiters which are additionally functions of the central difference.}
	The explicit choice of the limiter is up to the user and valid limiters are presented in \reffs\cite{RezzollaZanotti:2013,LeVeque:1992,wikiFluxLimiter}.
	In this work, we simply use the MinMod limiter,
	\begin{align}
		\label{eq:minmod_flux_limiter}
		f_{\mathrm{MinMod}}(a, b) =
		\begin{cases}
			\min(|a|, |b|) \, , & \text{if} \quad a \cdot b > 0 \, ,	\vdistance
			\\
			0 \, , & \text{otherwise} \, ,	\vdistance
		\end{cases}
	\end{align}
	see Ref.~\cite{KTO2-0}.

	At this point, we emphasize that we have to slightly adapt the original \gls{kt} scheme for our purposes. 
	During the benchmark tests of this work, but also in computations in a related work \cite{Rais:2024jio}, we experienced that using the limited derivatives in \cref{eq:gradient_estimation_x,eq:gradient_estimation_y} in the contributions to the diffusion fluxes in \cref{eq:diffusion_fluxes_discrete_kt_x,eq:diffusion_fluxes_discrete_kt_y} leads to incorrect results in some cases.
	Spurious oscillations can form in the solution which may originate from an underestimate of the gradients orthogonal to the direction of the diffusion flux.
	This leads to an artificial asymmetry of the fluxes and too little diffusion.
	We solve this by replacing \cref{eq:gradient_estimation_x,eq:gradient_estimation_y} with central difference stencils,
		\begin{align}
			& ( \partial_x u^\alpha  )_{j_x, j_y} = \frac{\bar{u}^{\, \alpha}_{j_x + 1, j_y} - \bar{u}^{\, \alpha}_{j_x - 1, j_y}}{2 \, \Delta x} \, ,	\label{eq:gradient_estimation_x_modified}	\Vdistance
			\\
			& ( \partial_y u^\alpha  )_{j_x, j_y} = \frac{\bar{u}^{\, \alpha}_{j_x, j_y + 1} - \bar{u}^{\, \alpha}_{j_x, j_y - 1}}{2 \, \Delta y} \, ,	\label{eq:gradient_estimation_y_modified}	\Vdistance
		\end{align}
	in the diffusion terms in \cref{eq:diffusion_fluxes_discrete_kt_x,eq:diffusion_fluxes_discrete_kt_y} only.
	However, for the reconstruction of the fluid fields on the cell interfaces in \cref{eq:reconstruction_x,eq:reconstruction_y}, we still use the limited derivatives in \cref{eq:gradient_estimation_x,eq:gradient_estimation_y} which ultimately enter the contributions to the advection fluxes \cref{eq:advection_fluxes_discrete_kt_x,eq:advection_fluxes_discrete_kt_y}.
	
	The final component of our scheme is the advection-velocity estimates of the cell boundaries,
		\begin{align}
			& a^x_{j_x + \frac{1}{2}, j_y} =	\Vdistance
			\\
			= \, & \max \Bigg[ \hat{\rho} \, \Bigg( \frac{\partial \vec{f}}{\partial \vec{u}} \, \Big[ \vec{u}^{\, +}_{j_x + \frac{1}{2}, j_y} \Big] \Bigg), \hat{\rho} \, \Bigg( \frac{\partial \vec{f}}{\partial \vec{u}} \, \Big[ \vec{u}^{\, -}_{j_x + \frac{1}{2}, j_y} \Big] \Bigg) \Bigg] \, ,	\Vdistance	\nonumber
			\\
			& a^y_{j_x, j_y + \frac{1}{2}} =	\Vdistance
			\\
			= \, & \max \Bigg[ \hat{\rho} \, \Bigg( \frac{\partial \vec{g}}{\partial \vec{u}} \, \Big[ \vec{u}^{\, +}_{j_x, j_y + \frac{1}{2}} \Big] \Bigg), \hat{\rho} \, \Bigg( \frac{\partial \vec{g}}{\partial \vec{u}} \, \Big[ \vec{u}^{\, -}_{j_x, j_y + \frac{1}{2}} \Big] \Bigg) \Bigg] \, ,	\Vdistance	\nonumber
		\end{align}
	where $\hat{\rho} ( A ) = \max \{ | \lambda_1 |, \ldots, | \lambda_\omega |\}$ is the spectral radius of the matrix $A$, with $\lambda_k$ being the eigenvalues.
	
	Apart from these formulae, no additional information is required to set up the core of our numerical scheme.
	However, for the numerical implementation on a compact domain, it is more suitable to present this scheme in a matrix-type formulation, as done in \cref{app:implementation_of_the_multi-dimensional_kt_central_scheme}, instead of the local formulation presented here.
	
\subsection{Adaptions of the KT scheme to our FRG problem(s)}
\label{sec:adaptions_of_the_kt_scheme_to_our_frg_problems}
	
	Next, we comment on some adaptions of the above presented scheme to the \gls{frg} flow equations of the effective potential.
	Therefore, we first identify the field-space variables with the spatial variables, thus $x = \varphi_1$ and $y = \varphi_2$, and the \gls{rg} time $t$ with the temporal parameter $t$.
	
	Anyhow, it is clear that
		\begin{enumerate}
			\item	a \gls{pde} problem is exclusively well-defined by specifying its boundary conditions and their numerical implementation, which we have not done yet,
			
			\item	typical fluxes within the \gls{frg} approach, such as \cref{eq:diffusion_flux}, are usually \gls{rg}-scale/time-dependent (in addition to the dependencis in \cref{eq:general_pd_kt_scheme}) and can comprise explicit dependences on the (field-space) position.
			Furthermore, in some truncations, one is confronted with additional coupled \glspl{ode},
			
			\item	our conservation law \labelcref{eq:conservation_law_u_v} matches the form of \cref{eq:general_pd_kt_scheme} for vanishing advection fluxes.
		\end{enumerate}
	Let us comment on these issues:
		\begin{enumerate}
			\item	Actually, the \gls{pde} in Eq.~\labelcref{eq:conservation_law_u_v} forms an initial value problem on the noncompact domain $\mathbb{R}^2$ and boundary conditions are not required to have a well-posed problem.
			Since we cannot compute numerically on noncompact domains, we restrict ourselves to the compact domain $[ - \varphi_{1, \max}, \varphi_{1, \max} ] \times [ - \varphi_{2, \max}, \varphi_{2, \max} ]$ and impose artificial boundary conditions at the domain boundaries.
			This is done in the \gls{kt} scheme by introducing two additional ghost cells at each domain boundary.
			In \reff\cite{Koenigstein:2021syz} we discussed in great detail that a linear extrapolation of the fluid field $\vec{u} = ( u, v )^T$ currently appears to be a decent choice, provided that the values of $\varphi_{i, \max}$ are large enough, such that fluxes are either suppressed or the in- and out-flux averages to net zero.
			In our present work we use this linear extrapolation from the last two physical cells to the ghost cells and choose $\varphi_{i, \max}$ according to the empirical knowledge that has been gained in \reff\cite{Koenigstein:2021syz}.
			For further details on the discussion of boundary conditions, we refer to this reference.

			For the \gls{pde} in Eq.~\labelcref{eq:conservation_law_u_v_on_om}, however, we have additional boundary conditions at $\varrho_{1/2} = 0 = \sigma_{1/2}$, whereas the boundary conditions at $\sigma_{1/2} = \pm \infty$ are the same as for the previous case and also handled in the same way.
			Using the formulation in $\sigma_{1/2}$, the new boundary conditions are easily derived, see again \reff\cite{Koenigstein:2021syz}.
			The functions $u$ and $v$ have the following symmetry properties with respect to the $\sigma_{1/2}$-axis,
				\begin{align}
					u ( \sigma_1, \sigma_2 ) = \, & - u ( - \sigma_1, \sigma_2 ) = u ( \sigma_1, - \sigma_2 ) \, ,	\vdistance
					\\
					v ( \sigma_1, \sigma_2 ) = \, & - v ( \sigma_1, - \sigma_2 ) = v ( - \sigma_1, \sigma_2 ) \, .	\vdistance
				\end{align}
			These can be directly derived from the symmetry properties of the effective potential $U$.
			On the level of the cell averages of the \gls{kt} scheme this is again implemented via the ghost cells which are given by (see also \cref{fig:finite_volume_discretization}):
				\begin{align}
					&	\bar{u}_{-1, j_y} = - \bar{u}_{1, j_y} \, ,	&&	\bar{u}_{-2, j_y} = - \bar{u}_{2, j_y} \, ,	\vdistance
					\\
					&	\bar{v}_{-1, j_y} = \bar{v}_{1, j_y} \, ,	&&	\bar{v}_{-2, j_y} = \bar{v}_{2, j_y} \, ,	\vdistance
					\\
					&	\bar{u}_{j_x, -1} = \bar{u}_{j_x, 1} \, ,	&&	\bar{u}_{j_x, -2} = \bar{u}_{j_x, 2} \, ,	\vdistance
					\\
					&	\bar{v}_{j_x, -1} = - \bar{v}_{j_x, 1} \, ,	&&	\bar{v}_{j_x, -2} = - \bar{v}_{j_x, 2} \, .	\vdistance
				\end{align}
			In addition, by symmetry, we know that the following cell averages are also fixed:
				\begin{align}
					&	\bar{u}_{0, j_y} = 0 \, ,	&&	\bar{v}_{j_x, 0} = 0 \, .	\vdistance
				\end{align}
			We emphasize that these boundary conditions pose a challenge:
			Consider for example the numerical advection flux in $x$-direction ($\sigma_1$ corresponds to $x$-direction) in \cref{eq:advection_fluxes_discrete_kt_x} with \cref{eq:advection_flux_u_v_on_om}.
			Due to the explicit position dependence, it is impossible to numerically evaluate this flux at the boundary cells with $j_y = 0$ because we would have to divide the cell averages $\bar{v}_{j_x, 0} = 0$ by $y_0 = 0$.\footnote{At $j_x = 0$, there is no problem for the flux in $x$-direction because the fluxes are only evaluated at the cell boundaries which are not located at $x_0 = 0$, see \cref{fig:finite_volume_discretization}.}
			The same happens along the $\sigma_2 = y$ axis for the flux in $y$-direction and the first term in \cref{eq:advection_flux_u_v_on_om}.
			We solve this issue by using the estimate
				\begin{align}
					\lim_{\sigma_1 \to 0} \tfrac{1}{\sigma_1} \, u \simeq \lim_{\sigma_1 \to 0} \partial_{\sigma_1} u \, ,
				\end{align}
			motivated by l'H\^opital's rule.
			However, this converts the respective part of the advection fluxes into a diffusion flux.
			Hence, for all cells along the domain boundaries at $\sigma_{1/2} = 0$, we modified the diffusion and advection fluxes \labelcref{eq:diffusion_flux_u_v_on_om,eq:advection_flux_u_v_on_om} as follows:
				\begin{align}
					f^{x} \big|_{\sigma_2 = 0} = \, & - \frac{( \bar{N} - 1 ) \, \big( \frac{1}{2} \, \partial_t r \big)}{r + \frac{1}{\sigma_1} \, u} \, ,	\Vdistance
					\\
					Q^{x} \big|_{\sigma_2 = 0} = \, & \frac{\big( \frac{1}{2} \, \partial_t r \big) \big( 2 r + \partial_{\sigma_1} u + \partial_{\sigma_2} v \big)}{\big( r + \partial_{\sigma_1} u \big) \big( r + \partial_{\sigma_2} v \big) - \big( \partial_{\sigma_1} v \big) \big( \partial_{\sigma_2} u \big)}   + 	\Vdistance	\nonumber
					\\
					& 
					+ \frac{( \bar{M} - 1 ) \, \big( \frac{1}{2} \, \partial_t r \big)}{r + \partial_{\sigma_2} v} \, ,	\Vdistance
				\end{align}
			and similarly for the fluxes in $y$-direction. 
			Even though this seems cumbersome, we did not find another solution, \eg, by ``moving" the grid.
			Note that the sign in front of the last term in the diffusion flux stems from moving the term to the other side of the equation, see \cref{eq:conservation_law_u_v_on_om}.
			
			In general, the above formulation of the \gls{kt} scheme does not explicitly include the boundary conditions and ghost cells.
			Therefore, it is better for the numerical implementation to discuss the \gls{kt} scheme in a matrix formulation in terms of pseudo code, where the ghost cells are explicitly included, see \cref{app:implementation_of_the_multi-dimensional_kt_central_scheme}.
			
			\item	Indeed, the original \gls{kt} scheme was presented for fluxes which exclusively depend on the fluid fields and their derivatives.
			Though, there is no reason why the scheme could be spoiled, if the fluxes are $t$-dependent or additional \glspl{ode} are coupled to the \gls{pde}.
			However, if the (advection) fluxes gain explicit position dependences, the situation is different because fundamental properties of the \gls{kt} scheme, such as being \gls{tvd}/\gls{tvni} get formally lost.
			In our previous works~\cite{Koenigstein:2021syz,Koenigstein:2021rxj,Steil:2021cbu}, however, we experienced by explicit benchmark tests that even position dependent advection fluxes do not seem to invalidate the applicability of the scheme.
			Nevertheless, great caution and detailed testing is in order when numerical methods for nonlinear \glspl{pde} are run at the edge of their applicability.
			
			\item	Of course, the conservation law \labelcref{eq:conservation_law_u_v} is represented as the viscous limit of \cref{eq:general_pd_kt_scheme} without any advection fluxes.
			However, the scheme is still applicable, as is demonstrated by the benchmark tests in \reff\cite{Koenigstein:2021syz} and, \eg, tests for the heat equation.
		\end{enumerate}
	Finally we comment on the initial condition for the \glspl{pde}.
	From \cref{eq:limits_effective_average_action} we deduce that the initial condition of the coupled \glspl{pde} \labelcref{eq:diffusion_equation_u,eq:diffusion_equation_v} at $t = 0$ is given by the field-space derivatives of the classical action (the \gls{uv} potential).
	On the level of the cell averages in \cref{eq:cell_average}, this implies that the discretized initial condition of the \gls{fv} scheme is calculated as follows (\eg, for $v ( t = 0, \vec{\varphi} \, )$),
		\begin{align}
			& \bar{v}_{j_x, j_y} ( 0 ) =	\Vdistance\nonumber
			\\
			= \, & \tfrac{1}{\Delta x \, \Delta y} \int_{x_{j_x - \frac{1}{2}}}^{x_{j_x + \frac{1}{2}}} \mathrm{d} x \int_{y_{j_y - \frac{1}{2}}}^{y_{j_y + \frac{1}{2}}} \mathrm{d} y \, v ( 0, x, y ) =	\Vdistance	\nonumber
			\\
			= \, & \tfrac{1}{\Delta x \, \Delta y} \int_{x_{j_x - \frac{1}{2}}}^{x_{j_x + \frac{1}{2}}} \mathrm{d} x \int_{y_{j_y - \frac{1}{2}}}^{y_{j_y + \frac{1}{2}}} \mathrm{d} y \, \partial_y U ( 0, x, y ) =	\Vdistance	\nonumber
			\\
			= \, & \tfrac{1}{\Delta x \, \Delta y} \int_{x_{j_x - \frac{1}{2}}}^{x_{j_x + \frac{1}{2}}} \mathrm{d} x \, \big[ U ( 0, x, y_{j_y + \frac{1}{2}} ) - U ( 0, x, y_{j_y - \frac{1}{2}} ) \big] \, ,	\Vdistance	\label{eq:initial_condition_discrete}
		\end{align}
	and analogously for $u ( t = 0, \vec{\varphi} \, )$.
	Hence, we can and should make use of the fact that we have direct access to $U ( t = 0, \vec{\varphi} \, )$, see also \reff\cite{Koenigstein:2021syz}.
	Usually, the remaining integral in \cref{eq:initial_condition_discrete} has to be evaluated with high precision which is however not a substantial problem.

\section{General test setup}
\label{sec:testing_setup}

	\begin{figure*}
		\includegraphics[width=\textwidth]{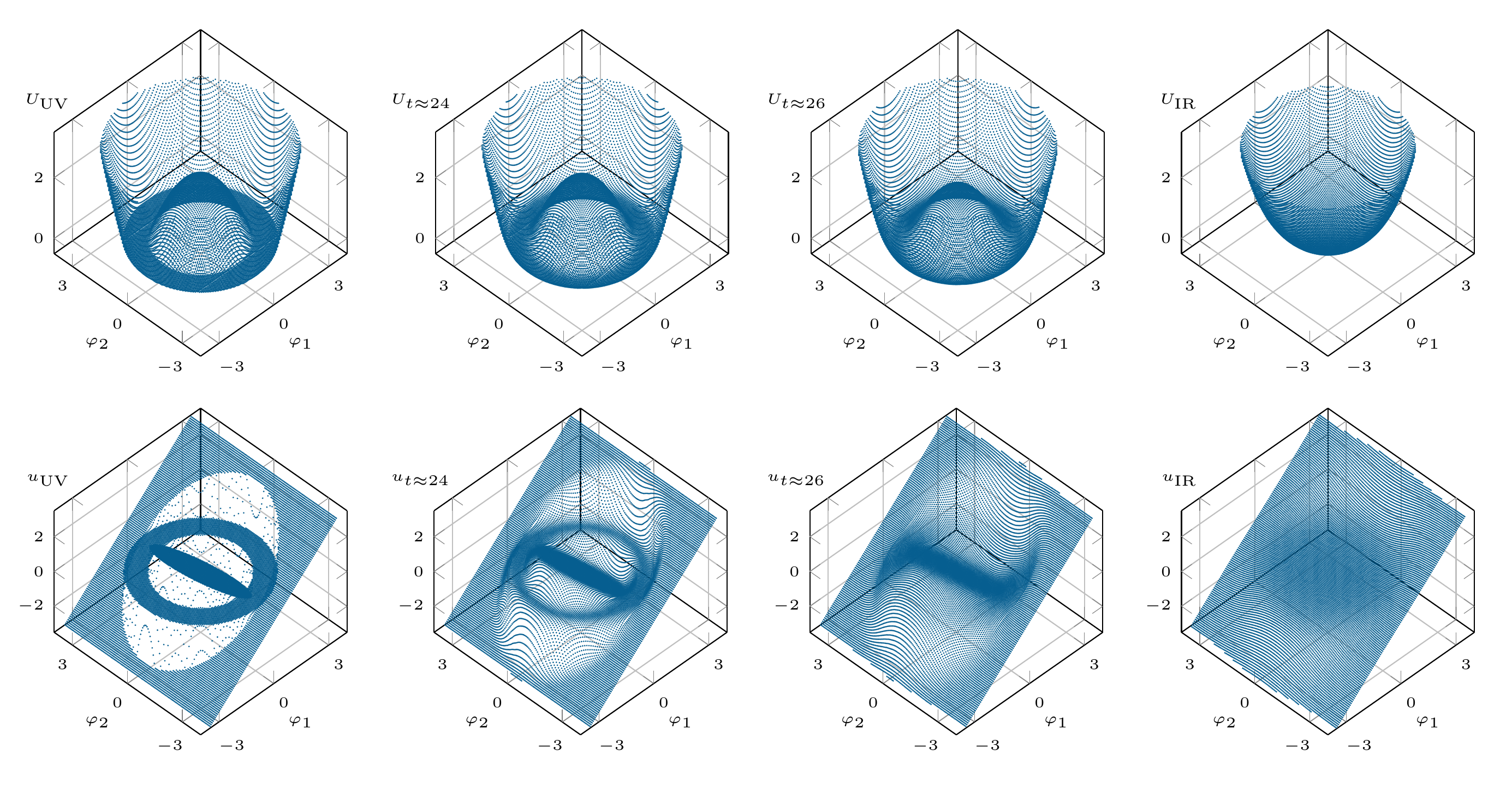}
		\caption{\label{fig:plot_case=1}%
			The \gls{rg} time evolution of the potential $U ( t, \vec{\varphi} \, )$ (upper row) and its $\varphi_1$-derivative $u ( t, \vec{\varphi} \, )$ (lower row) from the \gls{uv} (left column) to the \gls{ir} (right column) and selected intermediate times for test case I, \cref{eq:test_case_uv_potential_i}.
		}
	\end{figure*}

	Our present work deals with the development and tests of a fluid dynamic formulation of \gls{frg} problems which are two-dimensional in field space.
	Explicitly, we want to present a proof of concept that the numeric approach, which was presented in \reff\cite{Koenigstein:2021syz}, namely the application of the 1D \gls{kt} central scheme to the $O(N)$-symmetric flow equation of the effective potential, can be generalized and applied to flow equations of effective potentials that are two-dimensional in field space.
	In complete analogy to \reff\cite{Koenigstein:2021syz}, we therefore need benchmark tests which demonstrate that the spatial discretization scheme is applicable and performs well but also clearly shows its limitations at the same time.
	
	By taking over some findings from our earlier works concerning decent \gls{uv} and \gls{ir} cutoffs and appropriate sizes of the artificial computational domain in field space, our present study can mostly focus on the discretization errors stemming from the \gls{kt} scheme.
	These errors are directly caused by the finite resolution $\Delta x$ and $\Delta y$ in field space as well as a possible artificial breaking of a continuous symmetry in field space by our Cartesian discretization scheme.
	
	Therefore, we basically consider three testing scenarios:
		\begin{enumerate}
			\item	We study our field theory of two scalar fields under the additional assumption of $O(2)$ symmetry in field space.
			Still, we stick to the flow equations \labelcref{eq:diffusion_equation_u,eq:diffusion_equation_v}, which are two-dimensional in field space and solve the full two-dimensional \gls{pde} problem.
			From the \gls{ir} solution we can then extract $\Gamma^{(2)}$ at the \gls{ir} minimum and compare this result for different grid spacings $\Delta x$, $\Delta y$ against the exact result from the path integral as well as against the results from the one-dimensional reduction, see \cref{subsec:one-dimensional_reduction}.
			This is particularly interesting because we know that there is no dynamical symmetry breaking in zero-dimensional \glspl{qft}, such that the \gls{ir} minimum is always trivial, \ie, $\vec{\varphi} = 0$, see \reff\cite{Koenigstein:2021syz}.
			Hence, the results are not contaminated by errors which emerge from a location of the minimum in field space.
			Furthermore, it can be shown that the \gls{ir} potential has to be convex and smooth and should of course still be globally $O(2)$-symmetric.	
			Whether this is still (approximately) the case on a rectangular grid is subject of our investigations.
			Note that all these aspects are totally independent of the specific choice of the \gls{uv} action/potential.
			
			More specifically, if we choose the \gls{fv} grid such that there is a cell with cell center exactly at $\varphi_1 = \varphi_2 = 0$, we can extract $\Gamma^{(2)}$ from the solution via the finite difference stencil
				\begin{align}
					& \Gamma^{(2)} =	\vdistance	\label{eq:gamma_2_kt}
					\\
					\quad \qquad & = 
					\begin{cases}
						\partial_{\varphi_1} u ( t, \vec{\varphi} \, ) \big|_{t \to \infty, \vec{\varphi}_{\min} = 0} \simeq \frac{\bar{u}_{1, 0} - \bar{u}_{0, 0}}{\Delta x} + \mathcal{O} ( \Delta x ) \, ,	
						\\
						\partial_{\varphi_2} v ( t, \vec{\varphi} \, ) \big|_{t \to \infty, \vec{\varphi}_{\min} = 0} \simeq \frac{\bar{v}_{0, 1} - \bar{v}_{0, 0}}{\Delta y} + \mathcal{O} ( \Delta y ) \, .	
					\end{cases}	\nonumber
				\end{align}
			The results can then be compared with the exact results as well as the results from the one-dimensional formulation of the problem, see \cref{subsec:one-dimensional_reduction} below.
			
			The quality of the $O(2)$ symmetry is tested with the help of the $L^1$ and $L^\infty$ norms/errors,
				\begin{align}
					\qquad \mathcal{O}_{U, L^1} = \, & \tfrac{1}{\# ( i, j )} \sum_{i,j} \big(|\mathrm{rot}_{90^{\circ}}(\bar U)_{i,j} - \bar U_{i,j}|\big) \, ,	\vdistance	\label{eq:observable_U_L1}
					\\
					\qquad \mathcal{O}_{u, L^1} = \, & \tfrac{1}{\# ( i, j )} \sum_{i,j} \big(|\mathrm{rot}_{90^{\circ}}(\bar u/x)_{i,j} - \bar u_{i,j}/x_{i}|\big) \, ,	\vdistance	\label{eq:observable_u_L1}
				\end{align}
			and
				\begin{align}
					\quad \mathcal{O}_{U, L^\infty} = \, & \max_{i,j}\big(|\mathrm{rot}_{90^{\circ}}(\bar U)_{i,j} - \bar U_{i,j}|\big) \, ,	\vdistance	\label{eq:observable_U_Linfinity}
					\\
					\quad \mathcal{O}_{u, L^\infty} = \, & \max_{i,j}\big(|\mathrm{rot}_{90^{\circ}}(\bar u/x)_{i,j} - \bar u_{i,j}/x_{i}|\big) \, ,	\vdistance	\label{eq:observable_u_Linfinity}
				\end{align}
			respectively. 
			Here, $\bar U_{i,j}$ is calculated in the \gls{ir} via a simple Riemann sum\footnote{%
				We use the following algorithm for the Riemann summation for the calculation of $\bar U_{i, j}$:
				(i) Set $\bar U^{u}_{0,j} = 0$ and $\bar U^{v}_{i,0} = 0$ for all $i,j$.
				(ii) Define
				\begin{align}
					\bar U^{u}_{i, j} =
					\begin{cases}
						\bar U^{u}_{i - 1, j} + \frac{\Delta x}{2} (\bar u_{i - 1, j} + \bar u_{i, j}) & \text{for $i > 0$} \, ,
						\\
						\bar U^{u}_{i + 1, j} - \frac{\Delta x}{2} (\bar u_{i + 1, j} + \bar u_{i, j}) & \text{for $i < 0$} \, ,
					\end{cases}
				\end{align}
			and
				\begin{align}
					\bar U^{v}_{i,j} =
					\begin{cases}
						\bar U^{v}_{i, j - 1} + \frac{\Delta y}{2} (\bar v_{i, j - 1} + \bar v_{i, j}) & \text{for $j > 0$} \, ,
						\\
						\bar U^{v}_{i, j + 1} - \frac{\Delta y}{2} (\bar v_{i, j + 1} + \bar v_{i, j}) & \text{for $j < 0$} \, .
					\end{cases}
				\end{align}
			(iii) Combine $\bar U_{i, j} = \bar U^{u}_{i, j} + \bar U^{v}_{0, j}$ and $\bar U_{i, j} = \bar U^{u}_{i, 0} + \bar U^{v}_{i, j}$.
			Both procedures are equivalent and the results are identical.
			} from the cell averages $\bar u_{i,j}$ and $\bar v_{i,j}$ and $\mathrm{rot_{90^\circ}}(\cdot)$ means a $90^\circ$ rotation around the origin, \ie, $i=j=0$.
			Furthermore, $\# ( i, j )$ is the number of cells included in the sum.
			Since we are using a Cartesian discretization, we are restricted to $90^{\circ}$ rotations.
			The first observable is based on the potential, whereas the second one is based on the cell averages $\bar u_{i,j}$.
			Therefore, the latter observable is slightly more interesting as it only comprises errors from the violation of the $O(2)$ symmetry from the \gls{kt} method. 
			Additional errors from the numeric integration can be excluded.
			For details, we refer to the discussion in \cref{subsubsec:quantitative_benchmark_tests}.
			
			All these tests are based on some conventional and some uncommon choices of \gls{uv} initial potentials which were presented in \reff\cite{Koenigstein:2021syz} as hard tests for the spatial discretization scheme.
			For comparison, we also show results from the dimensionally reduced formulation of the two-dimensional problem as presented in \reff\cite{Koenigstein:2021syz}.
			
			\item	We study the field theory with minimal additional symmetry and simply consider \gls{uv} initial potentials of two interacting scalar fields.
			The potentials are asymptotically still at least $\mathbb{Z}_2 \times \mathbb{Z}_2$ symmetric to ensure well-posedness of the problem (\ie, well-behaved initial conditions), \cf\ \cref{eq:initial_conditions}, but contain also nonsymmetric interactions for small~$| \vec{\varphi} \, |$.

			One of the initial potentials is constructed such that we have a nonvanishing expectation value for~$\langle \vec{\phi} \, \rangle$.
			Here, we test the following aspects of the two-dimensional \gls{pde} approach to our \gls{frg} framework:
			(i) The precision of computations of the location of the \gls{ir} minimum $\vec{\varphi}_{\min}$ of the potential in field space.
			To that end, we use $\vec{\varphi}_{\min} = \langle \vec{\phi} \, \rangle$ from the path integral as reference.
			(ii) We again test the precision of our numerical scheme by calculating the two-point functions $\Gamma^{(2)}_{\varphi_1 \varphi_1}$, $\Gamma^{(2)}_{\varphi_1 \varphi_2} = \Gamma^{(2)}_{\varphi_2 \varphi_1}$, and $\Gamma^{(2)}_{\varphi_2 \varphi_2}$ at the minimum of the potential and benchmark these values against the exact result from \cref{eq:two-point_vertex}.
			
			Another example is a \gls{uv} potential, which has again a trivial \gls{ir} minimum and a global $\mathbb{Z}_2 \times \mathbb{Z}_2$ symmetry.
			However, by misaligning the discretization axes with the symmetry axes, we can test the violation of such symmetries by the \gls{kt} scheme.
			Furthermore, we test how the \gls{kt} scheme performs in the presence of a nonanalytic \gls{uv} potential.

			\item Both of the previous test setups are based on the ``diffusion-only'' setup, \ie, the advection fluxes are zero for the case with two scalar fields.
			To test the application of the full \gls{kt} scheme, we therefore also consider a problem with $O(\bar{N}) \times O(\bar{M})$ symmetry, see \cref{sec:generalization_to_on_om}.
			Here, we again know that the \gls{ir} minimum is trivial, but for $\bar{N} > 1$ and/or $\bar{M} > 1$ we have nontrivial advection fluxes.
			This benchmark test also includes the calculation of $\Gamma^{(2)}$ at the trivial \gls{ir} minimum and the comparison against the exact result from the path integral.
			Here, we can again use \cref{eq:gamma_2_kt} to extract $\Gamma^{(2)}$ from the solution.
		\end{enumerate}
	In summary, these are valuable tests for assessing the quality of numerical results that will be obtained in future applications of our scheme to higher-dimensional models.
	In particular, our error estimates will help to rank the relevance of systematic errors from truncation schemes, approximations, cutoffs, the time stepping, and the numerical errors which are exclusively linked to the (spatial) discretization.
	Note that we shall already present a first application of our scheme to a two-dimensional problem in three-dimensional spacetime, which is closer to realistic applications.
	
	Before we present the results from all these tests, let us introduce our explicit test models in the next section. 

\section{Zero-dimensional test models with \texorpdfstring{$O(2)$}{O(N)} symmetry} 
\label{sec:o_2_symmetric_model}

	\begin{figure*}
		\includegraphics[width=\textwidth]{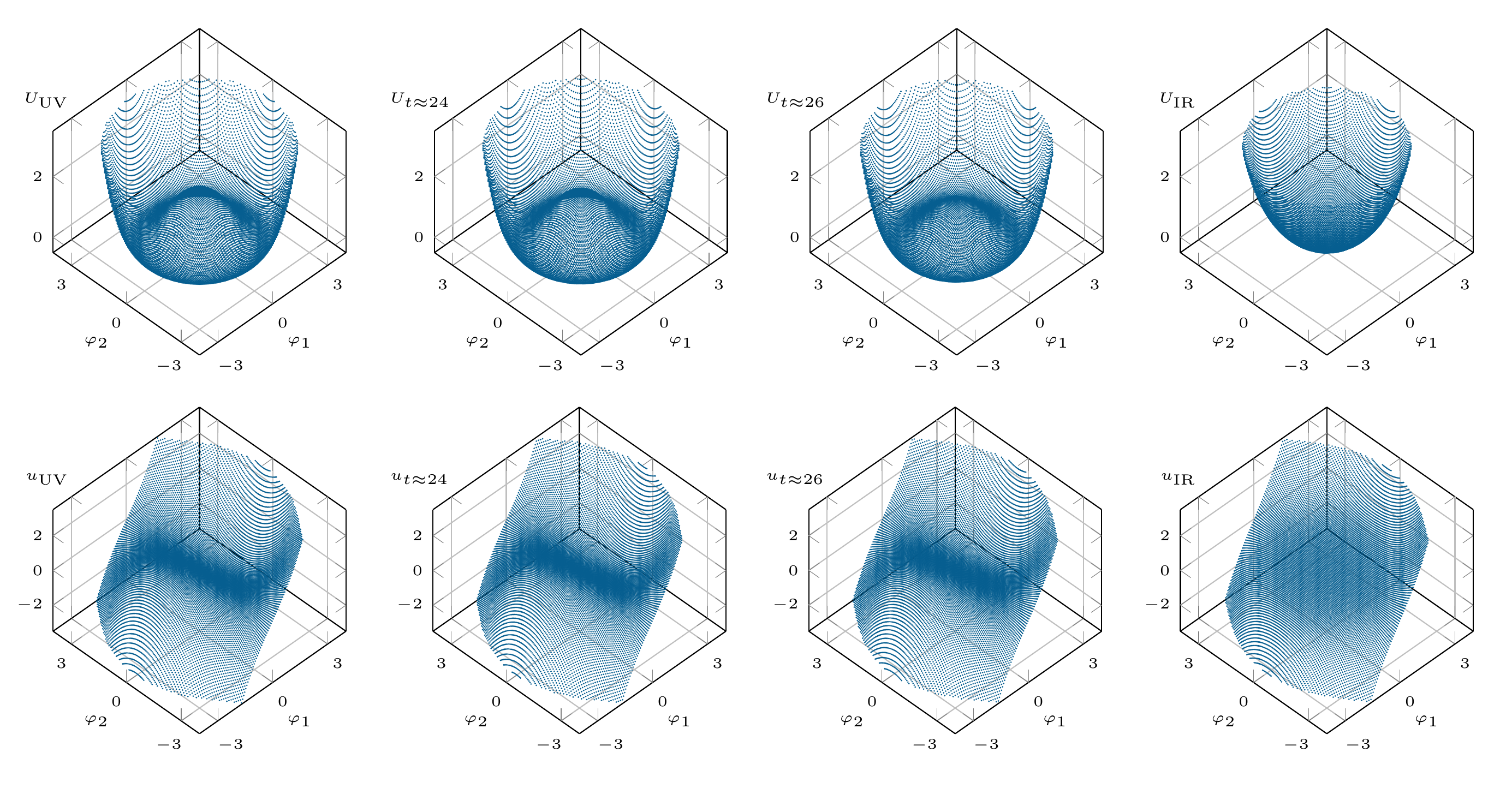}
		\caption{\label{fig:plot_case=2}%
			The \gls{rg} time evolution of the potential $U ( t, \vec{\varphi} \, )$ (upper row) and its $\varphi_1$-derivative $u ( t, \vec{\varphi} \, )$ (lower row) from the \gls{uv} (left column) to the \gls{ir} (right column) and selected intermediate times for test case II, see \cref{eq:test_case_uv_potential_ii}.
		}
	\end{figure*}

	In this section we introduce our first test environment, the $O ( 2 )$-symmetric model.
	
\subsection{Reference values from the path integral formalism}

	Assuming that the classical action $\mathcal{S} ( \vec{\phi} \, )$ and the partition function \labelcref{eq:generating_funcitonal_of_correlation_funcitons} at vanishing source fields $\vec{J} = 0$ are invariant under $O(2)$ transformations of the quantum fields,
		\begin{align}
			\vec{\phi} \mapsto \vec{\phi}^{\, \prime} = O \, \vec{\phi} \, 
		\end{align}
	with
		\begin{align}
			&	O =
			\begin{pmatrix}
				\cos ( \alpha )	&	\sin ( \alpha )
				\\
				- \sin ( \alpha )	&	\cos ( \alpha )
			\end{pmatrix} \, ,
			&&	\alpha \in [ 0, 2 \uppi ) \, ,
		\end{align}
	it is clear that the classical action has to be a function of the $O(2)$-invariant $\rho = \tfrac{1}{2} \, \vec{\phi}^{\, 2}$ rather than the independent fields $\phi_1$ and $\phi_2$:
		\begin{align}
			\mathcal{S} ( \vec{\phi} \, ) = \tilde{\mathcal{S}} ( \rho ) \, .
		\end{align}
	This implies that, using the $O(2)$-symmetry, all correlation functions \labelcref{eq:correlation_functions} can be expressed in terms of one-dimensional integrals:
		\begin{align}
			\langle ( \vec{\phi}^{\, 2} )^n \rangle = \frac{2^n \int_{0}^{\infty} \mathrm{d} \rho \, \rho^{n} \, \ee^{- \tilde{\mathcal{S}} ( \rho )}}{\int_{0}^{\infty} \mathrm{d} \rho \, \ee^{- \tilde{\mathcal{S}} ( \rho )}} \, ,	\label{eq:correlation_function_o_n}
		\end{align}
	see \reffs\cite{Fraboulet:2021amf,Keitel:2011pn,Koenigstein:2021syz} for details.
	In particular, we find that all correlation functions \labelcref{eq:correlation_functions} with an odd number of fields vanish.
	For our present work, the only relevant correlation function is
		\begin{align}
			\langle \phi_i \, \phi_j\rangle = \delta_{ ij} \, \tfrac{1}{2} \, \langle \vec{\phi}^{\, 2} \rangle \, .
		\end{align}
	Using \cref{eq:two-point_vertex}, we have
		\begin{align}
			\Gamma^{( 2 )} \equiv \Gamma^{( 2 )}_{\varphi_i \varphi_i} = \frac{2}{\langle \vec{\phi}^{\, 2} \rangle}	\label{eq:gamma_2_reference_o_n}
		\end{align}
	for the two-point vertex function.
	We can now simply solve \cref{eq:conservation_law_u_vec} numerically for $O(2)$-symmetric initial conditions, extracting $\Gamma^{( 2 )}$, and comparing this to the reference result from \cref{eq:gamma_2_reference_o_n}.
		
	However, due to the $O(2)$ symmetry, we could also reduce the spatial domain of the \gls{pde} in Eq.~\labelcref{eq:flow_equation_effective_potential} to a one-dimensional domain and solve this one-dimensional fluid-dynamical system numerically as is usually done in \gls{frg} literature, \cf\ Sec.\ $\mathrm{V}$ of \reff\cite{Koenigstein:2021syz}.
	Results from this formulation provide us with reference values and additional benchmark.

\subsection{\texorpdfstring{$O(2)$}{O(2)}-symmetric test cases} 

	In the following, we briefly recapitulate some explicit test cases for our comparisons which were developed in \reff\cite{Koenigstein:2021syz} and comment on the reasoning behind their choice.
	The corresponding reference values for $\Gamma^{(2)}$ for each potential are listed in \cref{tab:exact_gamma_2_o_2}.
		\begin{table}[b]
			\caption{\label{tab:exact_gamma_2_o_2}%
				The table lists the (up to numerical-integration errors) exact results for the $\Gamma^{(2)}$ of the $O ( 2 )$-model with the initial \gls{uv} potentials as given in \cref{eq:test_case_uv_potential_i,eq:test_case_uv_potential_ii,eq:test_case_uv_potential_iii,eq:test_case_uv_potential_iv}.
				These values have been obtained by a high-precision one-dimensional numerical integration of the expectation values using \cref{eq:correlation_function_o_n,eq:gamma_2_reference_o_n} for $n = 1$ with Mathematica's numerical integration routine \texttt{NIntegrate} \cite{Mathematica:12.2} with a \texttt{PrecisionGoal} and \texttt{AccuracyGoal} of 10.
				Here, we present the first ten digits only.
			}
			\begin{ruledtabular}
				\setlength\extrarowheight{2pt}
				\begin{tabular}{c c}
					test case	&	$\Gamma^{(2)}$
					\\
					I			&	0.295702274(6)
					\\
					II			&	0.316367789(4)
					\\
					III			&	0.178669819(6)
					\\
					IV			&	0.321461336(0)
				\end{tabular}
			\end{ruledtabular}
		\end{table}

\subsubsection{Test case I: nonanalytic initial condition}
\label{subsubsec:test_case_i}

	The \gls{uv} initial potential of test case I, which is plotted in \cref{fig:plot_case=1} (upper left panel), reads
		\begin{align}
			U ( \vec{\phi} \, ) =
			\begin{cases}
				- \tfrac{1}{2} \, \vec{\phi}^{\, 2} \, ,						&	\text{if} \quad | \vec{\phi} \, | \leq 2 \, ,	\vdistance
				\\
				- 2 \, ,														&	\text{if} \quad 2 < | \vec{\phi} \, | \leq 3 \, ,	\vdistance
				\\
				\tfrac{1}{2} \, ( \vec{\phi}^{\, 2} - 13 ) \, ,	&	\text{if} \quad 3 < | \vec{\phi} \, | \, ,	\vdistance
			\end{cases}	\label{eq:test_case_uv_potential_i}
		\end{align}
	and was chosen for the following reasons:
		\begin{enumerate}
			\item	Because of the quadratic asymptotic behavior of this potential, the linear extrapolation at large $| \vec{\phi} \, |$ for its derivative at the boundary of the computational domain does not generate any errors.
			Therefore one can completely focus on the small-$| \vec{\phi} \, |$ region.
			
			\item	The nonanalytic points $| \vec{\phi} \, | = 2$ and $| \vec{\phi} \, | = 3$ give rise to jump discontinuities in the derivatives of the potential, see \cref{fig:plot_case=1} (lower left panel), which are the fluid fields in the two \glspl{pde} \labelcref{eq:diffusion_equation_u,eq:diffusion_equation_v} and therefore represent challenging but manageable tests for modern schemes from \gls{cfd}. All schemes that cannot cope with discontinuities are bound to fail.
			
			Furthermore, the \gls{ir} potential of a zero-dimensional \gls{qft} has to be smooth, such that discontinuities in all derivatives need to be smeared out during the flow without spurious oscillations.
			
			\item	The potential comes with an infinite number of nontrivial degenerate minima in the \gls{uv}, whereas the minimum in the \gls{ir} has to be unique and the potential needs to become convex~\cite{Keitel:2011pn,Moroz:2011thesis,Koenigstein:2021syz}.
			Thus, symmetry restoration has to be handled numerically in a stable way.
		\end{enumerate}
	
\subsubsection{Test case II: \texorpdfstring{$\phi^4$}{phi 4}-theory}
\label{subsubsec:test_case_ii}

	Test case II is the zero-dimensional version of a $\phi^4$-theory with negative mass term, \ie,
		\begin{align}
			U ( \vec{\phi} \, ) = - \tfrac{1}{2} \, \vec{\phi}^{\, 2} + \tfrac{1}{4!} \, ( \vec{\phi}^{\, 2} )^2 \, ,	\label{eq:test_case_uv_potential_ii}
		\end{align}
	see \cref{fig:plot_case=2} (upper left panel) for an illustration. 
	From the perspective of the powerful \gls{kt} scheme, this case is not a challenge, but -- being the standard (toy) model of theoretical physics -- it has to be included in our analysis.
	In \reff\cite{Koenigstein:2021syz}, however, it was demonstrated that, although the \gls{uv} potential is smooth, truncations of the potential in terms of Taylor expansions with fixed expansion point do not converge and a full solution of the \gls{pde} is required to obtain small relative errors of vertex functions in the \gls{ir} limit. 

	This test case can be used to study symmetry restoration in the \gls{rg} flow.
	Furthermore, this potential allows to test the applicability of the linear extrapolation at the artificial large-$| \vec{\phi} \, |$ boundary of the computational domain (despite the cubic asymptotic behavior of the derivative of the potential, see lower left panel of \cref{fig:plot_case=2}) and to find a reasonable size of the computational domain.
	Here, we do not need to repeat this part of the analysis and simply take suitable values from \reff\cite{Koenigstein:2021syz}.

\subsubsection{Test case III: \texorpdfstring{$\phi^6$}{phi 6}-theory} 
\label{subsubsec:test_case_iii}

	\begin{figure*}
		\includegraphics[width=\textwidth]{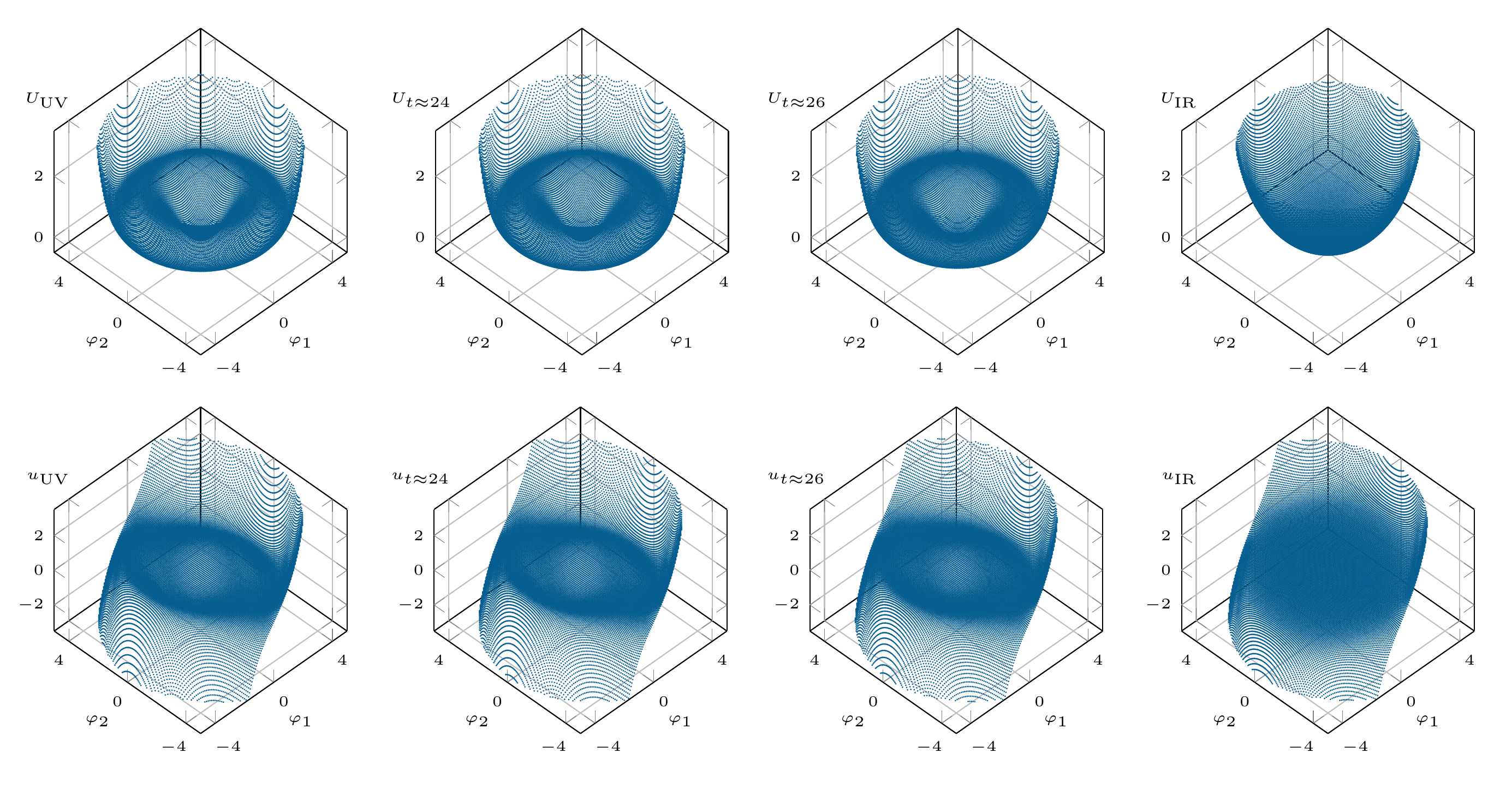}
		\caption{\label{fig:plot_case=3}%
			The \gls{rg} time evolution of the potential $U ( t, \vec{\varphi} \, )$ (upper row) and its $\varphi_1$-derivative $u ( t, \vec{\varphi} \, )$ (lower row) from the \gls{uv} (left column) to the \gls{ir} (right column) and selected intermediate times for test case III, see \cref{eq:test_case_uv_potential_iii}.
		}
	\end{figure*}
	The \gls{uv} potential of test case III reads
		\begin{align}
			U ( \vec{\phi} \, ) = \tfrac{1}{2} \, \vec{\phi}^{\, 2} - \tfrac{1}{20} \, ( \vec{\phi}^{\, 2} )^2 + \tfrac{1}{6!} \, ( \vec{\phi}^{\, 2} )^3	\label{eq:test_case_uv_potential_iii}
		\end{align}
	and is illustrated in \cref{fig:plot_case=3} (upper left panel).
	With this potential  the failure of the Taylor expansion of the effective potential can be demonstrated, see also \reff\cite{Koenigstein:2021syz}. 
	A potential of this type, which exhibits several local minima separated by a potential barrier, is expected to describe the dynamics in the vicinity of first order phase transitions.
	Therefore, an analysis of  its \gls{rg} flow and the merging of all minima in $\vec{\phi} = 0$ in the \gls{ir} is of great relevance also for higher-dimensional applications.

\subsubsection{Test case IV: pole in the derivative}
\label{subsubsec:test_case_iv}

		\begin{figure*}
			\includegraphics[width=\textwidth]{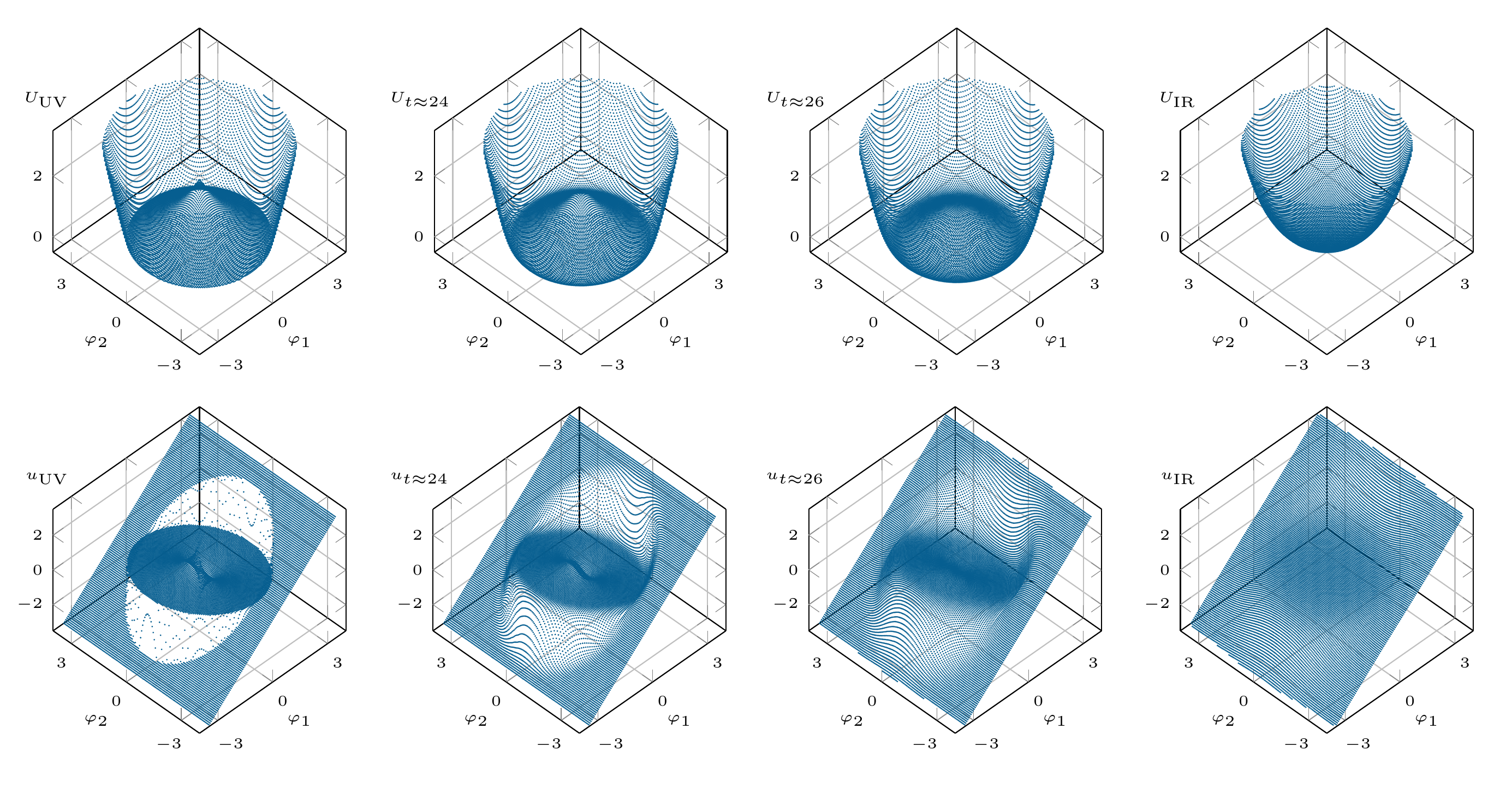}
			\caption{\label{fig:plot_case=4}%
				The \gls{rg} time evolution of the potential $U ( t, \vec{\varphi} \, )$ (upper row) and its $\varphi_1$-derivative $u ( t, \vec{\varphi} \, )$ (lower row) from the \gls{uv} (left column) to the \gls{ir} (right column) and selected intermediate times for test case IV, see \cref{eq:test_case_uv_potential_iv}.
			}
		\end{figure*}
	The fourth test case is given by the \gls{uv} potential
		\begin{align}
			U ( \vec{\phi} \, ) =
			\begin{cases}
				- ( \vec{\phi}^{\, 2} )^{\frac{1}{3}} \, ,	&	\text{if} \quad | \vec{\phi} \, | \leq \sqrt{8} \, ,	\vdistance
				\\
				\tfrac{1}{2} \, \vec{\phi}^{\, 2} - 6 \, ,	&	\text{if} \quad \sqrt{8} < | \vec{\phi} \, | \, .	\vdistance
			\end{cases}	\label{eq:test_case_uv_potential_iv}
		\end{align}
	For a visualization of this potential, we refer to \cref{fig:plot_case=4} (upper left panel). 
	By inspecting its field-space derivatives, see \cref{fig:plot_case=4} (lower left panel), we observe that both, $u = \partial_{\phi_1} U$ and $v = \partial_{\phi_2} U$ have a pole at $\phi_{1/2} = 0$.
	Poles in the fluid field certainly represent another challenge for our numerical setup in terms of a resolving nonanalytic structures and jump discontinuities.
	
\section{Zero-dimensional test models without \texorpdfstring{$O(2)$}{O(2)}~symmetry}
\label{sec:non-symmetric_model}

	We also employ two test models without $O(2)$ symmetry.
	Both can simply be understood as field theories of two scalar fields $\phi_1$ and $\phi_2$ which are (self-)interacting in a single point via some very complicated mechanism or simply as some statistical models with two random variables.
	The reason, why we define the corresponding initial potentials in terms of piecewise functions, each with quartic asymptotic behavior, is that the artificial boundary conditions of the computational domain at large $| \vec{\phi} \, |$ are then essentially irrelevant since linear extrapolation for its field-derivatives is justified.
	This allows us to focus on the small-$| \vec{\phi} \, |$ region and the handling of the dynamics of the fluid fields $u$ and $v$ during the \gls{rg} flow via the diffusive part of the \gls{kt} scheme.
	Numerical errors should therefore not stem from the boundary conditions but solely from the discretization of the \gls{pde} and the time stepping.

\subsection{Test case V: nonvanishing field expectation value} 
\label{subsec:test_case_v_nonsymmetric}

\begin{figure*}
			\includegraphics[width=\textwidth]{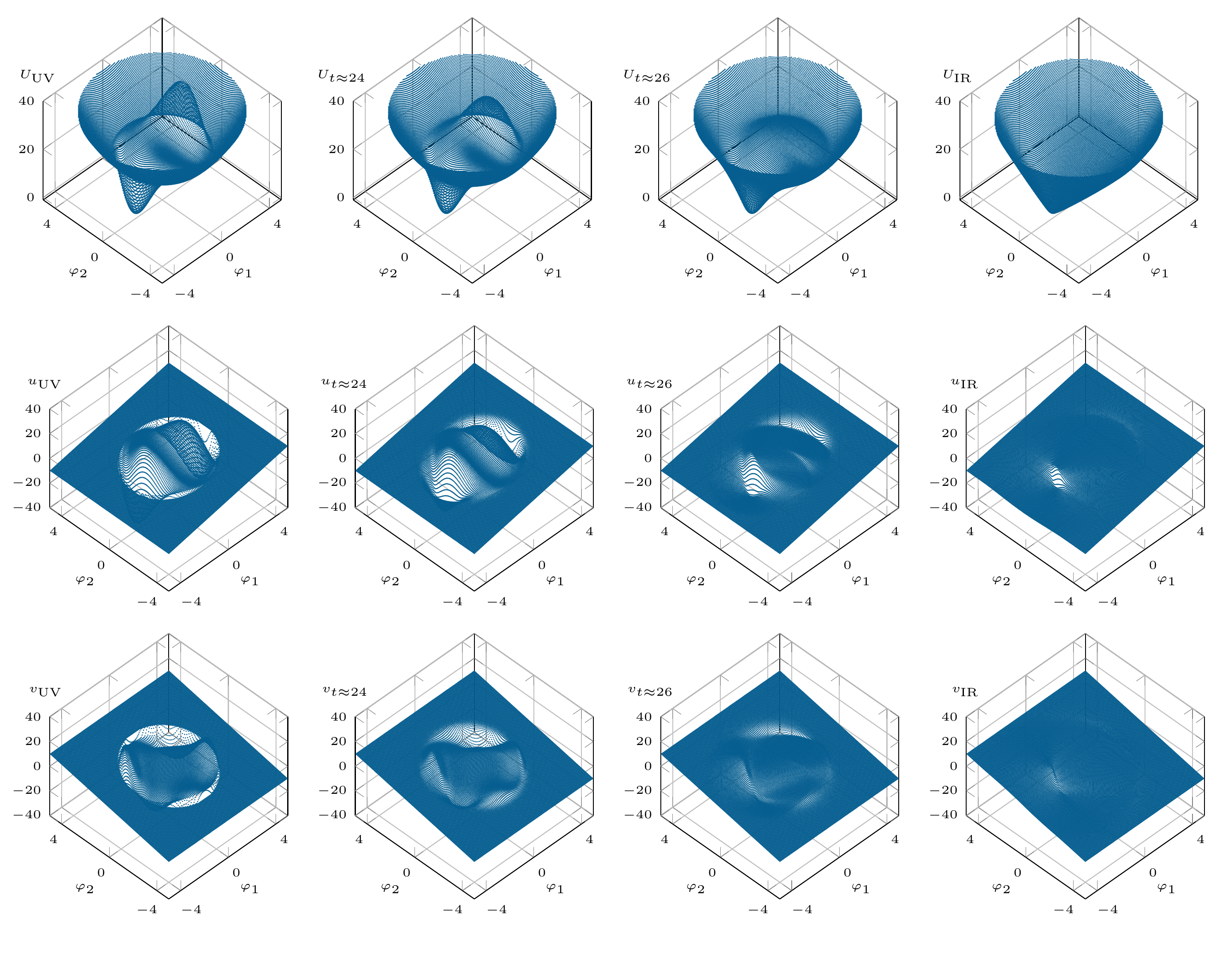}
			\caption{\label{fig:test_nonsymmetric}%
				The \gls{rg} time evolution of the potential $U ( t, \vec{\varphi} \, )$ (upper row), its $\varphi_1$-derivative $u ( t, \vec{\varphi} \, )$ (middle row), and $\varphi_2$-derivative $v ( t, \vec{\varphi} \, )$ (lower row) from the \gls{uv} (left column) to the \gls{ir} (right column) and selected intermediate times for test case V, see \cref{eq:test_nonsymmetric_uv_potential}.
			}
		\end{figure*}

	The corresponding \gls{uv} potential that we created reads
		\begin{align}
			& U ( \vec{\phi} \, ) =	\vdistance	\label{eq:test_nonsymmetric_uv_potential}
			\\
			= \, &
			\begin{cases}
				2 \, ( \phi_1^3 + \phi_1 \, \phi_2 ) \big[ \cos \big( \tfrac{\uppi \vec{\phi}^{\, 2}}{9} \big) + 1 \big]	&	\text{if} \quad | \vec{\phi} \, | \leq 3 \, ,	\vdistance
				\\
				\vec{\phi}^{\, 2} - 9	&	\text{if} \quad | \vec{\phi} \, | > 3 \, ,	\vdistance
			\end{cases}	\nonumber
		\end{align}
	and is illustrated in \cref{fig:test_nonsymmetric} (upper left panel).
	We constructed this potential for the following reasons:
	First, its asymptotic behavior ensures that the potential is bounded from below and the probability distribution is well defined.
	Furthermore, the asymptotic behavior allows for a linear extrapolation of the field derivatives at the boundary of the computational domain.
	The regime defined by  $| \vec{\phi} \, | \leq 3$ represents the actual challenge in calculations. 
	Here, the \gls{uv} potential has one global minimum at $\vec{\phi}_{\min} \simeq ( -2.07, 0.33 )^T$.
	Additionally, there is no symmetry in the small $| \vec{\phi} \, |$ region, which causes the \gls{ir} minimum to be nontrivial too.
	Inspecting the derivatives of the potential, see \cref{fig:test_nonsymmetric} (middle left and lower left panel), we also find that these contain cusps as well as rather large gradients which pose additional challenges in numerical calculations.

	Indeed, from the general formula for the correlation functions, \cref{eq:correlation_functions}, we find by direct numerical integration (using \texttt{NIntegrate} \cite{Mathematica:12.2} with a \texttt{PrecisionGoal} and \texttt{AccuracyGoal} of 12)
		\begin{align}
			\langle \vec{\phi} \, \rangle = \vec{\varphi}_{\min} =
			\begin{pmatrix}
				-2.040906130(6)
				\\
				\hphantom{-} 0.330687529(6)
			\end{pmatrix}	\label{eq:field_expectation_value}
		\end{align}
	for the expectation values of the fields.
	The \gls{ir} minimum of the potential computed with our \gls{frg} formalism must be identical to this result. 
	As expected from the \gls{uv} potential, see \cref{fig:test_nonsymmetric} (upper left panel), the field expectation value is at negative $\varphi_1$ and minimally shifted towards positive $\varphi_2$ and therefore almost identical to the \gls{uv} minimum.
	Hence, the question is whether the field space resolution of the \gls{kt} scheme suffices to correctly locate this minimum and therefore to correctly reproduce the result in \cref{eq:field_expectation_value}.
	From \cref{eq:correlation_functions}, we can also directly calculate the two-point correlation functions (using the same numerical integration routine as above)
		\begin{align}
			& \langle \phi_i \, \phi_j \rangle =	\Vdistance
			\\
			= \, &
			\begin{pmatrix}
				\hphantom{-} 4.184919188(1)	&	- 0.661996231(7)
				\\
				- 0.661996231(7)		&	\hphantom{-} 0.188507337(2)
			\end{pmatrix}_{ i j} \, .	\Vdistance	\nonumber
		\end{align}
	Using \cref{eq:two-point_vertex}, this leads us to the following values of the two-point vertex functions at the \gls{ir} minimum:
		\begin{align}
			\Gamma^{(2)}_{i j} =
			\begin{pmatrix}
				57.087319617(0)	&	- 9.308132636(9)
				\\
				- 9.308132636(9)	&	14.151443165(5)
			\end{pmatrix}_{i j} \, .	\label{eq:gamma_2_nonsymmetric_exact}
		\end{align}
	Again, the question is whether it is possible to extract these values with high precision at the minimum of the \gls{ir} potential which was obtained from the \gls{kt} scheme in our fluid-dynamical \gls{frg} approach.
	Both, the field expectation value and the two-point vertex functions will serve as benchmarks for the \gls{kt} scheme.

\subsection{Test case VI: misalignement of symmetry axes and discretization axes}
\label{subsec:test_case_vi_pyramid}

	As another nontrivial test case we consider a potential that is only asymptotically invariant under $O(2)$ transformations of the fields, but not in the small-$| \vec{\phi} \, |$ region, for the same reasons as in the previous test case.
	However, the potential we consider below comes with a global $\mathbb{Z}_2 \times \mathbb{Z}_2$ symmetry, which should not be broken by the \gls{rg} flow.
	In consequence, we know from \cref{eq:correlation_functions} that the \gls{ir} minimum of the potential is at $\langle \vec{\phi} \, \rangle = \vec{\varphi} = 0$ due to the symmetry.
	Furthermore, the two-point correlation and vertex functions have to be diagonal.
	
\begin{figure*}
			\includegraphics[width=\textwidth]{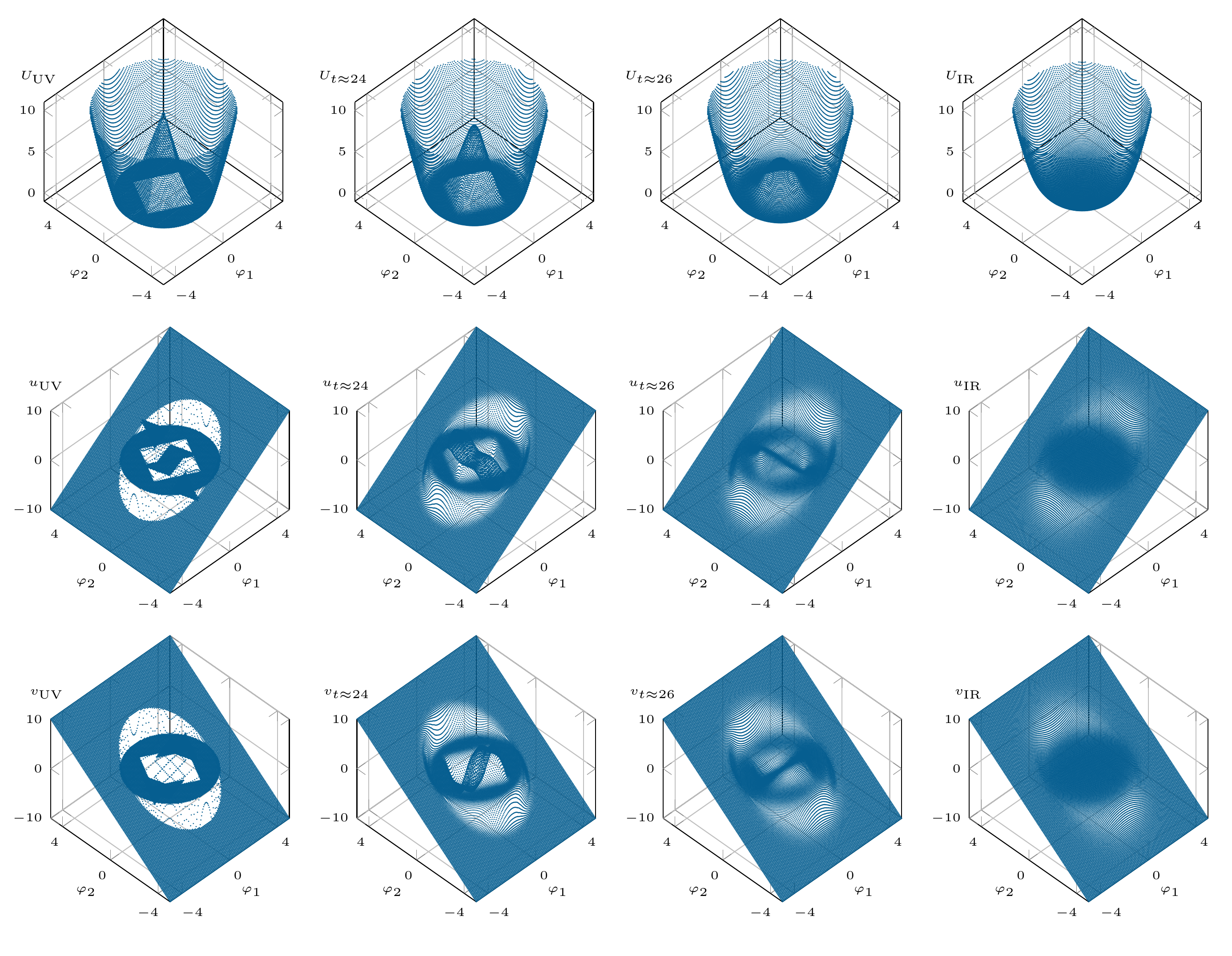}
			\caption{\label{fig:test_pyramid}%
				The \gls{rg} time evolution of the potential $U ( t, \vec{\varphi} \, )$ (upper row), its $\varphi_1$-derivative $u ( t, \vec{\varphi} \, )$ (middle row), and $\varphi_2$-derivative $v ( t, \vec{\varphi} \, )$ (lower row) from the \gls{uv} (left column) to the \gls{ir} (right column) and selected intermediate times for test case VI, \cref{eq:test_pyramid_uv_potential}, for $\alpha = 0.3$.
			}
		\end{figure*}

	The way we use this potential to challenge our numerical setup is by misaligning the $\mathbb{Z}_2 \times \mathbb{Z}_2$ symmetry axes of the potential with the discretization axes of the computational domain.
	On top of that, we introduce jump discontinuities in the derivatives of the potential, which additionally complicates the problem.
	A particular \gls{uv} potential which satisfies these specifications is given by
		\begin{align}
			U ( \vec{\phi} \, ) = \, & 5 \cdot ( 2 - | \theta_1 | - | \theta_2 | ) \cdot \Theta ( 2 - | \theta_1 | - | \theta_2 | ) +	\vdistance	\label{eq:test_pyramid_uv_potential}
			\\
			& + \Theta ( \vec{\phi}^{\, 2} - 9) \cdot ( \vec{\phi}^{\, 2} - 9) \, ,	\vdistance	\nonumber
		\end{align}
	where $\Theta$ is the Heaviside step function and
		\begin{align}
			\begin{pmatrix}
				\theta_1
				\\
				\theta_2
			\end{pmatrix}
			=
			\begin{pmatrix}
				\cos( \alpha )	&	\sin( \alpha )
				\\
				- \sin( \alpha )	&	\cos( \alpha )
			\end{pmatrix}
			\begin{pmatrix}
				\phi_1
				\\
				\phi_2
			\end{pmatrix} \, .	\label{eq:misalignement_matrix}
		\end{align}
	For a visualization of the potential we refer to \cref{fig:test_pyramid} (upper left panel).
	In the middle and lower left panels of the same figure we present the field derivatives of the potential in the \gls{uv}, the fluid fields $u$ and $v$, which exhibit several jumps.
	The angle $\alpha$ is used to rotate the symmetry axes of the potential relative to the discretization axes of the computational domain.
	Below, we shall perform tests for different values of $\alpha \in [ 0, \frac{\uppi}{4} ]$ by comparing the two-point vertex function extracted from the \gls{kt} scheme with the exact values which are
		\begin{align}
			\langle \phi_{1/2}^2 \, \rangle = \, & 3.041541448(7) \, , \vdistance
			\\
			\Gamma^{(2)}_{11/22} = \, & 0.328780658(4) \, , \vdistance	\label{eq:two-point_vertex_pyramid_exact}
		\end{align}
	independent of the choice of $\alpha$.

\section{Zero-dimensional test model with \texorpdfstring{$O(\bar{N}) \times O(\bar{M})$}{O(N) x O(M)} symmetry}
\label{sec:a_test_model_with_on_om_symmetry} 

\begin{figure*}
			\includegraphics[width=\textwidth]{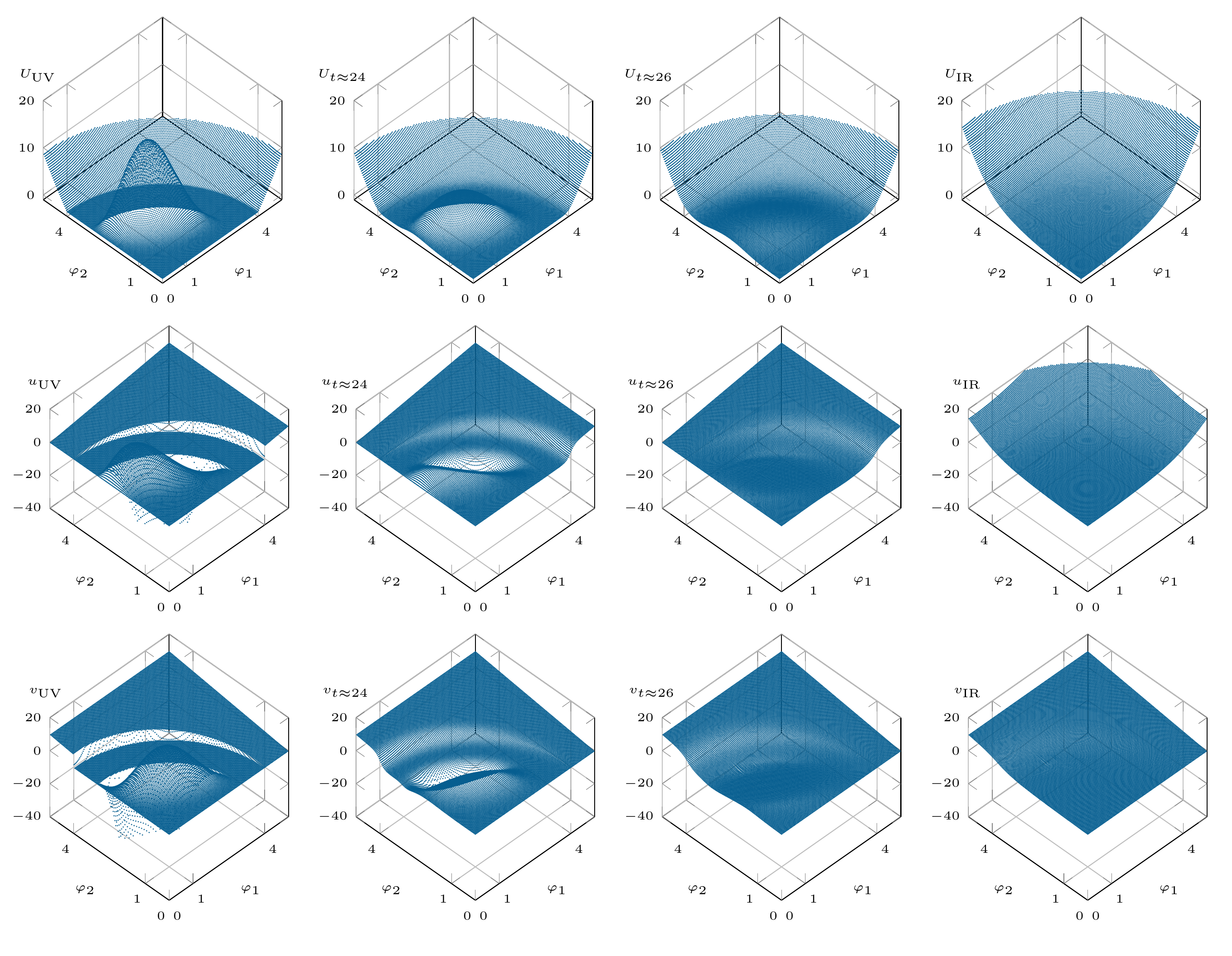}
			\caption{\label{fig:test_advection}%
				The \gls{rg} time evolution of the potential $U ( t, \sigma_1, \sigma_2 )$ (upper row), its $\sigma_1$-derivative $u ( t, \sigma_1, \sigma_2 )$ (middle row), and $\sigma_2$-derivative $v ( t, \sigma_1, \sigma_2 )$ (lower row) from the \gls{uv} (left column) to the \gls{ir} (right column) and selected intermediate times for test case VII, \cref{eq:test_advection_uv_potential}.
			}
		\end{figure*}

	The last zero-dimensional test model is a potential with $O(\bar{N}) \times O(\bar{M})$ symmetry where $\bar{N}$ and $\bar{M}$ are arbitrary integers.
	We consider a model of this type because, on the level of the fluid-dynamical formulation of the \gls{frg}, it also introduces advection terms in the \glspl{pde} for the fluid fields $u$ and $v$ for $\bar{N} > 1$ and/or $\bar{M} > 1$, see \cref{eq:conservation_law_u_v_on_om}. 
	To be specific, we consider
		\begin{align}
			& \tilde{U} ( \rho_1, \rho_2 ) =	\vdistance	\label{eq:test_advection_uv_potential}
			\\
			= \, & 4 \, \rho_1 \, \rho_2^2 \, \sin \big( \tfrac{2 \uppi}{9} \, ( \rho_1 + \rho_2 ) \big) \, \Theta ( 4.5 - \rho_1 - \rho_2 ) +	\vdistance	\nonumber
			\\
			& + 2 \, ( \rho_1 + \rho_2 - 8 ) \, \Theta ( \rho_1 + \rho_2 - 8 ) \, ,	\vdistance	\nonumber
		\end{align}
	where $\rho_{1/2}$ are the field invariants.
	For a visualization of the potential we refer to \cref{fig:test_advection} (upper left panel).
	However, note that the potential is shown as a function of the background field configuration $\sigma_{1/2}$, see \cref{sec:generalization_to_on_om}.
	The figure also shows the derivatives of the potential \gls{wrt} $\sigma_{1/2}$ in the \gls{uv}, \ie, the fluid fields $u$ and $v$, see \cref{fig:test_advection} (middle and lower left panel).
	Again, we defined the potential piecewise to avoid problems with the boundaries of the computational domain at large $| \vec{\phi} \, |$.
	On the other hand, for small $\rho_{1/2}$, the potential comprises nontrivial $O( \bar{N} ) \times O( \bar{M} )$ invariant interactions which are, however, not $O(N = \bar{N} + \bar{M})$ invariant as it is the case for the asymptotic behavior.
	In contrast to the previous test cases, we require additional boundary conditions at $\rho_{1/2} = 0 = \sigma_{1/2}$.
	Hence, our study of this model implicitly also tests the correct handling of the boundary conditions at $\rho_{1/2} = 0$.
	In order to benchmark our numerical framework, we again employ the two-point vertex functions since the global minimum of the potential is at $\vec{\varphi} = 0$, as a consequence of the $\mathbb{Z}_2 \times \mathbb{Z}_2$ symmetry, see \cref{sec:adaptions_of_the_kt_scheme_to_our_frg_problems}.
	Here and in the following, \gls{wlog}, we consider $\bar{N} = 2$ and $\bar{M} = 3$. 
	The exact reference values, which we obtained from a direct numerical integration, read
		\begin{align}
			& \langle \phi_i^2 \rangle = 2.700144567(9) \, ,	&&	\Gamma^{(2)}_{ii} = 0.370350540(4)  \, ,	\label{eq:two-point_vertex_advection_exact_on}
		\end{align}
	for $i \in \{ 1, \ldots, \bar{N} \}$ and
		\begin{align}
			& \langle \phi_i^2 \rangle = 2.683149014(3) \, ,	&&	\Gamma^{(2)}_{ii} = 0.372696408(1) \, ,	\label{eq:two-point_vertex_advection_exact_om}
		\end{align}
	for $i \in \{ \bar{N} + 1, \ldots, N = \bar{N} + \bar{M} \}$.
	Cross-correlations are zero, as expected from the symmetry of the potential.

\section{FRG results for zero-dimensional models}
\label{sec:results_of_tests_in_zero_dimensions} 

	In this section, we present our \gls{frg} results of the zero-dimensional models introduced above and compare them to the exact results. 
	We start the discussion with the $O(2)$-symmetric models from \cref{sec:o_2_symmetric_model} and then proceed with the models without $O(2)$ symmetry from \cref{sec:non-symmetric_model}.
	Finally, we discuss the $O(\bar{N}) \times O(\bar{M})$-symmetric model as introduced in \cref{sec:a_test_model_with_on_om_symmetry}.

\subsection{Zero-dimensional test models with \texorpdfstring{$O(2)$}{O(2)}~symmetry}

	Before we dive into the detailed quantitative analysis, we start with a brief qualitative discussion of the \gls{rg} flows of the $O(2)$-symmetric test cases I-IV.
	Furthermore, we already provide the numerical parameters used for all tests in \cref{tab:numerical_parameters}.
		\begin{table}[b]
			\caption{\label{tab:numerical_parameters}%
				Numerical control parameters used for the various test models in zero spacetime dimensions.
				For the integration of the flow equations, we have used \textit{RK45}.
			}
			\begin{ruledtabular}
				\setlength\extrarowheight{2pt}
				\begin{tabular}{l c c c c c}
					test case	&	$\varphi_\mathrm{max}$	&	$\Lambda$	&	$t_\mathrm{IR}$	&	$r_{\mathrm{tol}}$	&	$a_{\mathrm{tol}}$
					\\
					I \labelcref{eq:test_case_uv_potential_i} 		&	$10$					&	$10^{12}$	&	$60$			&	$10^{-10}$			&	$10^{-12}$
					\\
					II \labelcref{eq:test_case_uv_potential_ii} 		&	$10$					&	$10^{12}$	&	$60$			&	$10^{-10}$			&	$10^{-12}$
					\\
					III \labelcref{eq:test_case_uv_potential_iii} 		&	$10$					&	$10^{12}$	&	$60$			&	$10^{-10}$			&	$10^{-12}$
					\\
					IV \labelcref{eq:test_case_uv_potential_iv} 		&	$10$					&	$10^{12}$	&	$60$			&	$10^{-10}$			&	$10^{-12}$
					\\
					V \labelcref{eq:test_nonsymmetric_uv_potential}	&	6				&	$10^{12}$	&	$60$			&	$10^{-12}$			&	$10^{-12}$
					\\
					VI \labelcref{eq:test_pyramid_uv_potential}	&	10					&	$10^{12}$	&	$60$			&	$10^{-10}$			&	$10^{-12}$
					\\
					VII \labelcref{eq:test_advection_uv_potential}	&	7					&	$10^{12}$	&	$60$			&	$10^{-12}$			&	$10^{-12}$
				\end{tabular}
			\end{ruledtabular}
		\end{table}

\subsubsection{Qualitative discussion of RG flows in two field space dimensions}

	In \cref{fig:plot_case=1,fig:plot_case=2,fig:plot_case=3,fig:plot_case=4}, we show $u_{\mathrm{UV}} = u_{t = 0}$, $u_{\mathrm{IR}} = u_{t = 60}$, $U_{\mathrm{UV}} = U_{t = 0}$ and $U_{\mathrm{IR}} = U_{t = 60}$ together with the corresponding figures for two intermediate \gls{rg} times for the different test cases I-IV with $n_{\mathrm{cells}} = 175$.
	For all these test cases we find that the \gls{ir} potential is convex, smooth, and that the global minimum is at the origin, as it should be for every $O(2)$-symmetric zero-dimensional model.
	No spurious oscillations or discontinuities are found in the \gls{ir}.
	Furthermore, we see that the \gls{kt} scheme is indeed able to handle the nonanalytic points $| \vec{\phi} \, | = 2$ and $| \vec{\phi} \, | = 3$ of test case I as well as the poles of $u$ and $v$ at $\phi_{1/2} = 0$ of test case IV, see \cref{fig:plot_case=1,fig:plot_case=4}, respectively.
	In addition, at first glance, it seems that the \gls{ir} potentials of all $O(2)$ symmetric test cases are again $O(2)$ invariant.
	Numerically, however, this is not perfectly the case for the initial condition as we shall discuss in detail in the following paragraph.
		\begin{figure*}
			\subfloat[\label{subfig:rel_error_o2_sym}%
			Adapted 2D \gls{kt} scheme as presented in \cref{sec:adaptions_of_the_kt_scheme_to_our_frg_problems} for different test cases (T.C.).]{%
				\includegraphics{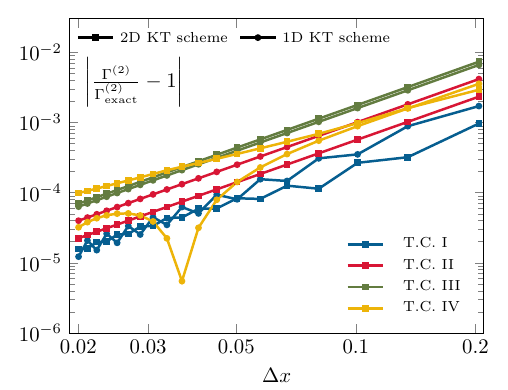}%
			}\hfill
			\subfloat[\label{subfig:rel_error_o2_sym_original_diffusion}%
			Original 2D \gls{kt} scheme as presented in \reff\cite{KTO2-0} for different test cases (T.C.).]{%
				\includegraphics{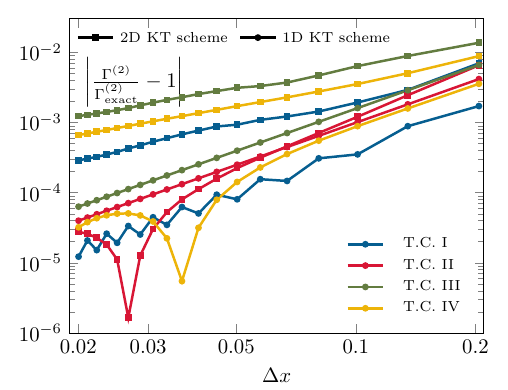}%
			}
			\caption{\label{rel_error_o2_sym}%
				The relative error \labelcref{eq:relative_error_gamma_2} of $\Gamma^{(2)}$ for test cases I-IV associated with the \gls{uv} potentials \labelcref{eq:test_case_uv_potential_i,eq:test_case_uv_potential_ii,eq:test_case_uv_potential_iii,eq:test_case_uv_potential_iv} as a function of the numerical resolution $\Delta x$.
				The two-point function $\Gamma^{(2)}$ has been obtained from \cref{eq:gamma_2_kt} with the solution of the one-dimensional \gls{pde} in Eq.~\labelcref{eq:one_dim_advection_diffusion} (1D) and the two-dimensional \gls{pde} system in Eqs.~\labelcref{eq:diffusion_equation_u,eq:diffusion_equation_v} (2D) with the 2D \gls{kt} scheme. The exact results for $\Gamma^{(2)}$ can be found in \cref{tab:exact_gamma_2_o_2}.
			}
		\end{figure*}

\subsubsection{Quantitative benchmark tests}
\label{subsubsec:quantitative_benchmark_tests}

\paragraph{Error scaling of the two-point vertex function}

		\begin{figure*}
			\subfloat[\label{subfig:o2_symmetry_IR_L1}$L^1$-norm, \cref{eq:observable_U_L1,eq:observable_u_L1}.]{%
				\includegraphics{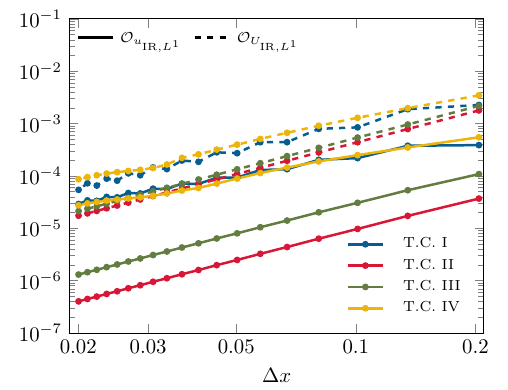}%
			}\hfill
			\subfloat[\label{subfig:o2_symmetry_IR_Linfinity}$L^\infty$-norm, \cref{eq:observable_U_Linfinity,eq:observable_u_Linfinity}.]{%
				\includegraphics{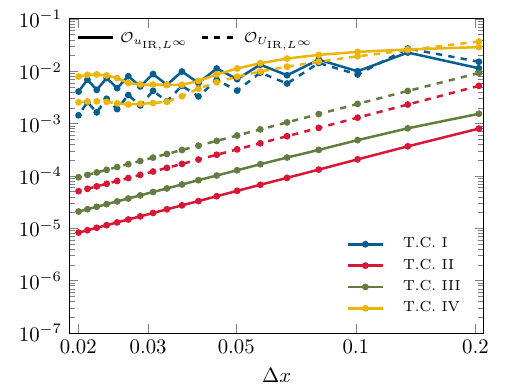}
			}
			\caption{\label{fig:o2_symmetry_IR}%
				$O(2)$-symmetry observables $\mathcal{O}_{u_{\mathrm{IR}}}$ (solid lines) and $\mathcal{O}_{U_{\mathrm{IR}}}$ (dashed lines) as functions of $\Delta x$ for test cases (T.C.) I-IV, \cref{subsubsec:test_case_i,subsubsec:test_case_ii,subsubsec:test_case_iii,subsubsec:test_case_iv}.
			}
		\end{figure*}
	In order to estimate discretization errors from the two-dimensional \gls{kt} scheme we follow \reff\cite{Koenigstein:2021syz} and study the relative error of the two-point vertex functions for test cases I-IV, \ie, we study
		\begin{align}
			\bigg| \frac{\Gamma^{( 2 )}}{\Gamma^{( 2 )}_{\mathrm{exact}}} - 1 \bigg|	\label{eq:relative_error_gamma_2}
		\end{align}
	as depicted in \cref{rel_error_o2_sym}. Here, $\Gamma^{( 2 )}$ is the two-point vertex function computed from the \gls{kt} scheme via the finite difference stencil \labelcref{eq:gamma_2_kt} and $\Gamma^{( 2 )}_{\mathrm{exact}}$ is the two-point vertex function which is determined by \cref{eq:correlation_function_o_n,eq:gamma_2_reference_o_n} for $n=2$.
	The exact values of $\Gamma^{( 2 )}$ are listed in \cref{tab:exact_gamma_2_o_2}.
	Furthermore, for comparison, we also include the relative errors of the 1D \gls{kt} scheme for the same test cases in \cref{rel_error_o2_sym}.
		\begin{figure*}
			\subfloat[\label{subfig:o2_symmetry_while_RG_flow_compare_500_L1}$L^1$-norm, \cref{eq:observable_U_L1,eq:observable_u_L1}.]{%
				\includegraphics{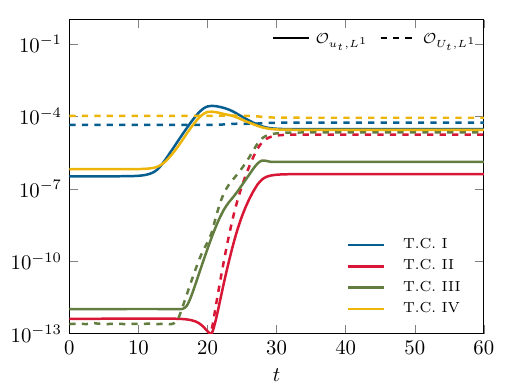}%
			}\hfill
			\subfloat[\label{subfig:o2_symmetry_while_RG_flow_compare_500_Linfinity}$L^\infty$-norm, \cref{eq:observable_U_Linfinity,eq:observable_u_Linfinity}.]{%
				\includegraphics{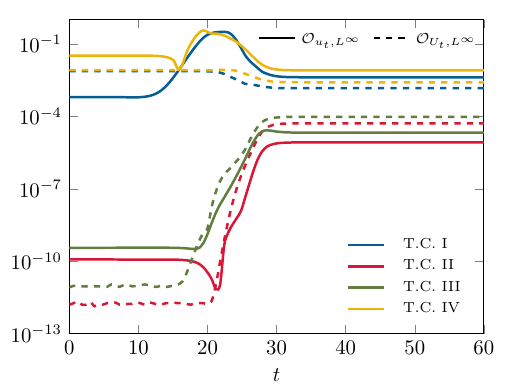}
			}
			\caption{\label{fig:o2_symmetry_while_RG_flow_compare_500}%
				$O(2)$-symmetry observables $\mathcal{O}_{u_t}$ (solid lines) and $\mathcal{O}_{U_t}$ (dashed lines) as a function of the \gls{rg} time for test cases I-IV, \cref{subsubsec:test_case_i,subsubsec:test_case_ii,subsubsec:test_case_iii,subsubsec:test_case_iv}, with fixed number of cells $n_{\mathrm{cells}} = 500$.
			}
		\end{figure*}
	For both,  the one- and two-dimensional calculations, we have used the parameters summarized in \cref{tab:numerical_parameters}.
	Looking at \cref{subfig:rel_error_o2_sym}, we observe that the relative errors decrease as we increase the number of cells.
	As one would expect, the relative error of the two-dimensional method slightly deviates from the one-dimensional one for the same field-space resolution $\Delta x$, due to a different Taylor coefficient in front of the error scaling but has approximately the same scaling exponent.
	To be explicit, we find an error scaling for the cases I-IV as listed in \cref{tab:error-scaling-I-IV}, respectively.
	These error scalings are in good agreement with the expected error scaling of $\Delta x^2$.
	Note that one usually does not find perfect agreement of the error scaling with the expected one, since the error scaling is only an approximation and the error scaling exponent is not exactly $2$.
	Especially in the presence of, \eg, nonanalyticities the error scaling can be of lower order.
	Interestingly, we observe that the \gls{kt} scheme in its original form as presented in \reff\cite{KTO2-0} systematically leads to an error scaling below the expected one, also for smooth potentials, see \cref{subfig:rel_error_o2_sym_original_diffusion}.
	We shall even see that the computations with the original \gls{kt} scheme do not converge at all and spuriously oscillations pop up in the solutions in case of the other test models to be discussed below. 
	Hence, we suggest to use our adapted version of the \gls{kt} scheme as presented in \cref{sec:adaptions_of_the_kt_scheme_to_our_frg_problems} for \gls{frg} applications (and possibly also for other applications).\footnote{Note that these adaptations solely affect the diffusion part of the higher-dimensional version of the \gls{kt} scheme, which was not the main focus of \reff\cite{KTO2-0}, and that the advection part is not affected.
	Furthermore, in Ref.~\cite{Rais:2024jio}, where almost linear \glspl{pde} have been considered and the code underlying our present studies has been employed, negligible differences between our and the original version of the two-dimensional diffusion has been observed.}
		\begin{table}[b]
			\caption{\label{tab:error-scaling-I-IV}%
				Error-scaling exponent $n$ extracted from the error scaling $\Delta x^n$ from \cref{rel_error_o2_sym}.
				(2D${}^\star$ denotes the original version of the two-dimensional \gls{kt} scheme as presented in \reff\cite{KTO2-0}, see \cref{subfig:rel_error_o2_sym_original_diffusion}.)
			}
			\begin{ruledtabular}
				\setlength\extrarowheight{2pt}
				\begin{tabular}{l c c c c}
					&	T.C.\ I	&	T.C.\ II	&	T.C.\ III	&	T.C.\ IV
					\\
					$n (\text{1D})$	&2.00 - 2.04&2.00&2.00&--
					\\
					$n (\text{2D})$	&1.94 - 1.95&2.00&2.00&1.47
					\\
					$n (\text{2D}^\star)$	&1.48&--&1.07&1.05
				\end{tabular}
			\end{ruledtabular}
		\end{table}
	Note that the number of coupled \glspl{ode} for the two-dimensional calculations in general increases quadratically as a function of $n_{\mathrm{cells}}$, namely by $8 \,n_{\mathrm{cells}}^2 \approx 8 \, (\frac{\varphi_{\mathrm{max}}}{\Delta x})^2$ for large $n_{\mathrm{cells}}$, as compared with the 1D \gls{kt} scheme where the number of coupled \glspl{ode} increases linearly with $n_{\mathrm{cells}}$.
	For this reason, we have only performed calculations down to the value $\Delta x = 0.02$. 
	In fact, already for $\Delta x = 0.02$ with $\varphi_{\mathrm{max}} = 10$, we have a system of two million coupled \glspl{ode} and hence the required computing time to solve this set of equations is rather long, and will even increase for smaller $\Delta x$.
	For future applications, one might therefore consider parallelization or adaptive mesh refinement with similar discretization schemes.
	For an estimate of the required run times, see \cref{app:computational_times}.

	Overall, the behavior of the relative error, as depicted in \cref{subfig:rel_error_o2_sym}, indicates that the two-dimensional \gls{kt} scheme is suitable for studies of $O(2)$-symmetric zero-dimensional models although the 1D \gls{kt} scheme, as discussed in great detail in Ref.~\cite{Koenigstein:2021syz}, performs much better \gls{wrt} runtime and slightly better for test case IV with the pole at $\vec{\varphi} = 0$.

\paragraph{Global error measures} 

		\begin{figure*}
			\subfloat[\label{subfig:o2_symmetry_while_RG_flow_test_id=1_L1}$L^1$-norm, \cref{eq:observable_U_L1,eq:observable_u_L1}.]{%
				\includegraphics{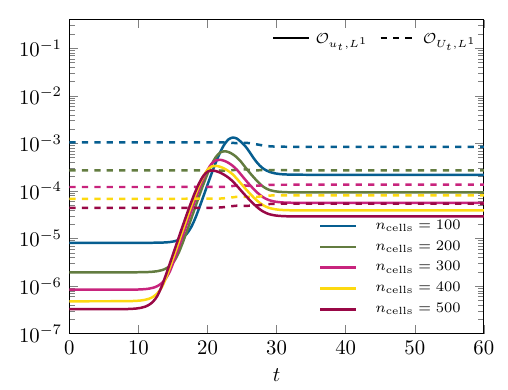}%
			}\hfill
			\subfloat[\label{subfig:o2_symmetry_while_RG_flow_test_id=1_Linfinity}$L^\infty$-norm, \cref{eq:observable_U_Linfinity,eq:observable_u_Linfinity}.]{%
				\includegraphics{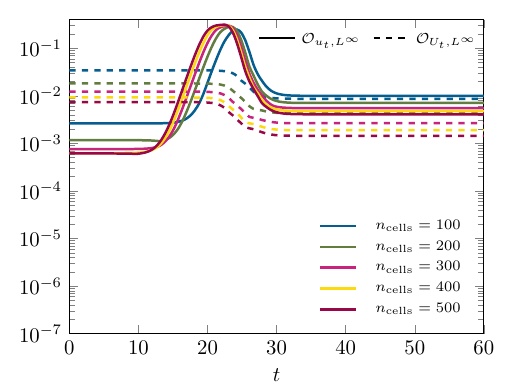}
			}
			\caption{\label{fig:o2_symmetry_while_RG_flow_test_id=1}%
				$O(2)$-symmetry observables $\mathcal{O}_{u_t}$ (solid lines) and $\mathcal{O}_{U_t}$ (dashed lines) as a function of the \gls{rg} time for various number of cells for test case I, \cref{subsubsec:test_case_i}.
			}
		\end{figure*}
	As already mentioned in the previous sections, there is another issue which we have to address, namely the artificial breaking of the $O(2)$ symmetry due to the Cartesian grid. 
	This is the subject of the following paragraphs.
		\begin{figure*}
			\subfloat[\label{subfig:o2_symmetry_while_RG_flow_test_id=2_L1}$L^1$-norm, \cref{eq:observable_U_L1,eq:observable_u_L1}.]{%
				\includegraphics{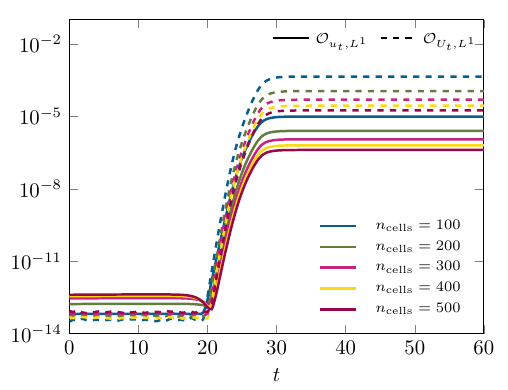}%
			}\hfill
			\subfloat[\label{subfig:o2_symmetry_while_RG_flow_test_id=2_Linfinity}$L^\infty$-norm, \cref{eq:observable_U_Linfinity,eq:observable_u_Linfinity}.]{%
				\includegraphics{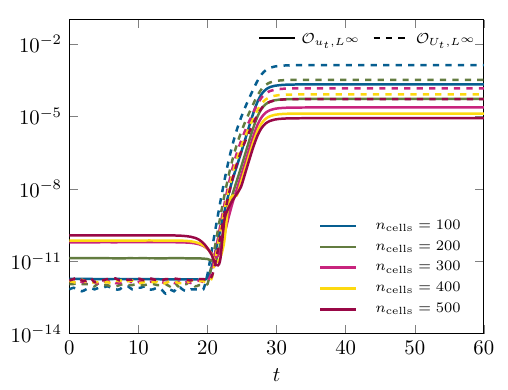}
			}
			\caption{\label{fig:o2_symmetry_while_RG_flow_test_id=2}%
				$O(2)$-symmetry observables $\mathcal{O}_{u_t}$ (solid lines) and $\mathcal{O}_{U_t}$ (dashed lines) as a function of the \gls{rg} time for various number of cells for test case II, \cref{subsubsec:test_case_ii}.
			}
		\end{figure*}
	Before we return to the observables, we briefly comment on the remnants of the $O(2)$ symmetry on the level of the cell averages of our Cartesian grid.
	First, we note that the diffusion fluxes on the \gls{rhs}\ of the \gls{rg} flow equations of $\partial_t u$ and $\partial_t v$ are identical, \cf, \cref{eq:diffusion_equation_u,eq:diffusion_equation_v}.
	Therefore, for $O(2)$-symmetric initial conditions we find a mirror symmetry \gls{wrt}\ the first bisector, namely
		\begin{align}
			&	\partial_t u ( x, y ) = \partial_t v (y, x) \, 	&&	\Leftrightarrow	&&	u ( x, y ) = v ( y, x) \, .	\label{eq:mirror_symmetry}
		\end{align}
	This symmetry is still present at the level of the cell averages in the \gls{kt} scheme and even the Riemann summation or the construction of the initial condition, \cref{eq:initial_condition_discrete}, do not alter it.
	Hence, the following equations are trivially fulfilled:
		\begin{align}
			\bar u_{i,j} = \, & \bar v_{j,i} \, ,	\vdistance
			\\
			y_j \, \bar u_{j,i} + x_i \, \bar v_{j,i} = \, & x_i \, \bar u_{i,j} + y_j\, \bar v_{i,j} \, ,	\vdistance
			\\
			\bar U^u_{i,j} = \, & \bar U^v_{j,i} \, ,	\vdistance
		\end{align}
	where $\bar U^{u/v}$ denotes the Riemann sum along $\bar u$ and $\bar v$, respectively.\footnote{
		In the matrix formulation as introduced in \cref{app:implementation_of_the_multi-dimensional_kt_central_scheme}, we find accordingly
		\begin{align*}
			\bm u^T_t = \, & \bm v_t \, ,
			\\
			(\bm x_C \cdot \bm u + \bm y_C \cdot \bm v)^T = \, & \bm x_C \cdot \bm u + \bm y_C \cdot \bm v \, ,
		\end{align*}	
		where $\bm u = \bm u [ 0 ]$, $\bm v = \bm u [ 1 ]$, $\bm x_C = \bm x_C [ 0 ]$, and $\bm y_C = \bm y_C [ 0 ]$.
	}
	Here, we have chosen the reference point for the integration at some point within the first bisector.
	Moreover, we find two additional symmetries which are still present in the two-dimensional \gls{kt} scheme: We have mirror symmetries \gls{wrt}\ the $x$- and $y$-direction for $\bar v$ and $\bar u$, respectively, \ie, 
		\begin{align}
			&	\bar u_{i,j} = \bar u_{i, -j}	&&	\text{and}	&&	\bar v_{i,j} = \bar v_{-i, j} \, ,
		\end{align}
	and reflection symmetries \gls{wrt}\ the $x$- and $y$-direction for $\bar u$ and $\bar v$, respectively, \ie, 
		\begin{align}
			&	\bar u_{i,j} = -\bar u_{-i, j}	&&	\text{and}	&&	\bar v_{i,j} = -\bar v_{i, -j} \, .
		\end{align}
	All three symmetries together imply that the whole information of the system is governed by a single quadrant of $\bar u_{i,j}$ or equivalent of $\bar v_{i,j}$.
	Furthermore, this analysis reduces the number of proper observables for measuring the deviation of the $O(2)$ symmetry.
	For example, the function
		\begin{align}
			\mathcal{O} = \, & \max_{i,j} \big( | \mathrm{rot}_{90^{\circ}} ( \bar v )_{i,j} - \bar u_{i,j} | \big)
		\end{align}
	is trivially fulfilled, meaning that $\mathcal{O} \approx 0$.
	Another possible observable could be of the form
		\begin{align}
			\mathcal{O} = \, & \max_{i,j} \big( | \mathrm{rot}_{90^{\circ}} ( A )_{i,j} - A_{i,j} | \big) \, ,
		\end{align}
	where $A_{i,j} = x_i \, \bar u_{i,j} + y_j \, \bar v_{i,j}$.
	However, this is not suited for measuring the failure of the $O(2)$ symmetry since it vanishes also for an elliptic shape of $A_{i,j}$.
	Therefore, the only observables which we are using are \cref{eq:observable_U_Linfinity,eq:observable_u_Linfinity},
		\begin{align}
			\mathcal{O}_{U, L^\infty} = \, & \max_{i,j} \big( |\mathrm{rot}_{90^{\circ}} ( \bar U )_{i,j} - \bar U_{i,j} | \big) \, , 	\vdistance \nonumber
			\\ 
			\mathcal{O}_{u, L^\infty} = \, & \max_{i,j} \big( |\mathrm{rot}_{90^{\circ}} ( \bar u/x )_{i,j} - \bar u_{i,j}/x_{i} | \big) \, ,	\vdistance \nonumber
		\end{align}
	and \cref{eq:observable_U_L1,eq:observable_u_L1},
		\begin{align}
			\mathcal{O}_{U, L^1} = \, & \tfrac{1}{\# ( i, j )} \sum_{i,j} \big(|\mathrm{rot}_{90^{\circ}}(\bar U)_{i,j} - \bar U_{i,j}|\big) \, ,	\vdistance\nonumber
			\\
			\mathcal{O}_{u, L^\infty} = \, & \tfrac{1}{\# ( i, j )} \sum_{i,j} \big(|\mathrm{rot}_{90^{\circ}}(\bar u/x)_{i,j} - \bar u_{i,j}/x_{i}|\big) \, ,	\vdistance\nonumber
		\end{align}
	respectively.
	The latter ones are of special interest since they only measure the deviation of the $O(2)$ symmetry induced by the use of the \gls{kt} scheme and not from the additional Riemann sum, which is required to reconstruct the potential.
	It is motivated by \cref{eq:chain_rule_U}:
	If the underlying potential $U ( t, \vec{\varphi} \, )$ is $O(2)$-symmetric then $\frac{1}{\varphi_1} u ( t, \vec{\varphi} \, ) = \partial_\varrho \bar{U} ( t, \varrho )$ has the same property.

	The reason for the implementation of both measures in terms of the $L^1$ and $L^\infty$-norms is that the $L^1$-norm is more sensitive to the average behavior of the solution, whereas the $L^\infty$-norm is more sensitive to the maximum deviation.
	If both observables are of the same size, then the deviation of the $O(2)$ symmetry is uniformly distributed over the whole grid.
	Note that for the calculation of both observables we have restricted ourselves to the inclusion of cells from the domain $[ - 9, 9 ] \times [ - 9, 9 ]$.
	Leaving out the outermost $10 \%$ of the cells ensures that boundary effects, which do not propagate into the inner region, are not overemphasized.
		\begin{table}[b]
			\caption{\label{tab:L1-Linf-norm-I-IV}%
				Scaling exponent $n$ extracted from the scaling $\Delta x^n$ from \cref{fig:o2_symmetry_IR}.
			}
			\begin{ruledtabular}
				\setlength\extrarowheight{2pt}
				\begin{tabular}{l c c c c}
					&	T.C.\ I	&	T.C.\ II	&	T.C.\ III	&	T.C.\ IV
					\\
					$n (\mathcal{O}_{u_{\mathrm{IR}}, L^{1}})$	&1.28 - 1.35&1.98&1.98&--
					\\
					$n (\mathcal{O}_{U_{\mathrm{IR}}, L^{1}})$	&1.78 - 1.83&2.00&2.00&--
					\\
					$n (\mathcal{O}_{u_{\mathrm{IR}}, L^{\infty}})$	&0.57 - 0.66&2.00&1.98&--
					\\
					$n (\mathcal{O}_{U_{\mathrm{IR}}, L^{\infty}})$	&1.19 - 1.31&2.00&2.00&--
				\end{tabular}
			\end{ruledtabular}
		\end{table}

	In \cref{fig:o2_symmetry_IR}, all four functions, evaluated in the \gls{ir}, are shown for several values of $\Delta x$.
	Both observables are in general nonzero, but decrease for decreasing $\Delta x$.
	This implies that the $O(2)$ symmetry is, as already anticipated, not perfectly conserved by our numerical scheme but becomes better approximated for smaller $\Delta x$.
	We extracted corresponding approximate error scalings (where appropriate), which we list in \cref{tab:L1-Linf-norm-I-IV}. 
	Again, the ideal expected error scaling is $\Delta x^2$ and we only find slight deviations from this scaling, especially for the $L^1$-norms, where we have the averaging over all cells.
	We conclude that the deviation from perfect $O(2)$ symmetry is indeed caused by the finite resolution $\Delta x$ of the Cartesian grid, similar as for spacetime lattices in lattice Monte-Carlo simulations.

	Still, let us study this effect even further and investigate, if symmetry violations actually increase or decrease during the \gls{rg} flow.

\begin{figure*}
			\subfloat[\label{subfig:o2_symmetry_while_RG_flow_test_id=3_L1}$L^1$-norm, \cref{eq:observable_U_L1,eq:observable_u_L1}.]{%
				\includegraphics{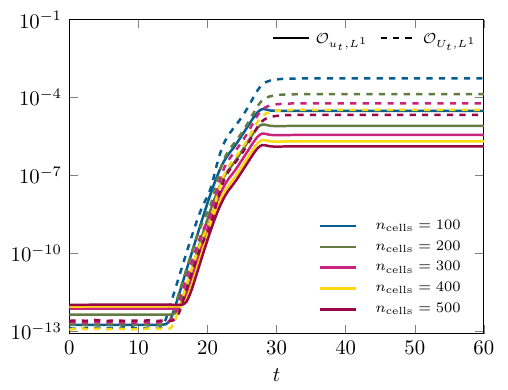}%
			}\hfill
			\subfloat[\label{subfig:o2_symmetry_while_RG_flow_test_id=3_Linfinity}$L^\infty$-norm, \cref{eq:observable_U_Linfinity,eq:observable_u_Linfinity}.]{%
				\includegraphics{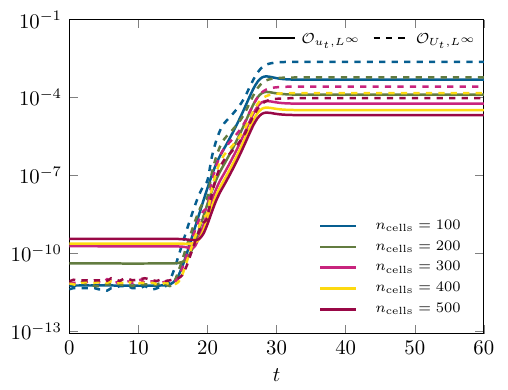}
			}
			\caption{\label{fig:o2_symmetry_while_RG_flow_test_id=3}%
				$O(2)$-symmetry observables $\mathcal{O}_{u_t}$ (solid lines) and $\mathcal{O}_{U_t}$ (dashed lines) as a function of the \gls{rg} time for various number of cells for test case III, \cref{subsubsec:test_case_iii}.
			}
		\end{figure*}

\begin{figure*}
			\subfloat[\label{subfig:o2_symmetry_while_RG_flow_test_id=4_L1}$L^1$-norm, \cref{eq:observable_U_L1,eq:observable_u_L1}.]{%
				\includegraphics{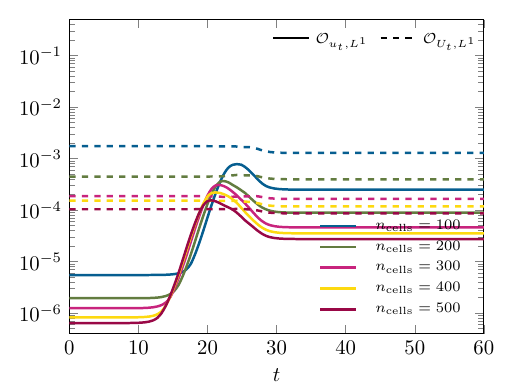}
			}\hfill
			\subfloat[\label{subfig:o2_symmetry_while_RG_flow_test_id=4_Linfinity}$L^\infty$-norm, \cref{eq:observable_U_Linfinity,eq:observable_u_Linfinity}.]{%
				\includegraphics{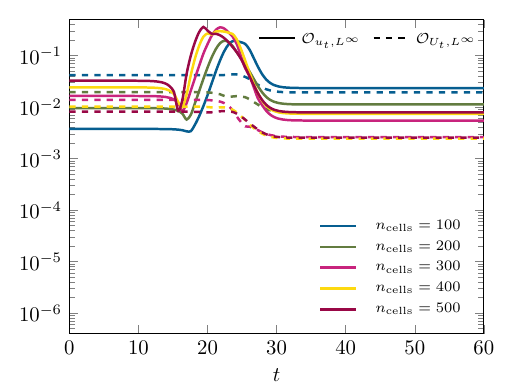}
			}
			\caption{\label{fig:o2_symmetry_while_RG_flow_test_id=4}%
				$O(2)$-symmetry observables $\mathcal{O}_{u_t}$ (solid lines) and $\mathcal{O}_{U_t}$ (dashed lines) as a function of the \gls{rg} time for various number of cells for test case IV, \cref{subsubsec:test_case_iv}.
			}
		\end{figure*}

	In fact, we find that the $O(2)$ symmetry is already spoiled in the \gls{uv} by the initial condition as can be seen in \cref{fig:o2_symmetry_while_RG_flow_compare_500} for test cases I and IV.
	The reason for this arises from the fact that we are assuming in all observables that the value of the cell average $\bar u_{i,j}$ is located at the cell center $(x_i,y_j)$.
	However, this is not always fulfilled. 
	In particular, this is not fulfilled for test cases that come with jumps or discontinuities of $u$ within a single cell, which is indeed the case, \eg, for test cases I and IV, see \cref{fig:o2_symmetry_while_RG_flow_compare_500}.
	This is also confirmed by \cref{fig:o2_symmetry_while_RG_flow_test_id=1,fig:o2_symmetry_while_RG_flow_test_id=4}, where we observe that, for increasing number of cells, the observables $\mathcal{O}_{U_t}$ only decrease during the \gls{rg} flow.
	For completeness, we also show these observables for test cases II and III in \cref{fig:o2_symmetry_while_RG_flow_test_id=2,fig:o2_symmetry_while_RG_flow_test_id=3}.
	Overall, we find that the deviation from perfect $O(2)$ symmetry slightly grows during the \gls{rg} flow, whereas it is in general possible to reduce the symmetry violation by increasing the number of cells.
	Hence, these tests confirm the reliability of the \gls{kt} scheme for $O(2)$-symmetric zero-dimensional models.
	Moreover, they suggest that our numerical scheme also works reliably for nonsymmetric situations as well as in higher spacetime dimensions because symmetry breaking artifacts solely introduced by the discretization can be systematically reduced.

\subsection{Zero-dimensional test models without \texorpdfstring{$O(2)$}{O(2)}~symmetry} 

	Now we turn to the zero-dimensional test cases with two field-space directions without $O(2)$ symmetry in their initial condition, see \cref{sec:non-symmetric_model} for their definition. 

\subsubsection{Test case V: nonvanishing field expectation value} 

	We begin with the discussion of the test case V, where the \gls{uv} potential has no global symmetry anymore. 
	As already discussed in \cref{subsec:test_case_v_nonsymmetric}, this also leads to a nonvanishing expectation value $\langle \vec{\phi} \, \rangle$, \ie, a nontrivial \gls{ir} minimum.
	Still, the \gls{ir} potential has in general to be convex and also to be smooth in zero spacetime dimensions.
	The result is a rather complicated \gls{rg} flow which is depicted in \cref{fig:test_nonsymmetric} for the potential and its derivatives, the ``fluid fields''.
	In general, this setup allows for another test of the \gls{kt} scheme, namely the correct location and extraction of the minimum of the \gls{ir} potential and its comparison with the exact expectation value of the field $\vec{\phi}$.
	Hence, as a first test, we consider the quantity
		\begin{align}
			&	\bigg| \frac{\varphi_{\mathrm{min},i}}{\langle \phi_i \rangle} - 1 \bigg| \, ,	&&	i \in \{ 1, 2 \} \, ,	\label{eq:relative_error_field_expectation_value}
		\end{align}
	where $\langle \phi_i \rangle$ is the exact expectation value of the field $\phi_i$, \cref{eq:field_expectation_value} and $\varphi_{\mathrm{min},i}$ is the position of the minimum of the \gls{ir} potential from the \gls{kt} scheme at resolution $\Delta x$.
	In addition, we analyze the relative error of the two-point vertex function $\Gamma^{(2)}_{ij}$ extracted from the \gls{kt} scheme at the \gls{ir} minimum $\vec{\varphi}_{\mathrm{min}}$ and compare it with the exact $\Gamma^{(2)}_{ij}$, see \cref{eq:gamma_2_nonsymmetric_exact}.
	These tests pose two challenges:
	
	First, the naive extraction of the minimum of the \gls{ir} potential is only as accurate as the numerical resolution~$\Delta x$.
	If $\langle \phi_i \rangle = \varphi_{\mathrm{min},i}$ is of the same order as $\Delta x$, the extraction of the minimum will naturally have a large error.
	Second, the error from extracting the minimum directly propagates into the error of the two-point vertex function $\Gamma^{(2)}_{ij}$ because this quantity is evaluated at the minimum of the \gls{ir} potential.
	
	Often, interpolation of the discrete \gls{ir} solution of grid-based methods is used to obtain a more accurate estimate of the minimum (and also of the two-point vertex function) or the methods are already based on interpolation right from the start.
	However, we remark that interpolation is usually based on splines which require a certain smoothness of the potential about the minimum.
	In zero spacetime dimensions, this requirement is formally fulfilled but this is not the case for effective potentials in the broken phase in higher-dimensional spacetimes in the \gls{ir}.
	In any case, we present both, the naive extraction of the minimum and results based on an interpolation with a bivariate spline approximation, \texttt{RectBivariateSpline} with spline degree $5$ \cite{2020SciPy-NMeth}.
	
\begin{figure*}
			\subfloat[\label{subfig:relative_error_minimum_test_id=5} Direct extraction.]{%
				\includegraphics{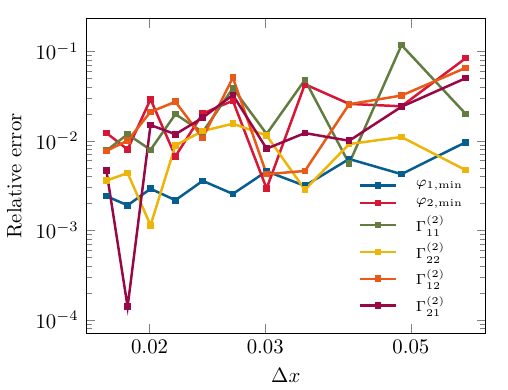}%
			}\hfill
			\subfloat[\label{subfig:relative_error_minimum_test_id=5_interpolation} Interpolation.]{%
				\includegraphics{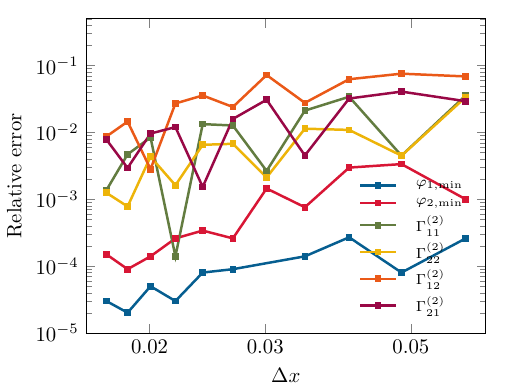}%
			}
			\caption{\label{fig:relative_error_minimum_test_id=5}%
				The relative error for the $\varphi_1$- and $\varphi_2$-position of the minimum of the \gls{ir} potential, \cref{eq:relative_error_field_expectation_value} with \cref{eq:field_expectation_value}, for test case V, \cref{subsec:test_case_v_nonsymmetric}, as a function of the numerical resolution $\Delta x$ as well as the relative error of $\Gamma^{(2)}_{ij}$, \cref{eq:relative_error_gamma_2} with \cref{eq:gamma_2_nonsymmetric_exact}, for the same test case.
				$\varphi_{i, \mathrm{min}}$ is extracted from the \gls{ir} solution of the \gls{kt} scheme at the cell center with the smallest value of the \gls{ir} potential, \cref{subfig:relative_error_minimum_test_id=5} or by interpolation, \cref{subfig:relative_error_minimum_test_id=5_interpolation}.
			}
		\end{figure*}

	In \cref{fig:relative_error_minimum_test_id=5}, we show the relative error of the minimum of the \gls{ir} potential and the relative error of the components of the two-point vertex function $\Gamma^{(2)}_{ij}$ as a function of the numerical resolution $\Delta x$.
	For the only quantity, where this is legitimate, namely $\varphi_{\mathrm{min}, 1}$, we can also extract the error scaling exponent. 
	Its value can be found in \cref{tab:rel-error-scaling-0d-no-sym}.
	Overall, we find that the relative errors of all quantities decrease slowly with increasing numerical resolution $\Delta x$.
	However, as expected for the error of the $\varphi_1$-component of the minimum, we find a consistent error scaling, which is however not as good as the expected $\Delta x^2$.
	For the $\varphi_2$-component we indeed observe that the absolute error given by the resolution $\Delta x$ is of the same order as $\langle \phi_2 \rangle$ itself, such that the relative error hardly decreases with increasing resolution.
	As a consequence, the relative errors of the two-point vertex functions $\Gamma^{(2)}_{ij}$ are rather large.
	We have improved our results, especially for the relative error of the position of the minimum, by using an interpolation method which is also depicted in \cref{fig:relative_error_minimum_test_id=5}.
	However, we would like to emphasize that this interpolation method is only safely applicable in zero spacetime dimensions where the effective potential is smooth about the minimum.

	In total, we conclude that the \gls{kt} scheme is working properly also for zero-dimensional models without $O(2)$ symmetry.
	Nevertheless, our tests clearly show that the finite resolution of the grid sets strong limits on the accuracy of observables even if the scheme itself is reliable.
	For computations in models in higher-dimensional spacetime this implies that one has to carefully compare the resolution of the grid with the expected scales of the observables.

\subsubsection{Test case VI: misalignment of symmetry axes} 
	
		\begin{figure}
			\includegraphics{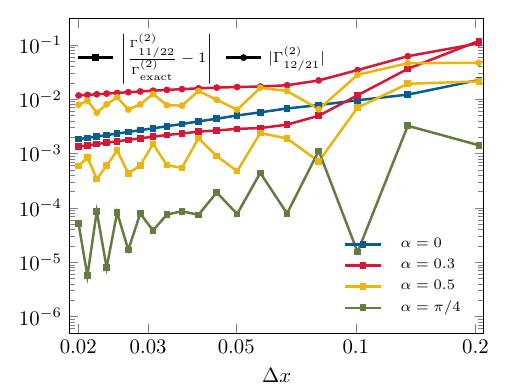}
			\caption{\label{fig:relative_error_gamma2_test_id=6}%
			Relative error \labelcref{eq:relative_error_gamma_2} of $\Gamma^{(2)}_{ii}$ for test case VI with the \gls{uv} potential \labelcref{eq:test_pyramid_uv_potential} as a function of the numerical resolution $\Delta x$ for different misalignments $\alpha$, see \cref{eq:misalignement_matrix}.
			$\Gamma^{(2)}_{ii}$ is obtained via \cref{eq:gamma_2_kt} from the solution of the \gls{pde} system with the \gls{kt} scheme, while the exact $\Gamma^{(2)}_{ii}$ are taken from \cref{eq:two-point_vertex_pyramid_exact}.
			For $\alpha \neq 0$, we also provide the absolute value of the off-diagonal component $\Gamma^{(2)}_{xy}$.
			}
		\end{figure}
	The test case VI, see \cref{subsec:test_case_vi_pyramid}, might seem extremely artificial for \gls{frg} practitioners because piecewise potentials with a pyramid-shaped small-$| \vec{\varphi} \, |$ region are not expected to appear in any physical situation.
	However, this test case is ideally suited to investigate how the results from the \gls{kt} scheme are influenced by a misalignment of the discretization axes associated with the \gls{fv} discretization and the symmetry axes of the potential.
	In addition, the chosen potential exhibits edges which cause jump discontinuities in the derivatives.
	Since these multiple jump discontiuities are also not aligned with the discretization axes, this also poses a test of our construction of the initial condition in terms of cell averages.
	The misalignement of the symmetry axes is parameterized by the angle $\alpha$, see \cref{eq:misalignement_matrix}, which we varied in the range~$[0, \pi/4]$.

	As can be seen in \cref{fig:test_pyramid}, the \gls{rg} flow associated with this test case is rather involved, while the \gls{ir} potential is smooth and convex.
	However, a qualitative analysis of the flow of the potential and its derivatives is not sufficient to judge the quality of the \gls{kt} scheme.
	Therefore, we present the relative deviation of the two-point vertex function of the \gls{kt} scheme from the exact result, see \cref{eq:two-point_vertex_pyramid_exact}, as a function of the numerical resolution $\Delta x$, see \cref{fig:relative_error_gamma2_test_id=6}.
	Due to the $\mathbb{Z}_2 \times \mathbb{Z}_2$ symmetry, the \gls{ir} minimum is trivial and this quantity is extracted at $\vec{\varphi} = 0$.
	Also by symmetry, only the diagonal components of the two-point vertex function are nonvanishing.
	Nevertheless, we also provide the absolute value of the off-diagonal component in \cref{fig:relative_error_gamma2_test_id=6} for the setup, where the symmetry of the pyramid is not aligned with the grid, $\alpha \neq 0$.
	We only show them because the off-diagonal components are numerically not exactly zero, which is a direct consequence of the misalignment and finite numeric resolution, but they have to tend to zero for $\Delta x \rightarrow 0$.
	
	In addition to \cref{fig:relative_error_gamma2_test_id=6}, we provide the error scaling exponents (where appropriate) in \cref{tab:rel-error-scaling-0d-no-sym}.
		\begin{table}[b]
			\caption{\label{tab:rel-error-scaling-0d-no-sym}%
			Error scaling exponent $n$ extracted from the scaling $\Delta x^n$ corresponding to \cref{fig:relative_error_minimum_test_id=5,fig:relative_error_gamma2_test_id=6,fig:relative_error_gamma2_test_id=7}
			}
			\begin{ruledtabular}
				\setlength\extrarowheight{2pt}
				\begin{tabular}{l c c c c c}
					T.C. &	$n(\varphi_1)$	&	$n(\varphi_2)$	&	$n(\Gamma^{(2)}_{11})$	&	$n(\Gamma^{(2)}_{22})$ & $n(\Gamma^{(2)}_{12})$
					\\
					V	&0.8(4) - 1.1(0)&--&--&--&--
					\\
					VI ($\alpha=0$)	& & &1.0(6)&1.0(6)& 0
					\\
					VI ($\alpha=0.3$)	& & &0.9(6)&0.9(6)&0.4(6)
					\\
					VI ($\alpha=0.5$)	& & &--&--&--
					\\
					VI ($\alpha=\pi/4$)	& & &--&--& 0
					\\
					VII	& & &1.9(0)&1.9(0)&--
				\end{tabular}
			\end{ruledtabular}
		\end{table}
	Overall, we find that the relative error of the two-point vertex function decreases with increasing numerical resolution $\Delta x$.
	However, the error scaling is reduced to approximately $\Delta x^1$, which is most likely a consequence of the jump discontinuities in the derivatives of the potential.
	Interestingly, the error scaling exponent is even lower for the off-diagonal component of the two-point vertex function.
	In general, this is not a problem but should be kept in mind when applying the \gls{kt} scheme to models with similar symmetry-breaking features.
	
	In summary, we observe that our numerical framework is still working properly, even under these difficult conditions, but it is very challenging to obtain a very high accuracy for every observable due to very slow convergence with the numerical resolution $\Delta x$. 
	However, we would like to emphasize that our results only differ from the exact results on a percentage level, already for the smallest resolution considered here.

\subsection{Zero-dimensional test models with \texorpdfstring{$O(\bar{N}) \times O(\bar{M})$}{O(N) x O(M)} symmetry and advection} 

	In our last zero-dimensional test case, which is test case VII, we consider a model with $O(\bar{N}) \times O(\bar{M})$ symmetry including advection in the \gls{pde}, see \cref{sec:a_test_model_with_on_om_symmetry}.
	Here, we have two invariants which are the $O(\bar{N})$ and $O(\bar{M})$ invariant and two background fields that span the physical and computational domain.
	For the \gls{uv} potential, we choose \cref{eq:test_advection_uv_potential}.
	We consider this test model since so far we have only considered models which are solely driven by the diffusion parts of the 2D \gls{kt} scheme.
	The model associated with this test case also has advective contributions from the terms in the \gls{rg} flow equations which correspond to the Goldstone modes in higher-dimensional systems, see \cref{sec:generalization_to_on_om}.
	
\begin{figure}
			\includegraphics{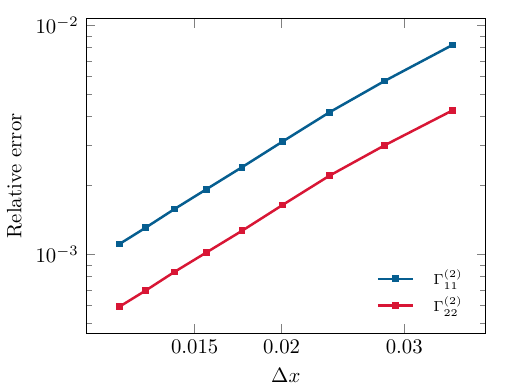}
			\caption{\label{fig:relative_error_gamma2_test_id=7}%
				The relative error \labelcref{eq:relative_error_gamma_2} of $\Gamma^{(2)}_{ii}$ for test case VII with \gls{uv} potential \labelcref{eq:test_advection_uv_potential} as a function of the numerical resolution $\Delta x$.
				$\Gamma^{(2)}_{ii}$ is obtained via \cref{eq:gamma_2_kt} from the solution of the \gls{pde} system with the \gls{kt} scheme, while the exact $\Gamma^{(2)}_{ii}$ are taken from \cref{eq:two-point_vertex_advection_exact_on,eq:two-point_vertex_advection_exact_om}.
			}
		\end{figure}

	As can be seen in \cref{fig:test_advection}, we again find that the \gls{rg} flow behaves as expected.
	The $O(\bar{N}) \times O(\bar{M})$ symmetry is restored in the \gls{ir} and the potential is smooth and convex.
	Also including Goldstone-like contributions (\ie, advection) in the \gls{rg} flow equation does not pose a general problem to the \gls{kt} scheme, even in the presence of jump discontinuities in the derivatives or huge gradients.
	However, without explicit numerical tests, a statement about the accuracy of the scheme is not possible.
	For this reason, we again consider the errors of the two-point vertex functions of the \gls{kt} scheme in relation to their exact values, see \cref{eq:two-point_vertex_advection_exact_on,eq:two-point_vertex_advection_exact_om}, as a function of the numerical resolution $\Delta x$ in \cref{fig:relative_error_gamma2_test_id=7}.
	Indeed, we find almost perfect agreement of our results with the expected $\Delta x^2$-scaling of the relative error as listed in the last line of \cref{tab:rel-error-scaling-0d-no-sym}.
	We find that this also holds for other initial conditions and other values for $\bar{N}$ and $\bar{M}$. 
	Therefore, we conclude that the 2D gls{kt} scheme is also perfectly suited to handle two-dimensional \gls{frg} problems that involve Goldstone-like modes in terms of advection in the fluid-dynamic reformulation.

\section{Selected examples in three-dimensional Euclidean spacetime} 
\label{sec:higher-dimensional_models}

	Having extensively tested the 2D \gls{kt} scheme for zero-dimensional \gls{frg} problems, we now turn to two selected sample applications in higher spacetime dimensions.

	First, we consider the $O(2)$ model in three Euclidean dimensions.
	Here, we compare the 2D \gls{kt} scheme with the 1D \gls{kt} scheme at very high resolution which represents another test for the two-dimensional scheme within a truncation.
	Similar to zero dimensions, this is a ``diffusion only'' problem.

	Second, we consider the three-dimensional version of the $O(\bar{N}) \times O(\bar{M})$ model from the last section.
	Here, we have advection included in the fluid-dynamical reformulation.
	However, we cannot benchmark our results against results from the 1D \gls{kt} scheme since the problem is inherently two-dimensional because of the two invariants.
	Nevertheless, we can study convergence of our results by comparing them to results from calculations with different resolutions.

\subsection{Example I: \texorpdfstring{$O(2)$}{O(2)}-model in three dimensions}
\label{subsec:example_1}

	We start with the well-known $O(N)$ model with $N = 2$ in $d = 3$ Euclidean spacetime dimensions in \gls{lpa} using the Litim regulator~\cite{Litim:2001up,Litim:2001fd}. 

\subsubsection{Setup}

	Our ansatz for the effective average action reads
		\begin{align}
			\bar{\Gamma}_k [ \vec{\varphi} \, ] = \int \dd^d x \, \big[ \tfrac{1}{2} \, ( \partial_\mu \vec{\varphi} \, )^2 + \tilde{U}_k ( \varrho ) \big] \, 
		\end{align}
	with the $O(N)$ invariant $\varrho = \frac{1}{2} \, \vec{\varphi}^{\, 2}$, the \gls{rg} scale $k ( t ) = \Lambda \, \ee^{- t}$, \gls{rg} time $t \in [ 0, \infty )$, and \gls{uv} cutoff $\Lambda$.
	
	In contrast to zero spacetime dimensions, this is of course a truncation of the full effective average action that only contains the scale-dependent effective potential $\bar{U}_k ( \varrho )$ and second-order derivative terms.
	Hence, we do not benchmark the results from the solution of the flow equation of the effective potential with the 2D \gls{kt} scheme against exact results but rather compare them with results from a solution computed with the 1D \gls{kt} scheme at very high resolution.

	To be specific, on the one hand, we use the \gls{rg} flow equation ($N = 2$, $d = 3$)
		\begin{align}
			\partial_t U = - A_d \, k^{d + 2} \, \bigg( \frac{N - 1}{k^2 + \frac{1}{\sigma} \, \partial_\sigma U} + \frac{1}{k^2 + \partial_\sigma^2 U} \bigg) \, .
		\end{align}
	To obtain this equation, we have made use of the symmetry in field space and projected onto the field configuration $\vec{\varphi} = ( 0, \sigma )^T$, such that $U = U ( t, \sigma )$.
	We add that $A_d = \frac{\Omega_d}{d ( 2 \uppi )^d}$ and $\Omega_d = \frac{2 \uppi^{d/2}}{\Gamma ( d/2 )}$. 
	In conservative form, defining $u = u ( t, \sigma ) = \partial_\sigma U$, this flow equation reads
		\begin{align}
			\partial_t u = \frac{\dd}{\dd \sigma} \bigg[ - A_d \, k^{d + 2} \, \bigg( \frac{N - 1}{k^2 + \frac{1}{\sigma} \, u} + \frac{1}{k^2 + \partial_\sigma u} \bigg) \bigg] \, .	\label{eq:flow_equation_test_case_o_n_3d_1d}
		\end{align}
	We solve this equation with the 1D \gls{kt} scheme  with $n_\mathrm{cells} = 2001$ on $\sigma \in [ 0, \varphi_\mathrm{max} ]$, see also \reff\cite{Koenigstein:2021syz}. 
	The solution then serves as the reference solution for our calculations based on the 2D \gls{kt} scheme.
	
	On the other hand, the flow equation can be kept two-dimensional in field space, \ie, $U = U ( t, \vec{\varphi} \, )$, similar to the zero-dimensional test cases from \cref{sec:o_2_symmetric_model}.
	The corresponding flow equation on the computational domain $[ - \varphi_\mathrm{max}, \varphi_\mathrm{max} ] \times [ - \varphi_\mathrm{max}, \varphi_\mathrm{max} ]$ is
		\begin{align}
			& \partial_t U =	\Vdistance	\label{eq:flow_equation_test_case_o_n_3d_2d}
			\\
			= \, & - \frac{A_d \, k^{d + 2} \, (2 k^2 + \partial_{\varphi_1}^2 U + \partial_{\varphi_2}^2 U)}{( k^2 + \partial_{\varphi_1}^2 U ) ( k^2 + \partial_{\varphi_2}^2 U ) - ( \partial_{\varphi_1} \partial_{\varphi_2} U ) ( \partial_{\varphi_2} \partial_{\varphi_1} U )} \, .	\Vdistance	\nonumber
		\end{align}
	 Introducing $u = \partial_{\varphi_1} U$ and $v = \partial_{\varphi_2} U$, we find the conservative form of the flow equation of our model:
		\begin{align}
			\partial_t
			\begin{pmatrix}
				u
				\\
				v
			\end{pmatrix}
			= \partial_{\varphi_1}
			\begin{pmatrix}
				Q
				\\
				0
			\end{pmatrix}
			+ \partial_{\varphi_2}
			\begin{pmatrix}
				0
				\\
				Q
			\end{pmatrix} \, ,
		\end{align}
	with the diffusion flux
		\begin{align}
			Q = - \frac{A_d \, k^{d + 2} \, (2 k^2 + \partial_{\varphi_1} u + \partial_{\varphi_2} v)}{( k^2 + \partial_{\varphi_1} u ) ( k^2 +\partial_{\varphi_2} v ) - ( \partial_{\varphi_1} v ) ( \partial_{\varphi_2} u )} \, .	\label{eq:diffusion_flux_test_case_o_n_3d_2d}
		\end{align}
	This set of equations can be solved with the 2D \gls{kt} scheme along the lines of the previous sections.
	
	Lastly, we choose the \gls{uv} potential to assume the form
		\begin{align}
			\tilde U ( \rho ) = 5 (\rho - 0.4) (\rho - 0.1) (\rho - 0.3) (\rho - 0.025) \label{eq:example_1_uv_potential}
		\end{align}
	at the cutoff scale $\Lambda = 1$.\footnote{Since we are mainly interested in the numerical aspects and comparison of the two \gls{kt} schemes we do not discuss any physical implications of the potential and scales.
	The same applies to the next example.}
	Note that we expressed all dimensionful quantities in units of $\Lambda$.
	The \gls{uv} potential is chosen such that the \gls{ir} minimum is nonzero and the \gls{rg} flow therefore ends in the symmetry broken phase. 
	This implies that we can use the position of the \gls{ir} minimum as well as the curvature mass at the minimum as observables to compare the 2D with the 1D \gls{kt} scheme.
	Furthermore, we introduced a complication by choosing a potential with a nontrivial shape which is not just the usual mexican hat potential, see \cref{fig:example_1} (upper left panel for the potential at the \gls{uv} scale).

	As a consquence of the symmetry breaking, it is no longer possible to flow arbitrarily far into the \gls{ir} without encountering numerical instabilities due to finite resolution close to the poles of the diffusion flux \labelcref{eq:diffusion_flux_test_case_o_n_3d_2d}.
	Because of that, we compare the 2D \gls{kt} scheme with the 1D \gls{kt} scheme at the same suitably small chosen \gls{ir} cutoff scale $k_\mathrm{IR} ( t_\mathrm{IR} )$.\footnote{Of course, by using the 1D \gls{kt} scheme, it is possible to flow to very small \gls{rg} scales because it is numerically cheaper to work at higher resolution.}
	Additionally, the structure of the denominator in the fluxes is simpler, see \cref{eq:flow_equation_test_case_o_n_3d_1d}, such that errors from mixed derivatives $\partial_{\varphi_i}\partial_{\varphi_j} U$ with $i\neq j$ are not present. 
	Still, we ensure in our numerical calculations that we are already in the deep \gls{ir} regime, where the potential is no longer smooth in the symmetry broken phase but tends to be flat in the small field region and approaches convexity.
	For a detailed discussion on convexity, we refer to \reffs~\cite{Litim:2006nn,Zorbach:2024zjx} and, for the related issue of time stepping, we refer to \reff\cite{Ihssen:2023qaq}. 
	In any case, the (numerical) parameters of our present study of this model can be found in \cref{tab:numerical_parameters_examples}.
		\begin{table}[b]
			\caption{\label{tab:numerical_parameters_examples}%
				Parameters used for the calculations of \cref{subsec:example_1}.
				For the integration we used \textit{RK45}.
			}
			\begin{ruledtabular}
				\setlength\extrarowheight{2pt}
				\begin{tabular}{l c c c c c}
					example	& $\sigma_\mathrm{max}$	& $\Lambda$	& $t_\mathrm{IR}$	&	$r_{\mathrm{tol}}$	&	$a_{\mathrm{tol}}$
					\\
					I		&	1 & 1	&	3.5			&	$10^{-12}$			&	$10^{-12}$
					\\
					II		&	6 & 40	&	3.69			&	$10^{-10}$			&	$10^{-12}$
				\end{tabular}
			\end{ruledtabular}
		\end{table}

\subsubsection{Discussion}

	Let us now discuss the results of our \gls{rg} flow study of the $O(2)$ model in three dimensions.
	\medskip

\paragraph{Qualitative discussion} 

		\begin{figure*}
			\includegraphics[width=\textwidth]{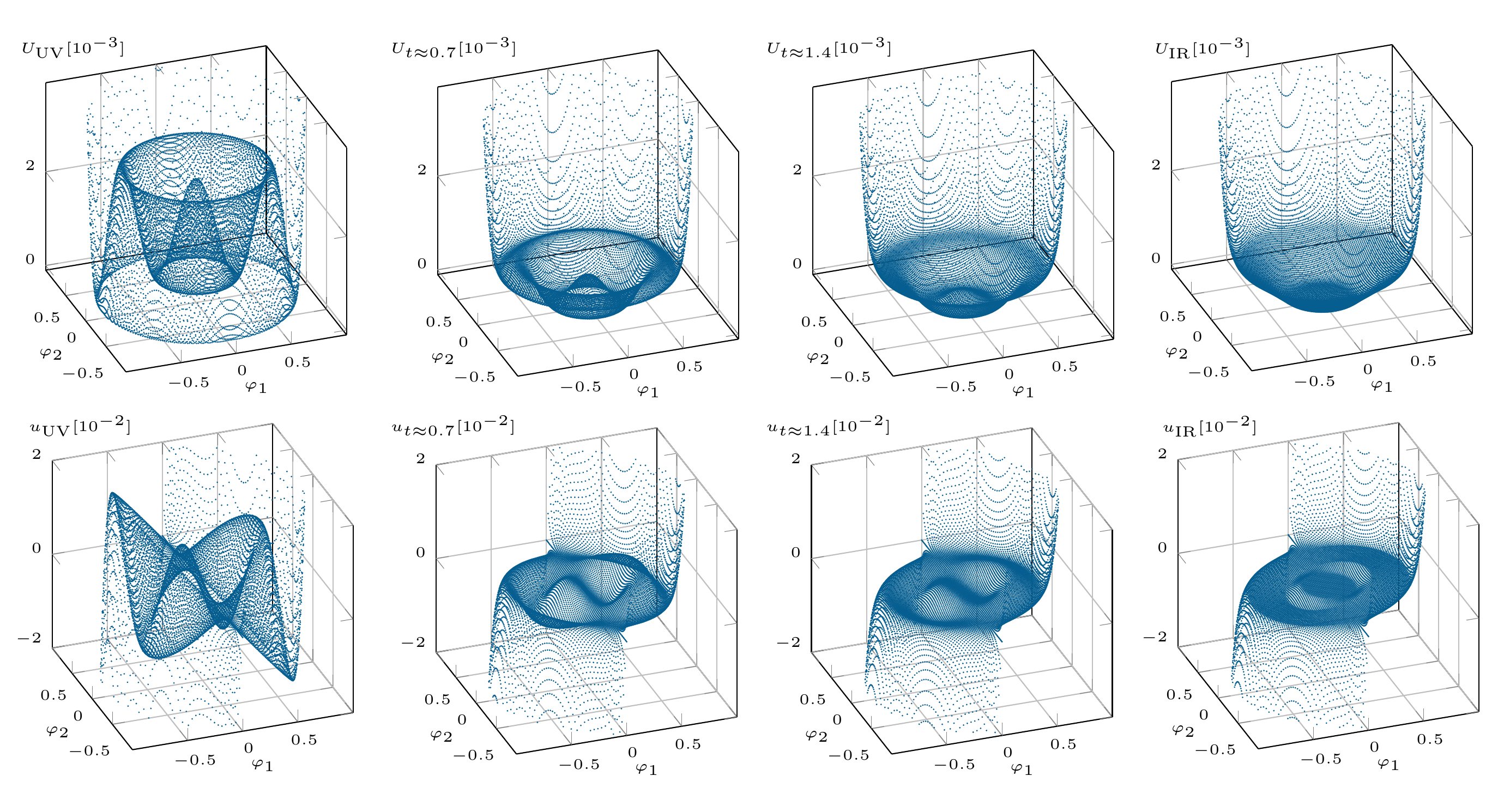}
			\caption{\label{fig:example_1}%
				The \gls{rg} time evolution of the of the potential $U ( t, \vec{\varphi} \, )$ (upper row) from the 2D \gls{kt} scheme and the corresponding $\varphi_1$-derivative of the potential $u ( t, \vec{\varphi} )$ (lower row) from the \gls{uv} (left column) to the \gls{ir} (right column) and selected intermediate times for the example I, \cref{eq:example_1_uv_potential}, with $n_{\mathrm{cells}} = 71$.
			}
		\end{figure*}
	At the beginning of the \gls{rg} flow associated with the \gls{uv} initial condition \labelcref{eq:example_1_uv_potential}, the potential well separating the outer ``ring'' of degenerate minima from the inner ``ring'' of degenerate minima starts to ``melt'', see \cref{fig:example_1}.
	The same happens for the maximum in the center of the potential.
	On the level of the derivative of the potential (lower panels), it is clearly visible that the nonlinear diffusion tends to ``equilibrate" the inner region(s) of the potential while the diffusion coefficients/fluxes are already smaller in the outer region.
		\begin{figure}
			\includegraphics{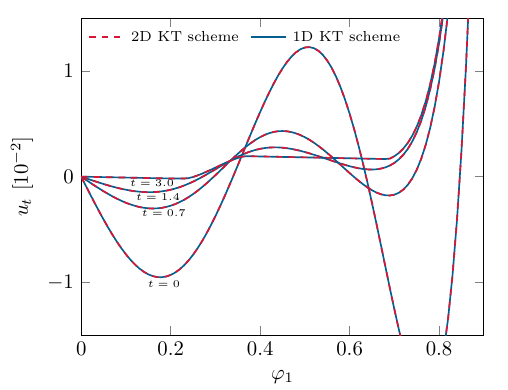}
			\caption{\label{fig:example_1_1d_cut}%
				The \gls{rg} flow of the derivative of the potential $u ( t, \sigma )$ from the 1D \gls{kt} scheme (red, dashed) and the $\varphi_{1}$-derivative of the potential $u ( t, \vec{\varphi} \, )$ from the 2D \gls{kt} scheme (blue solid) (evaluated at $\varphi_2 = 0$) at selected \gls{rg} times.
				The figure is essentially a section along the $\varphi_1$-axis for positive $\varphi_1$ of \cref{fig:example_1}.
			}
		\end{figure}
	When the flow approaches the \gls{ir} the competition between the $k^2$-terms and the gradients in the flux \labelcref{eq:diffusion_flux_test_case_o_n_3d_2d} sets in.
	Depending on the sign of the gradient, the diffusion is either enhanced or suppressed.
	This leads to the formation of the ``tilted plateaus'' in the derivative of the potential in the \gls{ir} and the formation of edges. 
	Usually, one would not expect such features in ordinary diffusion-type problems.
	Here, however, we are confronted with highly nonlinear diffusion with diffusion coefficients that depend on the derivatives of the fluid fields themselves as well as on the time.
	On the level of the potential itself, this causes the innermost region to be flat (within the numerical resolution) and the potential to approach convexity.
	A particularly interesting feature is the intermediate region between the flat region and the asymptotic part, which has constant slope in radial direction.
	Two ``rings'' (edges) are visible where the regions are connected.
	As a consequence, in the presence of the linear symmetry breaking term, one would observe phase transitions at these points. 
	The latter features are also clearly visible in  \cref{fig:example_1_1d_cut}, where we show a cut along the $\varphi_1$-axis for positive $\varphi_1$ of the derivative of the potential at selected \gls{rg} times.
	Note that this cut essentially corresponds to the computational domain in calculations using the 1D \gls{kt} scheme. 
	In any case, from a numerical standpoint, it is quite remarkable how the 2D version of the flow equation~\labelcref{eq:flow_equation_test_case_o_n_3d_2d}, which solely involves complicated nonlinear diffusion, leads to the same result as the calculations based on the 1D version of the flow equation, see Eq.~\labelcref{eq:flow_equation_test_case_o_n_3d_1d}, where the dynamics is driven by an advection-diffusion equation.

\paragraph{Quantitative discussion}

\begin{figure}
			\includegraphics{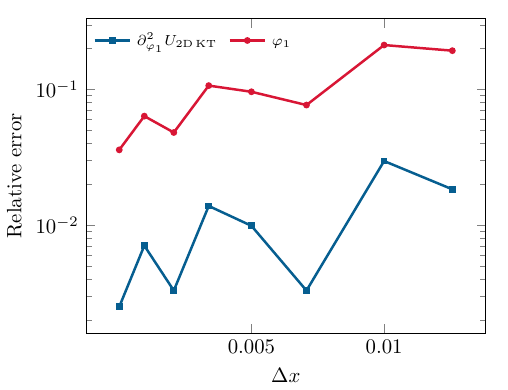}
			\caption{\label{fig:error_kt1_kt2_example_1}%
				The relative error between the 2D \gls{kt} scheme and the 1D \gls{kt} scheme ($n_{\mathrm{cells}} = 2001$) for the position of the minimum as well as $\partial_{\varphi_i}^2U$ for the higher-dimensional example~I (see \cref{subsec:example_1} with the \gls{uv} potential \labelcref{eq:example_1_uv_potential}) as a function of the numerical resolution $\Delta x$.
				The curvature mass $\partial_{\varphi_1}^2U_{\mathrm{1D\,KT/2D\,KT}}$ has been obtained via \cref{eq:gamma_2_kt} from the solution of the \gls{pde} system with the 1D/2D \gls{kt} scheme.
			}
		\end{figure}

	Returning now to \cref{fig:example_1_1d_cut}, we observe that the 1D \gls{kt} scheme and the 2D \gls{kt} scheme yield the same results for the derivative of the potential at the same \gls{ir} cutoff scale $k_\mathrm{IR} ( t_\mathrm{IR} )$.
	In order to quantitfy this observation, we also extracted the \gls{ir} position of the minimum from the 1D and 2D \gls{kt} scheme as well as the curvature mass $m^2 = \partial^2_{\varphi_i} U$ at the minimum.
	For the sake of the simplicity, we employed a sign change in the cell averages $\bar{u}_i$ to determine the position of the minimum and used the cell center of the positive cell average as the minimum position.
	The curvature mass is extracted as the right derivative at this cell center.
	While we extracted the 1D result with a high spatial resolution in order to use it as our ``exact'' reference, the 2D results are obtained at different numerical resolutions in order to observe correct error scaling.
	In \cref{fig:error_kt1_kt2_example_1}, we observe that the numerical error of the 2D \gls{kt} scheme indeed decreases systematically with increasing numerical resolution, see also \cref{tab:rel-error-scaling-examples} for the corresponding scaling exponents.	
	Hence, we conclude that the 2D \gls{kt} also performs satisfactorily in higher-dimensional models with $O(2)$ symmetry and leads to quantitative reliable results which are solely limited by the numerical resolution.

\subsection{Example II: \texorpdfstring{$O(\bar{N}) \times O(\bar{M})$}{O(N)xO(M)}-model in three dimensions} 
\label{subsec:example_2}

		\begin{figure*}
			\includegraphics[width=\textwidth]{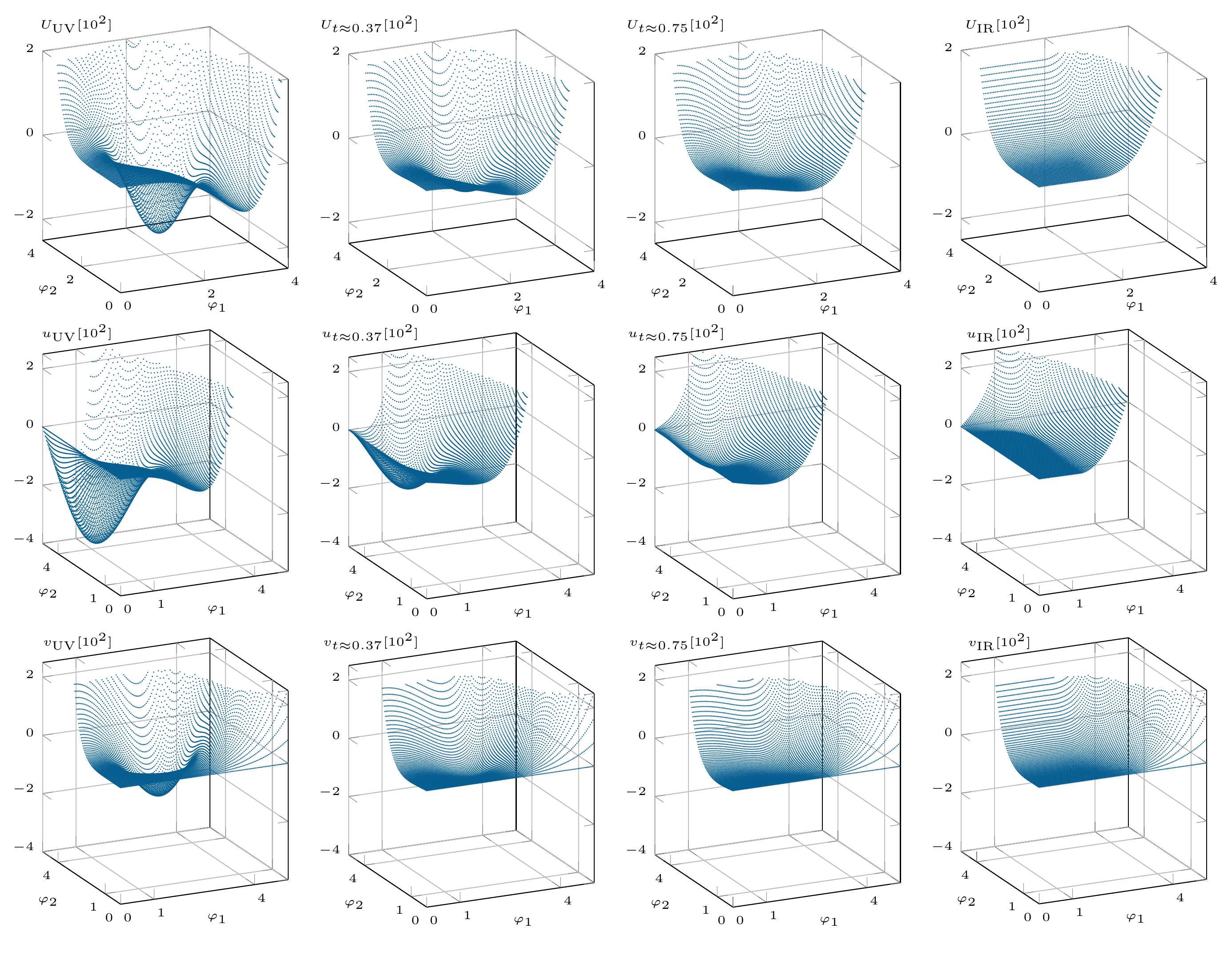}
			\caption{\label{fig:example_2}%
				\gls{rg} time evolution of the of the potential $U ( t, \sigma_1, \sigma_2 )$ (upper row) from the 2D \gls{kt} scheme and the corresponding $\sigma_1$- and $\sigma_2$-derivative of the potential $u ( t, \sigma_1, \sigma_2  )$ and $v ( t, \sigma_1, \sigma_2  )$ (middle/lower row) from the \gls{uv} (left column) to the \gls{ir} (right column) and selected intermediate times for the example II, \cref{eq:example_2_uv_potential}, with $n_{\mathrm{cells}} = 151$.
			}
		\end{figure*}
	We now turn to the $O(\bar{N}) \times O(\bar{M})$ model in three Euclidean dimensions, again considering the \gls{lpa} and using the Litim regulator.
	For concreteness, we shall set $\bar{N} = 2$ and $\bar{M} = 3$ in all our numerical studies.

\subsubsection{Setup}

	Our ansatz for the effective average action is given by
		\begin{align}
			& \bar{\Gamma}_k [ \vec{\varphi}_1, \vec{\varphi}_2 ] =	\Vdistance
			\\
			= \, & \int \dd^3 x \, \big[ \tfrac{1}{2} \, ( \partial_\mu \vec{\varphi}_1 \, )^2 + \tfrac{1}{2} \, ( \partial_\mu \vec{\varphi}_2 \, )^2 + \tilde{U}_k ( \varrho_1, \varrho_2 ) \big] \, ,	\Vdistance	\nonumber
		\end{align}
	where
		\begin{align}
			&	\varrho_1 = \tfrac{1}{2} \, \vec{\varphi}_1^{\, 2} \, ,	&&	\varrho_2 = \tfrac{1}{2} \, \vec{\varphi}_2^{\, 2} \, 
		\end{align}
	are the invariants of the $O(\bar{N})$ and $O(\bar{M})$ group, respectively, and
		\begin{align}
			\vec{\varphi}_1 = \, & ( \varphi_1, \varphi_2, \ldots, \varphi_{\bar{N}} )^T \, ,	\vdistance
			\\
			\vec{\varphi}_2 = \, & ( \varphi_{\bar{N}+1}, \varphi_{\bar{N}+2}, \ldots, \varphi_{\bar{N} + \bar{M}} )^T \, .	\vdistance
		\end{align}
	Using the same conventions as before, and evaluating the Wetterich equation for the background field configurations
		\begin{align}
			&	\vec{\varphi}_1 = ( 0, \ldots, 0, \sigma_1 )^T \, ,	&&	\vec{\varphi}_2 = ( \sigma_2, 0, \ldots, 0 )^T \, ,	\label{eq:background_field_configurations_on_om}
		\end{align}
	we find the following \gls{rg} flow equation of the effective potential (see, \eg, \reff\cite{Hawashin:2024dpp}):
		\begin{align}
			& \partial_t U =	\Vdistance
			\\
			= \, & - A_d \, k^{d + 2} \bigg( \frac{( \bar{N} - 1 )}{k^2 + \frac{1}{\sigma_1} \, \partial_{\sigma_1} U} + \frac{( \bar{M} - 1 )}{k^2 + \frac{1}{\sigma_2} \, \partial_{\sigma_2} U} +	\Vdistance	\nonumber
			\\
			& + \frac{\big( 2 k^2 + \partial_{\sigma_1}^2 U + \partial_{\sigma_2}^2 U \big)}{\big( k^2 + \partial_{\sigma_1}^2 U \big) \big( k^2 + \partial_{\sigma_2}^2 U \big) - \big( \partial_{\sigma_1} \partial_{\sigma_2} U \big) \big( \partial_{\sigma_2} \partial_{\sigma_1} U \big)} \bigg) \, .	\Vdistance	\nonumber
		\end{align}
	In complete analogy to \cref{sec:generalization_to_on_om}, this equation can be brought into the shape of a conservation law by introducing the fields $u = \partial_{\sigma_1} U$ and $v = \partial_{\sigma_2} U$.
	Here, we refrain from presenting this form of the flow equation as it can be easily derived from a comparison with \cref{eq:flow_equation_U_on_om_sigma,eq:conservation_law_u_v_on_om,eq:diffusion_flux_u_v_on_om,eq:advection_flux_u_v_on_om}.
	For the \gls{uv} potential in our numerical calculations, we simply choose
		\begin{align}
			\label{eq:example_2_uv_potential}
			\tilde U ( \rho_1, \rho_2 ) =\,& -10 \, \rho_1^2 - \rho_1 \rho_2 - 15 \, \rho_1 \rho_2^2 \, +  \\
			&\quad + \tfrac{1}{4} \, (\rho_1+\rho_2)^4 \nonumber \, .
		\end{align}
	Instead of expressing all quantities in terms of the cutoff scale $\Lambda$, we shall use arbitrary units here and set $\Lambda = 40$. 
	Note that our choice of the form of the \gls{uv} potential is not phenomenologically motivated.
	It is only constructed such that it comprises nontrivial dynamics in the \gls{rg} flow and ends uo in the symmetry broken regime with a residual $O ( \bar{N} - 1 ) \times O ( \bar{M} - 1 )$ symmetry in the \gls{ir}, see \cref{fig:example_2}.

\subsubsection{Discussion}

	Let us now discuss the results of the \gls{rg} flow of the $O(\bar{N}) \times O(\bar{M})$ model in three dimensions.

\paragraph{Qualitative discussion}

	In the present case, the \gls{uv} potential has a global $O ( \bar{N} ) \times O ( \bar{M} )$ symmetry.
	However, there is no $O ( \bar{N} + \bar{M} )$ symmetry and also the $O ( \bar{N} )$ and the $O ( \bar{M} )$ symmetries are separately broken by nontrivial minima.
	In \cref{fig:example_2} (upper left panel), one clearly observes a complicated nonconvex shape of the potential as a function of the two background fields.
	This can also be seen on the level of the derivatives with respect to the background field configurations, see the middle and lower left panels.
	Note that the computational domain is greater than the plot region, see \cref{tab:numerical_parameters_examples}, where we list all numerical control parameters.
	Of course, despite the large gradients, we also ensured that the initial condition is valid in the sense that it does not overshoot the poles of the propagator already at the \gls{uv} scale.

	In the \gls{rg} flow from the \gls{uv} to the \gls{ir} we find the usual overall behavior of purely bosonic systems.
	To be more specific, the minima of the potential equalize in their depth and the potential eventually starts to become convex.
	However, as can be seen from the derivatives of the potential in the lower panels of \cref{fig:example_2}, the dynamics in the different field/invariant directions sets in at different speeds and \gls{rg} times.
	This is due to the different slopes and curvatures of the potential in the different directions.
	Furthermore, the fact that $\bar{N} \neq \bar{M}$ also plays a role because there is more advection in the direction of the field with larger $\bar{N}$ or $\bar{M}$, respectively.

	We shall not discuss the details underlying the dynamics of this system further here. 
	The main motivation for considering this particular test case is to demonstrate the capabilities of our 2D \gls{kt} framework for problems of two invariant directions via discretization in terms of the background fields.

\paragraph{Quantitative discussion}

		\begin{figure}
			\includegraphics{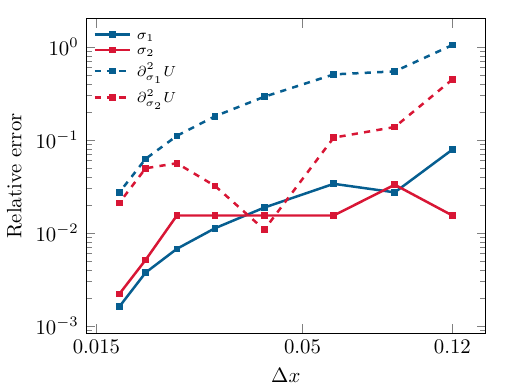}
			\caption{\label{fig:error_kt2_self_example_2}%
				Relative error of the $\sigma_{1/2}$-position of the minimum as well as $\partial_{\sigma_{1/2}}^2 U$ computed with $\Delta x$ and the same quantity at the highest tested resolution ($\Delta x = 0.015$) for \cref{subsec:example_2} with \gls{uv} potential \labelcref{eq:example_2_uv_potential}.
			}
		\end{figure}
	As already mentioned above, there are no benchmarks for our results for this model available, neither from the path integral nor from the 1D \gls{kt} scheme.
	Thus, the only way to assess the quality of the 2D \gls{kt} scheme is to study the convergence of the numerical results.
	To this end, we consider the position of the minimum and the curvature mass of the radial modes.
	In \cref{fig:error_kt2_self_example_2}, we show the relative error for these quantities between the calculation with our highest resolution $\Delta x = 0.015$ and calculations at lower resolution.
	Overall, we find that the relative deviation decreases systematically with increasing resolution.
	The corresponding scaling exponents can be found in \cref{tab:rel-error-scaling-examples}, which are in good overall agreement with the expected error scaling of the \gls{kt} scheme.
	However, we also remark that calculations at even higher resolution are required for an absolutely trustworthy result.

\begin{figure}
			\includegraphics{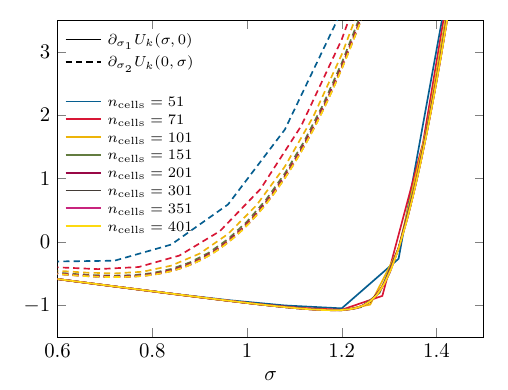}
			\caption{\label{fig:example_2_profiles}%
				Comparison of $u ( t, \sigma_1, \sigma_2 )$ and $v ( t, \sigma_1, \sigma_2 )$ in the \gls{ir} at $k_\mathrm{IR} ( t_\mathrm{IR })/\Lambda =0.025$ for different numerical resolutions (number of cells) for the model presented in \cref{subsec:example_2} with the \gls{uv} potential~\labelcref{eq:example_2_uv_potential}.
			}
		\end{figure}

	As a second convergence test, we compare the cuts through the potential at the \gls{ir} cutoff scale $k_\mathrm{IR} ( t_\mathrm{IR} )/\Lambda = 0.025$ for different numerical resolutions.
	To be specific, we consider $u ( t, \sigma_1, 0 )$ and $v ( t, 0, \sigma_2 )$, see \cref{fig:example_2_profiles}.
	Again, we find that the results converge systematically with increasing resolution, as it should be.	
	
	In view of our results for this test case and also the previous one, we conclude that it is promising from a phenomenological and numerical standpoint to generalize our 2D \gls{kt} framework to more complicated models with more than two invariants.
	
	\begin{table}[b]
		\caption{\label{tab:rel-error-scaling-examples}%
			Error scaling exponent $n$ extracted from the scaling of observables with $\Delta x^n$, corresponding to \cref{fig:error_kt1_kt2_example_1,fig:error_kt2_self_example_2}.
		}
		\begin{ruledtabular}
			\setlength\extrarowheight{2pt}
			\begin{tabular}{l c c c c c c}
				& $n(\varphi_1)$ & $n(\partial^2_{\varphi_1} U)$ &	$n(\sigma_1)$	&	$n(\sigma_2)$	&	$n(\partial^2_{\sigma_1} U)$	&	$n(\partial^2_{\sigma_2} U)$
				\\
				\cref{subsec:example_1}& 0.9(9)& 0.9(0) & & & & 
				\\
				\cref{subsec:example_2}& & & $1.7(0)$ & $0.8(5)$ & $1.6(6)$ & $1.2(4)$
			\end{tabular}
		\end{ruledtabular}
	\end{table}

\section{Summary, conclusions, and outlook}
\label{sec:conclusion_and_outlook}

\subsection{Summary}

	In this work, we discussed the numerical treatment of \gls{frg} flow equations in situations where the effective potential has to be resolved in more than one field or invariant direction.
	We have shown that the \gls{frg} flow equation of the effective potential of such systems can be reformulated as a fluid-dynamical system of advection-diffusion type for $O(N)$ and $O(\bar{N}) \times O(\bar{M})$ models in zero and higher-dimensional spacetime.
	We argued that this also generalizes to other systems which for example involve fermions.
	Furthermore, we have presented a numerical scheme from \gls{cfd}, which was developed to solve exactly such fluid-dynamical \glspl{pde}, namely the 2D \gls{kt} scheme, and we adapted it to the present problem, also noting a possible minor defect in the original implementation by Kurganov and Tadmor.
	In order to demonstrate the power of our approach, we constructed various models in zero spacetime dimensions and benchmarked the results from the \gls{kt} scheme against exact results from the underlying path integral.
	Most importantly, these tests comprised models that involved nonanalyticities in field space, symmetry breaking, multiple minima, and the misalignment of the symmetry axis from the field-space axes.
	Moreover, we applied our numerical framework to more realistic higher-dimensional models, namely the $O(2)$ and the $O(\bar{N}) \times O(\bar{M})$ model in three dimensions within the \gls{lpa}.
	Also for these cases, we carefully conducted convergence tests and found that the \gls{kt} scheme is capable of solving these models with satisfying accuracy.

\subsection{Conclusions}

	In total, we conclude that the \gls{kt} scheme can be used to numerically solve \gls{frg} flow equations in zero dimensions and also in higher-dimensional spacetimes.
	Our numerical framework represents a powerful blackbox solver that may also be included in existing \gls{frg} codes.
	All benchmark tests showed error scaling with a numerical resolution that is in agreement with the expected error scaling of the \gls{kt} scheme.
	However, we also demonstrated limitations of our approach.
	Especially in situations with involved symmetry breaking patterns, rather large numbers of finite volume cells are required to resolve the location of the minimum and the vertex functions to a satisfactory degree.
	The same is the case for situations with little symmetry or misalignment of the symmetry axis of the problem with the axes of the discretization.
	However, this poses a severe though solvable challenge since the number of degrees of freedom that is evolved effectively as \glspl{ode} grows quadratically with the number of cells.
	In the future, this may be tackled by exploiting high-performance computing techniques for an implementation of our our present framework.
	
	In addition to the presentation of our test setup and the explicit tests, we also discussed qualitative aspects of flow equations in higher-dimensional field space and details of the numerical implementation.
	For example, we described restrictions on the \gls{uv} potentials and hence the microscopic models that can be treated naively with the \gls{frg} approach by analyzing the pole structures of the propagators (the fluxes).

\subsection{Outlook}

	In our present technical work we showed that the fluid-dynamical approach to \gls{frg} flow equations with a field-space dependent potential and couplings is a viable and powerful approach which provides much more than just a benchmark for other numerical schemes.
	In fact, the present work indeed provides methods that can be used for a wide range of applications relevant for a huge variety of fields, ranging from condensed-matter physics to high-energy physics.
	One concrete application is the analysis of symmetry breaking patterns in the \gls{qcd} phase diagram.
	Here, the \gls{frg} is a very promising tool since it is capable of treating the nonperturbative regime of \gls{qcd} from first principles and can therefore be used to study the phase diagram in a systematic way, see, \eg, \reffs\cite{Braun:2019aow,Fu:2019hdw}.
	As the dynamics of \gls{qcd} is highly involved, the \gls{frg} flow equations are rather complicated and involve a large number of degrees of freedom.
	In particular at finite density, a reliable numerical framework is therefore required to map out the phase structure which may indeed be governed by multiple competing condensates (\eg, chiral and diquark condensates) and first-order phase transitions. 
	The framework developed in our present work may represent a valuable tool to provide a deeper qualitative and quantitative understanding of the dynamics of \gls{qcd} in this regime at low temperatures and high densities.
	
\acknowledgments

	N.Z., A.K., J.B.\ acknowledge support by the \textit{Deutsche Forschungsgemeinschaft} (DFG, German Research Foundation) through the CRC-TR 211 ``Strong-interaction matter under extreme conditions'' -- project number 315477589 -- TRR 211.

	N.Z.\ acknowledge support from the Helmholtz Graduate School for Hadron and Ion Research.

	A.K.\ thanks D.~Rischke and S.~Floerchinger for many discussions about flow equations and their great support and encouragement at the ITP in Frankfurt and TPI in Jena, respectively.

	The authors thank E.~Grossi, M.~Steil, J.~Stoll, N.~Wink for many discussions and comments concerning numerical methods for the FRG and multidimensional flow equations as well as collaboration in related projects.

	The authors also thank F.~Ihssen, F.~R.~Sattler, as well as M.~Scherer and his group for valuable comments and ideas during discussions at the ERG 2024 conference.
	
	The authors also thank A.~Sciarra for his advises concerning the visualization of the results, data management, as well as the preparation of the publication of the code.

	The authors of this work disclose the use of the \textit{GitHub Copilot} tool for the preparation of the manuscript.
	The underlying code is written in \textit{Python3} and reference values are calculated with \textit{Mathematica14.1}.

\appendix

\section{Computational times}
\label{app:computational_times}

	In this appendix we provide some estimates for the computational wall times required to perform the calculations in two-dimensional field space as presented in the present work.
	This may serve as reference for future studies to estimate the computational resources required for similar calculations.
	
	All calculations in the present work have been performed on a single core of an \texttt{AMD Ryzen Threadripper 3990X 64-Core Processor} CPU with $2.9\,\textrm{GHz}$ and with $128\,\mathrm{GB}$ of RAM.
	In \cref{tab:wall-time}, we list the wall times required to run the test cases at a resolution of $n_{\mathrm{cells}} = 200$ (without ghost cells) with the parameters given in the corresponding sections.
	As can be seen from the table, wall times can vary significantly between the different test cases and spacetime dimensions which has to be taken into account when applying the \gls{kt} scheme to other models.
	\begin{table}[b]
		\caption{\label{tab:wall-time}%
			Wall times required to run the test cases at a resolution of $n_{\mathrm{cells}} = 200$ (without ghost cells).
			For the tests in \cref{sec:o_2_symmetric_model,sec:non-symmetric_model,subsec:example_1}, this corresponds to a total number of $(400 \times 400) - 1$ cells, because $n_{\mathrm{cells}}$ defines the number of cells in positive $x$- and $y$-direction while, for \cref{sec:a_test_model_with_on_om_symmetry,subsec:example_2}, this corresponds to $200 \times 200$ cells in the positive quadrant only.
		}
		\begin{ruledtabular}
			\setlength\extrarowheight{2pt}
			\begin{tabular}{c c}
				test case	&	wall time (min)
				\\
				\cref{subsubsec:test_case_i}			&	$<6$
				\\
				\cref{subsubsec:test_case_ii}			&	$<4$
				\\
				\cref{subsubsec:test_case_iii}			&	$<5$
				\\
				\cref{subsubsec:test_case_iv}			&	$<4$
				\\
				\cref{subsec:test_case_v_nonsymmetric}	&	$<197$
				\\
				\cref{subsec:test_case_vi_pyramid}		&	$<12$
				\\
				\cref{sec:a_test_model_with_on_om_symmetry}  & $<39$ 
				\\
				\cref{subsec:example_1}					&	$<79$
				\\
				\cref{subsec:example_2}					 &	$<135$
			\end{tabular}
		\end{ruledtabular}
	\end{table}

\section{Implementation of the multi-dimensional KT central scheme}
\label{app:implementation_of_the_multi-dimensional_kt_central_scheme}

	In general, the \gls{kt} scheme is a \gls{fv} method tailored for solving fluid-dynamical \glspl{pde} of advection-diffusion type like \cref{eq:general_pd_kt_scheme}.
	In \cref{sec:numeric_approach}, the 2D \gls{kt} scheme is discussed locally, \ie, the equations are formulated in terms of cells and their adjacent cells.
	However, for the numerical implementation and in particular to get a better intuition on the functioning of the ghost cells, it is instructive to formulate the \gls{kt} scheme in a matrix formulation.
	Similar to \cref{sec:numeric_approach}, we present the two-dimensional semi-discrete version of the scheme meaning that we treat the temporal direction and the spatial directions as continuous and discrete, respectively.
	Using the semi-discrete version of the \gls{kt} scheme with $N_x$ cells in the $x$-direction and $N_y$ cells in the $y$-direction for the \gls{pde} \labelcref{eq:general_pd_kt_scheme}, it becomes a set of coupled ordinary differential equations which can be summarized in the matrix equation
		\begin{align}
			\partial_t \hat{\bm{u}} = \mathcal{F}_{\mathrm{KT}}(t, \hat{\bm{u}}) \, ,	\label{eq:semi_discrete_pd_kt_scheme_matrix_formulation}
		\end{align}
	where
		\begin{align}
			\hat{\bm{u}} = (\hat{\bm{u}}_{dyx})_{d = 0, \dots, \mathrm{dof} - 1; \, y = 0, \dots, N_y - 1, \, x = 0, \dots, N_x - 1} \, ,
		\end{align}
	\ie, $\hat{\bm{u}} \in \mathbb{R}^{\mathrm{dof} \times N_y \times N_x}$ and ``$\mathrm{dof}$'' denotes the number of degrees of freedom -- the number of fluid fields.
	The initial-value problem given by the \cref{eq:semi_discrete_pd_kt_scheme_matrix_formulation} and some initial condition $\hat{\bm{u}}(t=0) = \hat{\bm{u}}_0$ can then be solved numerically.
	
	This appendix is structured as follows:
	First, we introduce a ``slicing operator'' for the matrix formulation.
	Second, we define the two-dimensional spatial grid.
	Finally, in a third step, we discuss the matrix formulation of the \gls{kt} scheme which involves the specification of the \gls{rhs} of \cref{eq:semi_discrete_pd_kt_scheme_matrix_formulation} denoted as $\mathcal{F}_{\mathrm{KT}}$.

\subsection{The slicing operator}
	
	For the matrix formulation of the 2D \gls{kt} scheme, it is practical to introduce a so-called slicing operator which is used to systematically cut rows and columns of matrices.
	This enhances the readability of the formulation, particularly for readers familiar with the \textit{Python} package \textit{numpy} \cite{2020NumPy-Array}, as this slicing operator functions identically to that of \textit{numpy}.
	Consequently, all equations in the following sections can be directly implemented in \textit{Python}.
	
	Let $\bm{A} = ( \bm{A}_{i} )_{i = 0, 1, \dots, N-1} \in \mathbb{R}^{N}$, then we define the slicing operator $[ a \slice b]$ with $a, b \in \mathbb{Z}$ by
		\begin{align}
			&	\bm{A} [ a \slice b ] = ( \bm{A}_i )_{i = \bar a, \dots, \bar b - 1} \, ,	&&	\text{and}	&&	\bm{A} [ i ] = \bm{A}_{i} \, ,
		\end{align}
	where $\bar x = x$ if $x \geq 0$ else $\bar x = N + x$.
	For simplicity, if $a$ or $b$ is not set, we mean $a = 0$ or $b = N$, respectively.
	For example, we have $\bm{A} [ : ] = \bm{A}$.
	This slicing operator can be simply extended to higher dimensions.
	For example, let $\bm{A} = ( \bm{A}_{i_1 \dots i_M} )_{ i_1 = 0, \dots, N_1 - 1 ;\, \dots ; \, i_M = 0, \dots, N_M - 1} \in \mathbb{R}^{N_1 \cdots N_M}$, then we have 
		\begin{align}
			& \bm{A} [ a_1 \slice b_1,\dots, a_M \slice b_M ] =	\vdistance 
			\\
			= \, & ( \bm{A}_{i_1 \dots i_M} )_{ i_1 = \bar{a}_1,\dots, \bar{b}_1 - 1 ; \,\dots ; \, i_M = \bar{a}_M,\dots, \bar b_M - 1} \, 	\vdistance \nonumber
		\end{align}
	and $\bm{A}[i_1,\dots ,i_M] = \bm{A}_{i_1\dots i_M}$.

\subsection{The two-dimensional grid}

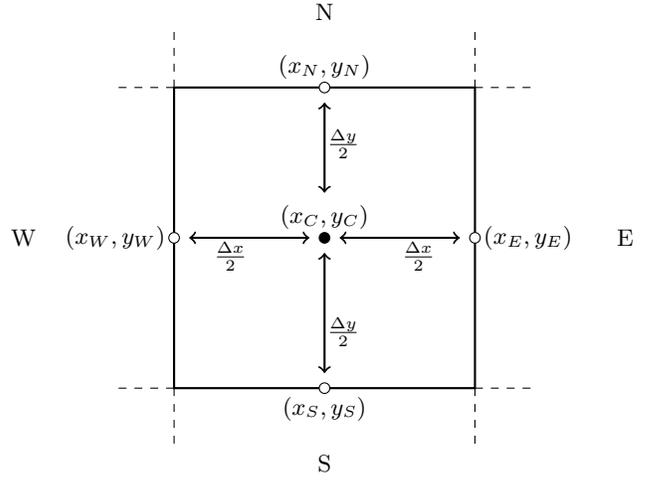
\begin{figure}
			\begin{tikzpicture}
				\draw[thick] (0,0) rectangle (4,4);
				
				\filldraw[black] (2,2) circle (2pt) node[anchor=south] {$(x_C, y_C)$};
				
				
				\node at (2,5) {N};
				\node at (2,-1) {S};
				\node at (-2,2) {W};
				\node at (6,2) {E};
				
				\filldraw[black,fill=white] (2,4) circle (2pt) node[anchor=south] {$(x_N, y_N)$};
				\filldraw[black,fill=white] (2,0) circle (2pt) node[anchor=north] {$(x_S, y_S)$};
				\filldraw[black,fill=white] (0,2) circle (2pt) node[anchor=east] {$(x_W, y_W)$};
				\filldraw[black,fill=white] (4,2) circle (2pt) node[anchor=west] {$(x_E, y_E)$};
				
				\draw[<->, thick] (2,2.6) -- (2,3.8);
				\draw[<->, thick] (2,0.2) -- (2,1.8);
				\draw[<->, thick] (2.2,2) -- (3.8,2);
				\draw[<->, thick] (0.2,2) -- (1.8,2);
				\draw[dashed] (4,4) -- (4,4.75);
				\draw[dashed] (4,4) -- (4.75,4);

				\draw[dashed] (4,0) -- (4,-0.75);
				\draw[dashed] (4,0) -- (4.75,0);

				\draw[dashed] (0,0) -- (0,-0.75);
				\draw[dashed] (0,0) -- (-0.75,0);

				\draw[dashed] (0,4) -- (0,4.75);
				\draw[dashed] (0,4) -- (-0.75,4);

				\node at (2.25,3.25) {$\tfrac{\Delta y}{2}$};
				\node at (2.25,0.75) {$\tfrac{\Delta y}{2}$};
				\node at (0.75,1.75) {$\tfrac{\Delta x}{2}$};
				\node at (3.25,1.75) {$\tfrac{\Delta x}{2}$};
			\end{tikzpicture}
			\caption{\label{fig:the-fluid-cell}%
				A sketch of a single fluid cell with the center point $( x_C, y_C)$ and the interfaces $(x_N, y_N)$, $(x_S, y_S)$, $(x_W, y_W)$, and $(x_E, y_E)$.
			}
		\end{figure}

	Since the fluid fields are summarized in the multidimensional matrix $\hat{\bm{u}} \in \mathbb{R}^{\mathrm{dof}\times N_y \times N_x}$ in the following, where $N_x$ and $N_y$ are the number of cells in $x$- and $y$-direction, respectively, we need grid objects of the same form/dimension.
	These are generated as follows:
	A general two-dimensional rectangular grid is completely specified by the locations of (the centers of) the edges parallel to the $x$- and $y$-direction.
	Let $x_{\mathrm{edges}}$ be the list of all $x$-coordinates of edges pointing in $x$-direction (analogously for $y_{\mathrm{edges}}$).
	Then, we can raise those one-dimensional lists to matrices $\bm{x}_{\mathrm{edges}}$ and $\bm{y}_{\mathrm{edges}}$ by
		\begin{subequations}
			\begin{align}
				\bm{x}_{\mathrm{edges}} [ i, j, : ] = \, & x_{\mathrm{edges}} \, ,	\vdistance
				\\
				\bm{y}_{\mathrm{edges}}[ i, :, k ] = \, & y_{\mathrm{edges}} \, ,	\vdistance
			\end{align}
		\end{subequations}
	for all $0 \leq i < \mathrm{dof}$, $0 \leq j < | y_{\mathrm{edges}}|$ and $0 \leq k < | x_{\mathrm{edges}}|$.
	Hence, the $\bm{x}_{\mathrm{edges}}$ object contains $\mathrm{dof} \cdot | y_{\mathrm{edges}} |$ copies of the $x_{\mathrm{edges}}$ list and the $\bm{y}_{\mathrm{edges}}$ object contains $\mathrm{dof} \cdot | x_{\mathrm{edges}} |$ copies of the $y_{\mathrm{edges}}$ list.
	This, however, will turn out to be useful for what follows:
		
	We can now determine the cell centers as well as the cell width in $x$- and $y$-direction.
	For the widths, we find
		\begin{subequations}
			\begin{align}
				\bm{\Delta}_{x} = \, & \bm{x}_{\mathrm{edges}} [ :, : \! - 1, 1 \! : ] - \bm{x}_{\mathrm{edges}} [ :, : \! - 1, : \! - 1 ] \, ,	\vdistance
				\\
				\bm{\Delta}_{y} = \, & \bm{y}_{\mathrm{edges}} [ :, 1 \! :, : \! - 1 ] - \bm{y}_{\mathrm{edges}} [ :, : \! - 1, : \! - 1 ] \, ,	\vdistance
			\end{align}
		\end{subequations}
	and for the cell centers we have
		\begin{subequations}
			\begin{align}
				\bm{x}_C = \, & \bm{x}_{\mathrm{edges}} [ :,: \! - 1, : \! - 1 ] + \tfrac{1}{2} \, \bm{\Delta}_x \, , \vdistance
				\\
				\bm{y}_C = \, & \bm{y}_{\mathrm{edges}} [ :, : \! - 1, : \! - 1 ] + \tfrac{1}{2} \, \bm{\Delta}_y \, .
			\end{align}
		\end{subequations}
	However, for the \gls{kt} scheme, we need further locations, \eg, the locations of the cell interfaces of each cell.\footnote{For higher reconstruction orders of the \gls{kt} scheme, additional positions for the corners of each cell are required, see for example \reff\cite{Kurganov2001:thirdorder}.}
	We define them analogously by
		\begin{subequations}
			\begin{align}
				&	\bm{x}_{W} = \bm{x}_C - \tfrac{1}{2} \, \bm{\Delta}_x \, ,	&&	\bm{x}_{E} = \bm{x}_C + \tfrac{1}{2} \, \bm{\Delta}_x \, ,	\vdistance
				\\
				&	\bm{y}_{W} = \bm{y}_{C} \, ,	&&	\bm{y}_{E} = \bm{y}_{C} \, ,	\vdistance
				\\
				&	\bm{x}_{S} = \bm{x}_C \, ,	&&	\bm{x}_{N} = \bm{x}_C \, ,	\vdistance
				\\
				&	\bm{y}_{S} = \bm{x}_C - \tfrac{1}{2} \, \bm{\Delta}_y \, ,	&&	\bm{y}_{N} = \bm{y}_C + \tfrac{1}{2} \, \bm{\Delta}_y \, ,	\vdistance
			\end{align} 
		\end{subequations}
	where we have used the $N$-$S$-$W$-$E$ convention as it is introduced in Ref.~\cite{Kurganov2002:2Dgas,Kurganov2001:thirdorder,Kurganov2000:HamJac}, see also \cref{fig:the-fluid-cell} for a sketch of the situation of a single cell.
	However, note that the above formulation allows for a parallel handling of all cells.
	
\subsection{The KT scheme in matrix formulation}

	With the definition of the two-dimensional grid as it is done in the previous section, we can now discuss the \gls{kt}-scheme implementation in a matrix formulation with an arbitrary number of degree of freedoms ($\mathrm{dof}$).
	This means that we have to define/construct the \gls{rhs}\ of \cref{eq:semi_discrete_pd_kt_scheme_matrix_formulation}, $\mathcal{F}_{\mathrm{KT}}$.
	For that we need at least two boundary layers on each side in the semi-discrete version of the \gls{kt} scheme, as discussed in detail in \cref{sec:adaptions_of_the_kt_scheme_to_our_frg_problems}.
	Here, the number of boundary layers is denoted by $\mathcal{B} \in \mathbb{N}_{\geq 2}$.
	Furthermore, let $N_x$ ($\bar N_x$) and $N_y$ ($\bar N_y$) be the number of cells excluding (including) the boundary cells in the $x$- and $y$-direction, respectively, meaning that $\bar N_{x} = |x_{\mathrm{edges}}| - 1$, $\bar N_{y} = |y_{\mathrm{edges}}| - 1$, $N_x = \bar N_x - 2 \mathcal{B}$ and $N_y = \bar N_y - 2 \mathcal{B}$.
	Hence, the dimensions of the grid objects are $\bm{\Delta}_x, \bm{\Delta}_y, \bm{x}_i, \bm{y}_i \in \mathbb{R}^{\mathrm{dof} \times \bar N_y \times \bar N_x}$ for $i \in \{N,S,W,E\}$.
	In the following, multiplication and division of matrices are understood component-wise.
	
	For a general \gls{pde} of the advection-diffusion type, \cf\ \cref{eq:general_pd_kt_scheme}, the \gls{rhs}\ of \cref{eq:semi_discrete_pd_kt_scheme_matrix_formulation} has two contributions in the end, one for the advection part and one for the diffusion part, such that the \gls{rhs}\ of \cref{eq:semi_discrete_pd_kt_scheme_matrix_formulation} reads
		\begin{align}
			\mathcal{F}_{\mathrm{KT}} = -\bm{dH} + \bm{dQ} \, ,
		\end{align}
	where $\bm{dH}$ and $\bm{dQ}$ stand for the advection and diffusion part, respectively.
	Below, the terms $\bm{dH}$ and $\bm{dQ}$ are provided. 
	Note that they must have the same dimension as $\hat{\bm{u}}$: $\bm{dH}, \bm{dQ} \in \mathbb{R}^{\mathrm{dof} \times N_y \times N_x}$.

\subsubsection{The boundary conditions}
			
	First of all, we have to extend the dimension of $\hat{\bm{u}}$ to the dimension of the grid objects.
	For that, we have to choose some boundary conditions, meaning that we extend $\hat{\bm{u}} \in \mathbb{R}^{\mathrm{dof} \times N_y \times N_x}$ to $\bm{u} \in \mathbb{R}^{\mathrm{dof} \times \bar N_y \times \bar N_x}$ by the boundary layers where the boundary cells are filled according to the boundary conditions.

\subsubsection{The flux limiter}
	
	As already explained in \cref{sec:KT-central-scheme}, we have to reconstruct the values of the fluid fields at the interfaces of each cell.
	To this end, one has to estimate a gradient/slope for each fluid field in both directions in every cell.
	However, since we only have access to cell averages, the estimate can only be based on those.
	Using a flux limiter\footnote{For example, the MinMod limiter is defined by $f_{\mathrm{MinMod}}(a, b) = \min(|a|, |b|)$ if $a\cdot b > 0$ else $f_{\mathrm{MinMod}}(a, b) = 0$, see \cref{eq:minmod_flux_limiter}.}, $f_{\mathrm{limiter}}$, the slopes are given by\footnote{The division as well as the evaluation of the function $f_{\mathrm{limiter}}$ are meant componentwise.}
		\begin{subequations}
			\begin{align}
				\bm{f_x} = \, & f_{\mathrm{limiter}} \bigg( \frac{\Delta^{x}_{1, 0} ( \bm{u} )}{\Delta^{x}_{1, 0} ( \bm{x}_C )}, \frac{\Delta^{x}_{0, - 1} ( \bm{u} )}{\Delta^{x}_{0, - 1} ( \bm{x}_C )} \bigg) \, ,	\Vdistance	\label{eq:limited-slope-x}
				\\
				\bm{f_y} = \, & f_{\mathrm{limiter}} \bigg( \frac{\Delta^{y}_{1, 0} ( \bm{u} )}{\Delta^{y}_{1, 0} ( \bm{y}_C )}, \frac{\Delta^{y}_{0, - 1} ( \bm{u} )}{\Delta^{y}_{0, - 1} ( \bm{y}_C )} \bigg) \, ,	\Vdistance	\label{eq:limited-slope-y}
			\end{align}
		\end{subequations}
	where we have used the difference operators:
		\begin{subequations}
			\begin{align}
				& \Delta^{x}_{i, j} ( \bm{A} ) =	\vdistance
				\\
				= \, & \bm{A} [ :, \mathcal{B} \! - \! 1 \slice  - \mathcal{B} \! + \! 1, \mathcal{B} \! - \! 1 \! + \! i \slice \mathcal{B} \! + \! 1 \! + \! i \! + \! N_x ] -	\vdistance	\nonumber
				\\
				& \quad - \bm{A} [ :, \mathcal{B} \! - \! 1 \slice - \mathcal{B} \! + \! 1, \mathcal{B} \! - \! 1 + \! j \slice \mathcal{B} \! + \! 1 \! + \! j \! + \! N_x ] \, ,	\vdistance	\nonumber
				\\
				& \Delta^{y}_{i, j} ( \bm{A} ) =	\vdistance
				\\
				= \, & \bm{A} [ :, \mathcal{B} \! - \! 1 \! + \! i \slice \mathcal{B} \! + \! 1 \! + \! i \! + \! N_y, \mathcal{B} \! - \! 1 \slice - \mathcal{B} \! + \! 1 ] -	\vdistance	\nonumber
				\\
				& \quad - \bm{A} [ :,  \mathcal{B} \! - \! 1 \! + \! j \slice \mathcal{B} \! + \! 1 \! + \! j \! + \! N_y, \mathcal{B} \! - \! 1 \slice - \mathcal{B} \! + \! 1 ] \, ,	\vdistance	\nonumber
			\end{align}
		\end{subequations}
	with $- \mathcal{B} \leq i, j \leq \mathcal{B}$ such that $\Delta^{x/y}_{i, j} ( \bm{A} ) \in \mathbb{R}^{\mathrm{dof} \times ( N_x + 1 ) \times ( N_y + 1 )}$.
	Roughly speaking, the flux limiter compares the left and right gradients at each cell and returns the estimated gradient with which we set the value of the fluid fields at the interfaces.
	Hence, defining
		\begin{align}
			& \bm{u}_{\bm{x}/\bm{y}} =	\Vdistance
			\\
			= \, & \bm{f}_{\bm{x}/\bm{y}} \cdot \tfrac{1}{2} \, \bm{\Delta}_{x/y} [ :, \mathcal{B} \! - \! 1 \slice - \mathcal{B} \! + \! 1, \mathcal{B} \! - \! 1 \slice - \mathcal{B} \! + \! 1 ]\,,	\Vdistance	\nonumber
		\end{align}
	we find:
		\begin{subequations}
			\begin{align}
				\bm{u}_C = \, & \bm{u} [ :, \mathcal{B} \! - \! 1 \slice - \mathcal{B} \! + \! 1, \mathcal{B} \! - \! 1 \slice - \mathcal{B} \! + \! 1 ] \, ,	\vdistance
				\\
				\bm{u}_E = \, & \bm{u}_C + \bm{u_{x}} \, ,	\vdistance
				\\
				\bm{u}_W = \, & \bm{u}_C - \bm{u_{x}} \, ,	\vdistance
				\\
				\bm{u}_N = \, & \bm{u}_C + \bm{u_{y}} \, ,	\vdistance
				\\
				\bm{u}_S = \, & \bm{u}_C - \bm{u_{y}} \, ,	\vdistance
			\end{align}
		\end{subequations}
	such that $\bm{u}_C, \bm{u}_E, \bm{u}_W, \bm{u}_N, \bm{u}_S \in \mathbb{R}^{\mathrm{dof} \times ( N_x + 1 ) \times ( N_y + 1 )}$.\\
	
\subsubsection{The advection term}
			
	For the advection term we completely follow \reff\cite{KTO2-0} and apply the dimension-by-dimension approach for the reconstruction of the fluxes, where only information from two cells north/south and two cells east/west of a respective cell is used.
	For a higher-order reconstruction of fluxes like the second-order or third-order genuinely multidimensional central scheme, we refer the reader to \reffs\cite{Kurganov2002:2Dgas,Kurganov2001:thirdorder}.
	There, also information from the cells that are diagonally adjacent to the cell of interest is used.

	For the dimension-by-dimension approach we need to estimate the advection-velocities on the cell interfaces in both directions. 
	They read\footnote{Note that by comparing $\bm{u}_W [ :, :, 1 \! : ]$ and $\bm{u}_E [ :, :, : \! - 1 ]$, the locations of the interfaces are identical, \ie, $\bm{x}_W [ :, :, 1 \! : ] = \bm{x}_E [ :, :, : \! - 1 ]$ and $\bm{y}_W [ :, :, 1 \! : ] = \bm{y}_E [ :, :, : \! - 1 ]$.
	The same is true for comparing the locations of $\bm{u}_S[:,1\!:,:]$ and $\bm{u}_N[:,:\!-1,:]$.}
		\begin{subequations}
			\begin{align}
				\bm{a}_x = \, & \max \big( \hat{\rho} ( t, \bm{u}_W [ :, :, 1 \! : ] ), \hat{\rho} ( t, \bm{u}_E[ :, :, : \! - 1 ] ) \big) \, ,	\vdistance
				\\
				\bm{a}_y = \, & \max \big( \hat{\rho} ( t, \bm{u}_S [ :, 1 \! :, : ] ), \hat{\rho} ( t, \bm{u}_N[ :, : \! - 1, : ] ) \big) \, .	\vdistance
			\end{align}
		\end{subequations}
	Here, we understand the function $\hat{\rho}$ componentwise in the $x$- and $y$-direction but not in the ``$\mathrm{dof}$-direction''.
	The latter direction is given by $\hat{\rho}(t, u_0, \dots, u_{\mathrm{dof}-1}) = (\lambda_{\max}, \ldots, \lambda_{\max})$, where  $\lambda_{\max}$ is the spectral radius of $\frac{\partial \vec{f}}{\partial \vec{u}}$. It is determined by $\lambda_{\max} = \max \{ | \lambda_1, \ldots, \lambda_\omega |\}$, where $\lambda_k$ are the eigenvalues of $\frac{\partial \vec{f}}{\partial \vec{u}}$ at $(u_0, \dots, u_{\mathrm{dof}-1})$.
	
	With the estimates of the advection-velocities on the cell interfaces at hand, we finally find the advection fluxes 
			\begin{subequations}
				\begin{align}
					\bm{H^x} = \, & \tfrac{1}{2} \, \Big( f^x ( t, \bm{u}_W [ :, :, 1 \! : ] ) + f^x ( t, \bm{u}_E [ :, :, : \! - 1 ] )\Big) + \Vdistance\nonumber
					\\
					&- \tfrac{1}{2} \, \bm{a}_x \Big( \bm{u}_W [ :, :, 1 \! : ] - \bm{u}_E [ :, :, : \! - 1 ] \Big) \, ,	\Vdistance
					\\
					\bm{H^y} = \, & \tfrac{1}{2} \, \Big( f^y ( t, \bm{u}_S [ :, :, 1 \! : ] ) + f^y ( t, \bm{u}_N [ :, :, : \! - 1 ] ) \Big) + \Vdistance\nonumber 
					\\
					&- \tfrac{1}{2} \, \bm{a}_y \Big( \bm{u}_S [ :, :, 1 \! : ] - \bm{u}_N [ :, :, : \! - 1 ] \Big) \, ,	\Vdistance
				\end{align}
			\end{subequations}
			Eventually, we have for the advection contribution of the \gls{rhs} of \cref{eq:semi_discrete_pd_kt_scheme_matrix_formulation}
			\begin{align}
				\bm{dH} = \, & \frac{\bm{H^x} [ :, 1 \! : \! - 1, 1 \! : ] - \bm{H^x} [ :, 1 \! : \! - 1, : \! - 1 ]}{\bm{\Delta}_x [ :, \mathcal{B} \! : \! - \mathcal{B}, \mathcal{B} \! : \! - \mathcal{B} ]} + \Vdistance
				\\
				&+ \frac{\bm{H^y} [ :,  1 \! :, 1 \! : \! - 1 ] - \bm{H^y} [ :, : \! - 1, 1 \! : \! - 1 ]}{\bm{\Delta}_y [ :, \mathcal{B} \! : \! - \mathcal{B}, \mathcal{B} \! : \! - \mathcal{B} ]} \, . \Vdistance \nonumber
			\end{align}

\subsubsection{The diffusion term}

	In the original paper of the \gls{kt} scheme \cite{KTO2-0}, the two-dimensional implementation of the diffusion fluxes makes use of the limited slopes/gradients $\bm{f_x}$ and $\bm{f_y}$ from \cref{eq:limited-slope-x,eq:limited-slope-y}.
	Hence, the diffusion fluxes are specified as follows
		\begin{widetext}
			\begin{subequations}
				\begin{align}
					\bm{P^x} = \, & \tfrac{1}{2} \Big(
					Q^x ( t, \bm{u}_C [ :, :, : \! - 1 ], \bm{d_x u} [ :, :, : \! - 1 ], \bm{f_y} [ :, :, : \! - 1 ] ) + Q^x ( t, \bm{u}_C [ :, :, 1 \! : ], \bm{d_x u} [ :, :, : \! - 1 ], \bm{f_y} [ :, :, 1 \! : ] )\Big) \, ,	\Vdistance	\label{eq:original-diffusion-flux-x}
					\\
					\bm{P^y} = \, & \tfrac{1}{2} \Big( Q^y ( t, \bm{u}_C [ :, : \! - 1, : ], \bm{f_x} [ :, : \! - 1, : ], \bm{d_y u} [ :, : \! - 1, : ] ) + Q^y ( t, \bm{u}_C [ :, 1 \! :, : ], \bm{f_x} [ :, 1 \! :, : ], \bm{d_y u} [ :, : \! - 1, : ] ) \Big) \, ,	\Vdistance	\label{eq:original-diffusion-flux-y}
				\end{align}
			\end{subequations}
		\end{widetext}
	where we used the abbreviates
		\begin{align}
			&	\bm{d_x u} = \frac{\Delta^{x}_{1, 0} ( \bm{u} )}{\Delta^{x}_{1, 0} ( \bm{x}_C )} \, ,
			&&	\bm{d_y u} = \frac{\Delta^{y}_{1, 0} ( \bm{u} )}{\Delta^{y}_{1, 0} ( \bm{y}_C )} \, .
		\end{align}
	As we have already discussed in \cref{sec:KT-central-scheme}, this implementation leads to stability issues and a wrong error scaling behavior for certain systems, see \cref{rel_error_o2_sym}.
	We therefore tested several minor modifications of the original implementation and suggest the following solution:
	Instead of using the limited slopes $\bm{f_x}$ and $\bm{f_y}$ in \cref{eq:original-diffusion-flux-x} and \cref{eq:original-diffusion-flux-y}, we simply approximate these contributions by central difference stencils
		\begin{align}
			&	\bm{d^c_x u} = \frac{\Delta^{x}_{1, - 1} ( \bm{u} )}{\Delta^{x}_{1, - 1} ( \bm{x}_C )} \, ,
			&&	\bm{d^c_y u} = \frac{\Delta^{y}_{1, - 1} ( \bm{u} )}{\Delta^{y}_{1, - 1} ( \bm{y}_C )} \, .	\label{eq:central-difference-stencils}
		\end{align}
	Note that these are gradients of the fluid orthogonal to the respective diffusion fluxes.
	Thus, by inserting \cref{eq:central-difference-stencils} into \cref{eq:original-diffusion-flux-x} and \cref{eq:original-diffusion-flux-y}, we obtain the following diffusion fluxes:
		\begin{widetext}
			\begin{subequations}
				\begin{align}
					\bm{P^x} = \, & \tfrac{1}{2} \Big( Q^x ( t, \bm{u}_C [ :, :, : \! - 1 ], \bm{d_x u} [ :, :, : \! - 1 ], \bm{d^c_y u}[:, :, : \! - 1 ] ) + Q^x ( t, \bm{u}_C [ :, :, 1 \! : ], \bm{d_x u} [ :, :, : \! - 1 ], \bm{d^c_y u} [ :, :, 1 \! : ] ) \Big) \, ,	\Vdistance	\label{eq:correct-diffusion-flux-x}
					\\
					\bm{P^y} = \, & \tfrac{1}{2} \Big( Q^y ( t, \bm{u}_C [ :, : \! - 1, : ], \bm{d^c_x u} [ :, : \! - 1, : ], \bm{d_y u} [ :, : \! - 1, : ] ) + Q^y( t, \bm{u}_C [ :, 1 \! :, : ], \bm{d^c_x u} [ :, 1 \! :, : ], \bm{d_y u} [ :, : \! - 1, : ] ) \Big) \, .	\label{eq:correct-diffusion-flux-y}
				\end{align}
			\end{subequations}
		\end{widetext}
	Eventually, combining \cref{eq:correct-diffusion-flux-x} and \cref{eq:correct-diffusion-flux-y}, the total diffusion contribution of the \gls{rhs} in \cref{eq:semi_discrete_pd_kt_scheme_matrix_formulation} reads
		\begin{align}
			\bm{dQ} = \, & \frac{\bm{P^x} [ :, 1 \! : \! - 1, 1 \! : ] - \bm{P^x} [ :, 1 \! : \! - 1, : \! - 1 ]}{\bm{\Delta}_x [ :, \mathcal{B} \! : \! - \mathcal{B}, \mathcal{B} \! : \! - \mathcal{B} ] } +	\Vdistance
			\\
			& \quad + \frac{\bm{P^y} [ :, 1 \! :, 1 \! : \! - 1 ] - \bm{P^y} [ :, : \! - 1, 1 \! : \! - 1 ]}{\bm{\Delta}_y [ :, \mathcal{B} \! : \! - \mathcal{B}, \mathcal{B} \! : \! - \mathcal{B} ]} \, .	\Vdistance	\nonumber
		\end{align}
	
	\bibliography{bib/general,bib/gn,bib/inhomo,bib/instanton,bib/lattice,bib/math,bib/numerics,bib/qcd,bib/rg,bib/software,bib/symmetries,bib/thermal_qft,bib/thies,bib/virasoro_algebra,bib/zero-dim-qft}

\end{document}